\pdfoutput=1
\documentclass[aps,prd,twocolumn,showpacs,eqsecnum,nofootinbib]{revtex4-1}

\usepackage{graphicx}
\usepackage{amsmath}
\usepackage{amssymb}
\usepackage{bm}
\usepackage{hyperref}

\DeclareMathOperator{\erfc}{erfc}
\DeclareMathOperator{\sinc}{sinc}
\DeclareMathOperator{\trace}{tr}

\newcommand{\avg}[2][]{#1\langle #2 #1\rangle}
\newcommand{\chisqr}{{\chi^2}}
\newcommand{\normal}{{\mathcal{N}}}
\newcommand{\rhob}{\bar{\rho}}
\newcommand{\rhoh}{\hat{\rho}}
\newcommand{\rhot}{\tilde{\rho}}
\newcommand{\ua}{_{\mathrm{fa}}}
\newcommand{\ud}{_{\mathrm{fd}}}
\newcommand{\us}{_{\mathrm{s}}}
\renewcommand{\vec}[1]{\bm{#1}}
\newcommand{\uvec}[1]{\vec{\hat{#1}}}
\newcommand{\mat}[1]{\bm{\mathsf{#1}}}
\newcommand{\ut}{_{\mathrm{t}}}
\newcommand{\up}{_{+}}
\newcommand{\ux}{_{\times}}
\newcommand{\dd}{\mathrm{d}}
\newcommand{\upp}{_{\perp}}
\newcommand{\ull}{_{\parallel}}
\newcommand{\Upr}{^{\prime}}
\newcommand{\Uprpr}{^{\prime\prime}}
\newcommand{\F}{\mathcal{F}}
\newcommand{\umax}{_{\mathrm{max}}}
\newcommand{\utrial}{_{\mathrm{trial}}}
\newcommand{\udinj}{_{\mathrm{fd,inj}}}
\newcommand{\uiso}{_{\mathrm{iso}}}
\newcommand{\umism}{_{\mathrm{mism}}}
\newcommand{\uisomism}{_{\mathrm{iso,mism}}}

\begin{document}

\title{Estimating the sensitivity of wide-parameter-space searches for \protect\\ gravitational-wave pulsars}
\author{Karl Wette}
\affiliation{%
Max-Planck-Institut f\"ur Gravitationsphysik (Albert-Einstein-Institut), Callinstra\ss e 38, D-30167 Hannover, Germany%
}
\email{karl.wette@aei.mpg.de}

\begin{abstract}
This paper presents an in-depth study of how to estimate the sensitivity of searches for gravitational-wave pulsars -- rapidly-rotating neutron stars which emit quasi-sinusoidal gravitational waves.
It is particularly concerned with searches over a wide range of possible source parameters, such as searches over the entire sky and broad frequency bands.
Traditional approaches to estimating the sensitivity of such searches use either computationally-expensive Monte Carlo simulations, or analytic methods which sacrifice accuracy by making an unphysical assumption about the population of sources being searched for.
This paper develops a new, analytic method of estimating search sensitivity which does not rely upon this unphysical assumption.
Unlike previous analytic methods, the new method accurately predicts the sensitivity obtained using Monte Carlo simulations, while avoiding their computational expense.
The change in estimated sensitivity due to properties of the search template bank, and the geographic configuration of the gravitational wave detector network, are also investigated.
\end{abstract}


\maketitle

\section{Introduction}\label{sec:intro}

The last decade saw the successful construction and operation of the first generation of large-scale gravitational-wave interferometric detectors, at the observatories of LIGO \cite{Abbott.etal.2009e} in the United States, and of Virgo \cite{Accadia.etal.2011} and GEO~600 \cite{Grote.etal.2010} in Europe.
Ongoing improvements to the sensitivities of these detectors (e.g.\ \cite{Harry.etal.2010}), combined with the construction of new large-scale gravitational-wave interferometers in Japan \cite{Kuroda.etal.2010} and potentially elsewhere, are widely anticipated to result in the first ground-based detection of gravitational waves within the next decade.
From these first detections will follow new tests of the fundamental physics of gravity, as well as new tools for exploring the astrophysics of compact objects.

Gravitational-wave pulsars are one class of sources which may be detected by the next generation of gravitational-wave interferometers.
They are rapidly-rotating neutron stars which emit long-lived, narrow-band, quasi-sinusoidal gravitational waves, and are often referred to as \emph{continuous} or \emph{periodic} sources.
The dominant gravitational-wave emission is expected to be due to non-axisymmetric distortions of the neutron star; other possible emission mechanisms are unstable oscillation modes such as $r$-modes, and free precession due to misaligned rotation and symmetry axes; see \cite{Prix.2009} for a review.
An isolated neutron star may have acquired a non-axisymmetric deformation during birth in a supernova, although it is uncertain for how long such a deformation might be retained.
For neutron stars in binary systems, non-axisymmetry may result from e.g.\ differential heating from accreted matter leading to differential density gradients \cite{Bildsten.1998}, or from the confinement of the accreted matter by the star's magnetic field in a magnetic mountain \cite{Melatos.Payne.2005,Vigelius.Melatos.2009}.

Signals from gravitational-wave pulsars are being actively searched for in data from the LIGO and Virgo detectors.
Although none have been found to date, many of these searches have reached sensitivities where (albeit under optimistic conditions) such signals could be detected.
The most recent search for gravitational waves from known radio- and X-ray pulsars \cite{Abbott.etal.2010} constrained the gravitational-wave power radiated by the Crab pulsar to be less than 2\% of the total power available from the loss of rotational energy.
A search for gravitational waves from the Vela pulsar \cite{Abadie.etal.2011} set energy constraints of 35--45\%, depending on assumptions about Vela's orientation.
Upper limits on the amplitude of gravitational waves from the neutron star in the supernova remnant Cassiopeia~A, set in \cite{Abadie.etal.2010a}, are below the level expected (optimistically) from the total conversion of rotational energy into gravitational waves, assuming that the (as-yet unknown) rotation period of the neutron star is within the searched frequency band.
Recent searches for undiscovered neutron stars which may be radiating gravitational waves \cite{Abbott.etal.2009b,Abadie.etal.2011d} have set gravitational-wave amplitude limits comparable to upper limits hypothesized for a population of such stars \cite{Knispel.Allen.2008}.
Searches for gravitational-wave pulsars in binary systems have so far focused on the most promising known target, the low-mass X-ray binary Scorpius X-1 \cite{Abbott.etal.2007c,Abbott.etal.2007d}.

To best assess the prospects of future searches for gravitational-wave pulsars, it is important to be able to accurately estimate the sensitivity such searches can achieve.
(What is meant here by \emph{sensitivity} is defined in Section~\ref{sec:gwave-sens}.)
In particular, designing searches for continuous gravitational waves which cover a wide range of possible signal parameters (e.g.\ searches for undiscovered neutron stars) commonly requires constructing a hierarchical pipeline comprised of different data analysis techniques, each with different trade-offs, such as better sensitivity but increased computational cost, or vice versa.
An accurate estimate of the overall sensitivity of such a pipeline is important, therefore, for identifying the optimal combination of its elements.

Obtaining the sensitivity of a search targeting a \emph{single} source, such as a known pulsar, is relatively straightforward (e.g.\ \cite{Prix.2009}).
The calculation becomes more difficult, however, for searches over wide signal parameter spaces, e.g.\ searches for undiscovered neutron stars, or searches targeting known neutron stars with unknown rotation periods.
This difficulty has resulted in two different approaches to sensitivity estimation.
Wide-parameter-space searches of LIGO and Virgo data \cite{%
Abbott.etal.2004c,Abbott.etal.2005a,Abbott.etal.2007c,Abbott.etal.2008,Abbott.etal.2009,Abbott.etal.2009d,%
Abbott.etal.2009b,Abadie.etal.2010a,Abadie.etal.2011d%
} set upper limits on gravitational-wave amplitude (which in turn characterize the sensitivity of the search) by performing Monte Carlo simulations, where the search is re-performed on computer-generated data containing simulated signals (see Section~\ref{sec:gwave-sens}).
While Monte Carlo simulations are appropriate for accurately computing the sensitivity of searches of real gravitational-wave detector data, which often contain e.g.\ non-Gaussian instrumental noise artifacts, they are usually too computationally expensive to be useful for theoretical studies of the sensitivities of different data analysis techniques.

Instead, theoretical studies, e.g.\ \cite{%
Jaranowski.etal.1998,Brady.Creighton.2000,Jaranowski.Krolak.2000,Krishnan.etal.2004,Cutler.etal.2005,Mendell.Landry.2005,%
Krishnan.Sintes.2007,Watts.etal.2008,Prix.Shaltev.2011%
}, commonly make certain assumptions about the distribution of the gravitational-wave signals being searched for, in order to simplify the sensitivity calculation (see Section~\ref{sec:estim-csnr}).
These assumptions, however, result in a measure of sensitivity that, as shown in Section~\ref{sec:estim-iso}, is quantitatively different from that arrived at using Monte Carlo simulations.
To date, there has been little published work in the gravitational-wave literature on the discrepancy between these two approaches.
Furthermore, an accurate, computationally cheap (i.e.\ suitable for theoretical studies) estimator of sensitivity, as obtained by Monte Carlo simulations, has yet to be proposed (although combinations of numerical and analytic sensitivity estimation methods have been developed; see \cite{Betzwieser.2007,Wette.2009}).
It is these two issues that the present work intends to address.

Section~\ref{sec:gwave} of this paper presents an overview of gravitational-wave pulsar searches, and describes the most common method by which their sensitivities are estimated.
Section~\ref{sec:estim} contains the main result of this paper: an analytic expression which may be used to quickly and accurately estimate the sensitivity of wide-parameter-space searches for gravitational-wave pulsars.
Section~\ref{sec:accur} verifies the accuracy of the analytic sensitivity estimator, and Section~\ref{sec:assum} assesses the validity of some assumptions that were made during its derivation.
Section~\ref{sec:discussion} discusses the results presented in this paper, as well as possible avenues for future research.

\section{Gravitational-wave pulsar searches}\label{sec:gwave}

This section is an overview of the signal model of gravitational-wave pulsars (Section~\ref{sec:gwave-signal}), the data analysis techniques used to search for them (Section~\ref{sec:gwave-search}), and the method by which search sensitivities are estimated (Section~\ref{sec:gwave-sens}).
See also \cite{Prix.2009} for a review of gravitational-wave pulsar data analysis, and \cite{LSC.VC.2011} for an overview of the current data-analysis activities of the LIGO and Virgo scientific collaborations.

\subsection{Signal model}\label{sec:gwave-signal}

The signal from a gravitational-wave pulsar is written as a time series $h(t)$ of the dimensionless strain amplitude $h$, which for a ground-based interferometric detector is proportional to the differential change in the length of its arms (which are assumed to be much shorter than the gravitational wave-length).
We assume that the signal contains only a single frequency component, although it is possible for it to contain multiple frequency components arising from free precession \cite{VanDenBroeck.2005}.
Following \cite{Jaranowski.etal.1998}, the time series $h(t)$ may be written as the summed products of four amplitudes $A_i$ and four time-dependent functions $h_i(t)$:
\begin{equation}
h(t) = \sum_{i=1}^4 A_i h_i(t)
\end{equation}
The amplitudes $A_i$ are related to the four \emph{amplitude parameters} of the signal: its overall strain amplitude $h_0$; its initial phase $\phi_0$; the inclination angle $\iota$ between the neutron star angular momentum and wave propagation vectors; and the polarization angle $\psi$, which fixes the orientation of the neutron star about the wave propagation vector.
The functions $h_i(t)$ are functions of the signal's remaining \emph{phase parameters}: its sky position, given by its right ascension $\alpha$ and declination $\delta$; and its frequency evolution, given by an initial frequency $f$, and frequency time-derivatives or \emph{spindowns} $\dot{f}$, $\ddot{f}$, $f^{(3)}$, etc.
The number of spindowns required generally depends on the age of the sources being targeted \cite{Brady.etal.1998,Jaranowski.Krolak.1999,Wette.etal.2008}.

The time series $h(t)$ may also be written in a form which illustrates the two polarizations, \emph{plus} and \emph{cross}, of a gravitational wave:
\begin{equation}
h(t) = A\up F\up(t) \cos \Phi(t) + A\ux F\ux(t) \sin \Phi(t) \,,
\end{equation}
where $A\up$ and $A\ux$ are the amplitudes of their respective polarizations, and $\Phi(t)$ is the signal phase.
The \emph{antenna-pattern} functions $F\up(t)$ and $F\ux(t)$ give the response of the detector to each polarization, and are modulated by the sidereal motion of the Earth.
Expressions for $F\up(t)$ and $F\ux(t)$ are given in \cite{Schutz.Tinto.1987,Jaranowski.Krolak.1994,Bonazzola.Gourgoulhon.1996,Jaranowski.etal.1998,Srivastava.Sahay.2002}, and in Appendix~\ref{apx:antenna}.

The signal-to-noise ratio (SNR) of a signal, $\rho$, is found by integrating $h(t)$ over the observation time $T$, which gives:
\begin{equation}
\label{eq:signal-to-noise-ratio}
\rho^2 = \frac{ h_0^2 T }{ S_h } \big( a\up^2 \avg{F\up^2}_{t} + a\ux^2 \avg{F\ux^2}_{t} \big) \,,
\end{equation}
where $\avg{F\up^2}_{t} = (1/T) \int_{-T/2}^{T/2} \dd t \, F(t)^2$ is the time average of $F\up(t)^2$ (similarly for $\avg{F\ux^2}_{t}$), $a\up = A\up/h_0$ (similarly for $a\ux$), and $S_h$ is the one-sided detector noise power spectral density.
The SNR is independent of the signal's phase modulation; when $T$ is longer than several days, the SNR's dependence on the signal's amplitude modulation also vanishes, and $\rho^2$ becomes a linear function of $T$.
Expressions for $\avg{F\up^2}_{t}$ and $\avg{F\ux^2}_{t}$ are given in Appendix~\ref{apx:antenna}; see also \cite{Jaranowski.etal.1998} for expressions for $\rho^2$.

\subsection{Search techniques}\label{sec:gwave-search}

Gravitational-wave pulsars are very weak sources; the amplitude of their signals is likely to be several orders of magnitude smaller than the noise amplitude of current- and even next-generation interferometric detectors.
Nevertheless, gravitational-wave pulsar signals can be recovered using the well-known technique of \emph{matched filtering}, where the data are correlated against a template which models the signals' amplitude and phase evolution over time.
Matched filtering was first applied to the detection of gravitational-wave pulsars in \cite{Jaranowski.etal.1998}, and extended to multiple detectors in \cite{Cutler.Schutz.2005}.
The detection statistic derived in these papers, known as the $\F$-statistic, maximizes the signal SNR over the four amplitude parameters, but requires values to be chosen for the phase parameters.

A search for gravitational-wave pulsars therefore consists of performing matched filtering against a bank of templates, whose phase parameters are chosen to cover the parameter space of interest, e.g. over the whole sky and a broad range of frequencies for a search for unknown neutron stars.
It is almost certain, however, that any signal in the data will possess parameters which are different from any one of the searched templates; consequentially, no template will perfectly match the signal, and the signal SNR will be degraded.
While some loss in SNR is unavoidable, template banks are constructed such that the fractional loss in SNR, also known as the \emph{mismatch}, can never be greater than some prescribed maximum.
To accomplish this, a \emph{metric} is often used to determine how closely the templates must be spaced in each parameter \cite{Brady.etal.1998,Prix.2007}.
How to construct a bank which minimizes the number of templates is known in theory \cite{Prix.2007a}, but is often difficult to accomplish in practice.

Unfortunately, the number of templates which must be matched filtered increases rapidly with the length of data being analyzed.
If $T$ denotes the time-span of the analyzed data, the number of templates which must be placed in each parameter dimension scales as follows: $T^{\sim 2}$ for the sky position $(\alpha,\delta)$, $T$ for frequency, and $T^{k+1}$ for each spindown parameter $f^{(k)}$.
For example, the number of templates for an all-sky search requiring one spindown scales as $T^{\sim 5}$.
In contrast, the sensitivity achievable by matched filtering increases only as $T^{1/2}$ (assuming that the data are contiguous in time).
Due to the rapid increase in the number of required templates, matched filtering quickly becomes too computationally expensive for searching long data sets and large parameter spaces \cite{Jaranowski.Krolak.2000}.

The solution is to resort to a hierarchical pipeline, where typically the data are broken into short segments, each of which are matched filtered separately.
The results from each segment are then combined using \emph{semi-coherent} analysis methods, which resemble matched filtering but do not require full amplitude and phase consistency of the signal template between data segments: typically, only the derivative of the phase (i.e.\ the frequency) must be consistent.
(To distinguish it from semi-coherent methods, matched filtering is also referred to as \emph{coherent} matched filtering.)
For fixed $T$, and assuming no limits on computational cost, semi-coherent methods are less sensitive than coherent matched filtering: their sensitivity scales roughly as $N\us^{1/4} T\us^{1/2}$, where $N\us$ and $T\us$ are the number and time-span of each segment, while a coherent search (assuming contiguous data) scales as $T^{1/2} = N\us^{1/2} T\us^{1/2}$.
On the other hand, wide-parameter-space gravitational-wave pulsar searches are almost always computationally limited, due to the large parameter spaces which must be searched.
The number of templates, and hence the computational cost, of semi-coherent methods scale with $T\us \ll T$, instead of $T$, making them computationally cheaper than a fully-coherent search.
This in turn permits a semi-coherent search to use more data, improving its sensitivity, while remaining computationally tractable.
Some hierarchical searches, e.g.\ \cite{Abbott.etal.2008,Abbott.etal.2009d}, use $T\us \sim 30$-minute segments, so that the coherent matched-filtering step closely resembles the computation of a power spectrum; other searches, e.g. \cite{Abbott.etal.2009,Abbott.etal.2009b} use segments of $T\us \sim 1$~day.

Examples of semi-coherent methods are the StackSlide \cite{Brady.Creighton.2000,Mendell.Landry.2005,Cutler.etal.2005}, Hough \cite{Krishnan.etal.2004,Krishnan.Sintes.2007}, PowerFlux \cite{Dergachev.2010a}, cross-correlation \cite{Dhurandhar.etal.2008}, and global correlation transform \cite{Pletsch.Allen.2009} methods.
Recently, methods which blend together aspects of semi- and fully-coherent methods have been developed \cite{Dergachev.2010,Pletsch.2011,Cutler.2011}.
The problem of how to optimise hierarchical searches is studied in \cite{Cutler.etal.2005,Prix.Shaltev.2011}.

\subsection{Search sensitivity}\label{sec:gwave-sens}

The sensitivity of a wide-parameter-space search for gravitational-wave pulsars has traditionally been characterized by the method presented in this section; two alternative methods of estimating sensitivity are discussed in Section~\ref{sec:discussion}.

The method follows from the canonical framework for statistical hypothesis testing formulated in \cite{Neyman.Pearson.1933,Neyman.1937}.
It is commonly referred to as the \emph{frequentist method} in the gravitational-wave literature.
In essence, the method provides the answer to the following question: if there were a population of gravitational-wave pulsar signals present in the searched data, each with the same amplitude $h_0$, how large would $h_0$ need to be before we would be confident of detecting a very large fraction of them, e.g. 95\%?

In order to make the above question more precise, we must first define what is meant here by \emph{detection}.
This is complicated by the unavoidable fact that the output of any real gravitational-wave detector contains noise, in addition to any signal.
When noisy data is analyzed, e.g.\ using matched filtering, the results of the analysis may be subject to \emph{false alarms} and \emph{false dismissals}.
A false alarm is when the results of the analysis falsely indicate the presence of a signal, which is instead simply due to spurious noise fluctuations;
a false dismissal is when a real signal present in the data is sufficiently corrupted by noise that it fails to be identified as a signal in the analysis results.
We define a \emph{detection statistic} to be a number quantifying a single search result, e.g. a correlation of the data against a single signal template.
Typically, the detection statistic increases with the probability that the data contains a signal which matches the template.
As an example, the $\F$-statistic represents the result of a matched-filtering analysis maximized over signal amplitude parameters.
We say that a signal has been \emph{detected}\footnote{%
It is important to note that, when performing a search of real gravitational-wave detector data, any signal thus detected would never be automatically claimed as a genuine gravitational-wave signal; extensive follow-up investigations would first be performed to e.g.\ consider possible contamination by instrumental artifacts.}
if the value of its detection statistic $s$ exceeds some threshold $s\ua$\footnote{%
The hierarchical searches in \cite{Abbott.etal.2009,Abbott.etal.2009b} employ a slightly different definition of \emph{detection}: the detection statistic must exceed a threshold in a given number of data segments.}.

Suppose that we perform a wide-parameter-space search of data which is known to contain no signal, e.g. the data may comprise only computer-generated Gaussian noise.
Because the data used to compute the detection statistic are combined differently for each template, the values of $s$ returned by the search will not be identical, but will instead follow a certain probability distribution.
The probability of a detection in this data, i.e.\ the probability of a false alarm, denoted $p\ua$, is the probability of the detection statistic $s$ exceeding the threshold $s\ua$, under the assumption that no signal is present in the data:
\begin{equation}
\label{eq:false-alarm-prob}
p\ua = p \big( s > s\ua \,\big|\, \text{no signal in data} \big) \,,
\end{equation}
where $p(A|B)$ denotes the probability of the statement $A$ being true, given that we already know that $B$ is true.

We now perform the same search using data which is known to contain a single signal, e.g. a simulated signal added to computer-generated Gaussian noise.
Assuming that the majority of the searched templates are insensitive to the signal (which would be true of a properly constructed template bank), the majority of the returned values of $s$ will follow the same probability distribution as before.
For templates whose parameters are close to those of the signal, however, the values of $s$ will follow a different probability distribution, which instead assumes that a signal is present.
Assume that the SNR of the signal is large enough that these templates, close to the signal, are distinguishable from the remaining search templates; this assumption is satisfied by requiring a low false dismissal probability.
The probability of this particular signal not being detected, i.e.\ the probability of its false dismissal, denoted $p\ud(\vec p)$, is the probability of $s$, in the neighborhood of the signal, falling below the threshold $s\ua$, under the assumption that a signal with parameters $\vec p$ is present in the data:
\begin{multline}
\label{eq:false-dism-prob-p}
p\ud(\vec p) = p \big( s \le s\ua \,\big|\, \text{signal(param.\ $\vec p$) in data} \big) \,.
\end{multline}
Note that we assume here that the search template is perfectly matched to the signal, whereas in reality there will be some mismatch between them.
(The loss of sensitivity due to mismatch between template and signal is considered in Section~\ref{sec:assum-mism}.)
We repeat the search for a large number of signals, each with different $\vec p$ drawn from a chosen distribution of signals parameters, $p(\vec p)$.
The overall false dismissal probability, denoted $p\ud$, is the average false dismissal probability obtained from each of the sampled signals:
\begin{equation}
\label{eq:false-dism-prob}
p\ud = \avg[\Big]{p\ud(\vec p)}_{\vec p} = \idotsint \dd\vec p \, p(\vec p) \, p\ud(\vec p) \,.
\end{equation}
Equation~\eqref{eq:false-dism-prob} is computed using Monte Carlo integration, i.e.\ by the repeated computation of Eq.~\eqref{eq:false-dism-prob-p}, with different parameters $\vec p$ drawn from the distribution $p(\vec p)$.

We now return to the question stated at the beginning of this section: how large would the amplitude $h_0$ of a population of gravitational wave signals need to be in order for them to be detected, e.g.\ 95\% of the time?
First, Eq.~\eqref{eq:false-alarm-prob} is solved for the threshold $s\ua$ that would result in a desired false alarm probability $p\ua$.
Then, given a target false dismissal probability $p\ud$, Eq.~\eqref{eq:false-dism-prob} is solved for the overall strain amplitude $h_0$.
Suppose $h_0^{95\%}$ is the value of $h_0$ which solves Eq.~\eqref{eq:false-dism-prob} for $p\ud = 5\%$; then, if a population of signals have constant amplitudes $h_0^{95\%}$, then a fraction $1 - p\ud = 95\%$ of them will be detected.
The amplitude $h_0^{95\%}$ thus characterizes the amplitude of signals a particular search method is able to confidently detect, and hence gives a useful measure of the search's sensitivity.
Where a search finds no credible gravitational-wave signal, this sensitivity is interpreted as an upper limit, with confidence $1 - p\ud$, on the amplitude of signals present in the searched data.

The above procedure is used to set upper limits on the amplitude of gravitational waves for wide-parameter-space searches of LIGO and Virgo data (see Section~\ref{sec:intro} for references).
The search parameter space is typically partitioned into small frequency bands, and upper limits are set separately for each frequency band.
One important difference to the procedure described in this section is that $s\ua$ is determined by the largest value of $s$ returned by the search (after instrumental noise artifacts have been removed); an effective false alarm probability can then be determined from $s\ua$ using Eq.~\eqref{eq:false-alarm-prob}.

\section{Analytic sensitivity estimation}\label{sec:estim}

Theoretical studies of gravitational-wave pulsar search pipelines require a method of efficiently and accurately estimating the sensitivity achievable by such searches.
The procedure described in Section~\ref{sec:gwave-sens} is generally unsuited to this task, due to the computational expense of repeatedly generating and searching data containing simulated signals.
In this section, we construct an analytic expression which accurately reproduces sensitivity estimates computed using the frequentist method, using the procedure described in Section~\ref{sec:gwave-sens}, but is simpler to implement and computationally cheaper to calculate.
Expressions for the false alarm and false dismissal probabilities are presented in Sections~\ref{sec:estim-fap} and ~\ref{sec:estim-fdp} respectively.
Section~\ref{sec:estim-csnr} presents a commonly used, but inaccurate, analytic estimator of the sensitivity, and Section~\ref{sec:estim-iso} presents a new, more accurate expression.

We restrict our attention to detection statistics $s$ which follow $\chisqr$ distributions (e.g.\ \cite{Johnson.etal.1994}).
This implies that $s$ can be written as the sum of squares of some number of normally-distributed random variables.
Examples of such statistics are the StackSlide power \cite{Mendell.Landry.2005}, and the $\F$-statistic \cite{Jaranowski.etal.1998,Cutler.etal.2005}.
Our prototypical search is a single-stage hierarchical search of $N\us$ data segments, each of which span a time $T\us$.
(A fully-coherent search is then given by the special case $N\us = 1$.)
A coherent analysis is performed for each segment $i$, returning detection statistics $s_i$, which are then summed using a semi-coherent method to attain the final detection statistic $s = \sum_{i=1}^{N\us} s_i$.

In the absence of a signal, $s$ is distributed according to a central $\chisqr$ distribution, which takes a single parameter: the number of degrees of freedom of the statistic.
Since $s$ is the sum of the $N\us$ values $s_i$, its number of degrees of freedom is given by $N\us \nu$, where each of the $s_i$ has $\nu$ degrees of freedom.
A detection statistic derived from the power of a signal (e.g. the StackSlide power) has $\nu = 2$; the $\F$-statistic has $\nu = 4$.
We denote that $s$ is distributed according to a central $\chisqr$ distribution with $N\us \nu$ degrees of freedom by $s \sim \chisqr(N\us \nu, 0)$.

In the presence of a signal, $s$ is distributed according to a non-central $\chisqr$ distribution, which take two parameters: the number of degrees of freedom $N\us \nu$ as before, and the non-centrality parameter
\begin{equation}
\label{eq:signal-to-noise-ratio-Ns}
N\us \rho^2 = h_0^2 T \big( a\up^2 \avg{F\up^2}_{t} + a\ux^2 \avg{F\ux^2}_{t} \big) \sum_{i=1}^{N\us} \sum_{j=1}^{N_{\mathrm{det.}}} (S_h)_{ij}^{-1} \,.
\end{equation}
The non-centrality parameter is the accumulated SNR of a signal analyzed over $N\us$ data segments from $N_{\mathrm{det.}}$ detectors, assuming perfect match between signal and template.
The noise power spectral density of the $i$th data segment from the $j$th detector is labeled $(S_h)_{ij}$.
It is convenient to define $N\us S_h^{-1} = \sum_{i=1}^{N\us} \sum_{j=1}^{N_{\mathrm{det.}}} (S_h)_{ij}^{-1}$, whereupon we recover the expression for $\rho^2$ given in Eq.~\eqref{eq:signal-to-noise-ratio}.
For simplicity, therefore, we can ignore the summation of $(S_h)_{ij}$ over segments and detectors, and take $\rho^2$ to be given by Eq.~\eqref{eq:signal-to-noise-ratio} for some appropriate value of $S_h$.
For a real detector, $S_h$ also varies as a function of frequency; we therefore assume that we are considering the sensitivity of a search over a frequency band sufficiently narrow, so that $S_h$ can be assumed constant.

We denote that $s$ is distributed according to a non-central $\chisqr$ distribution with $N\us \nu$ degrees of freedom and non-centrality parameter $N\us \rho^2$ by $s \sim \chisqr(N\us \nu, N\us \rho^2)$.
Note that, as in Section~\ref{sec:gwave-sens}, we assume that the search template is perfectly matched to the signal, and delay considering the sensitivity lost due to template--signal mismatch until Section~\ref{sec:assum-mism}.

\subsection{False alarm probability}\label{sec:estim-fap}

We start by determining the threshold $s\ua$ on the detection statistic required to give a certain false alarm probability $p\ua$.

Assuming that no signal is present, the probability that a single value of the detection statistic $s$ falls below $s\ua$ is
$p \big( s \le s\ua \big| s \sim \chisqr[N\us \nu, 0] \big)$.
Suppose that the search returns $N\ut$ values of $s$, i.e. $N\ut$ templates\footnote{%
Note that we are referring to the number of templates searched in the semi-coherent stage of a hierarchical pipeline, which is distinct from the number of templates searched in each individual data segment; see e.g.\ \cite{Cutler.etal.2005}.}
are searched.
We assume that each value of $s$ is statistically independent, i.e.\ that the joint probability of obtaining any two values $s_1$ and $s_2$ is just the product of the probabilities of obtaining $s_1$ and $s_2$ individually.
(The validity of this assumption is examined in Section~\ref{sec:assum-statindp}.)
The probability that $N\ut$ values of $s$ fall below $s\ua$ is then given by $N\ut$ multiples of the single-value probability: $p(\cdots)^{N\ut}$.
Its complement, $1 - p(\cdots)^{N\ut}$, is the probability that in a search of $N\ut$ templates, one or more values of $s$ will be returned above the threshold, i.e.\ the probability of a false alarm, $p\ua$.
In short, we have
\begin{equation}
\label{eq:fap-1}
p\ua = 1 - \big[ p \big( s \le s\ua \big| s \sim \chisqr[N\us \nu, 0] \big) \big]^{N\ut} \,.
\end{equation}
Simple re-arrangement gives
\begin{equation}
\label{eq:fap-2}
( 1 - p\ua )^{1/N\ut} = p \big( s \le s\ua \big| s \sim \chisqr[N\us \nu, 0] \big) \,.
\end{equation}
Since typically either $N\ut = 1$ (e.g. for a search for a known pulsar) or $N\ut \gg 1$, the left-hand side of Eq.~\eqref{eq:fap-2} can be replaced with the first-order binomial expansion $1 - p\ua/N\ut$.
We now have
\begin{equation}
\label{eq:fap}
p\ua/N\ut = p \big( s > s\ua \big| s \sim \chisqr[N\us \nu, 0] \big) \,.
\end{equation}
Note that $p\ua/N\ut$ can be interpreted as the false alarm probability for a single template out of the $N\ut$ templates searched.

\begin{figure}
\includegraphics[width=\linewidth]{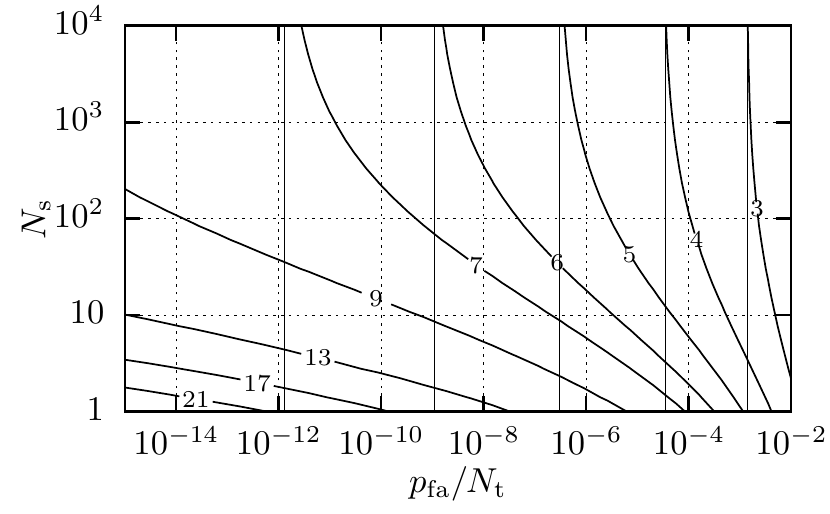}
\caption{\label{fig:za_thresh}
Normalized false alarm threshold $z\ua$, as a function of $p\ua/N\ut$ and $N\us$, with $\nu = 4$ (thick lines).
Its limiting values for large $N\us$, given by the normal distribution approximation, are plotted as thin vertical lines for $z\ua = 3$--7.
}
\end{figure}

We must now solve Eq.~\eqref{eq:fap} for $s\ua$, i.e.\ we must compute the inverse function of the central $\chisqr$ distribution.
For small values of $p\ua/N\ut$ (typically, $p\ua$ is chosen to be 1\%, and $N\ut \gg 1$), an analytic expression for $s\ua$ derived from \cite{Temme.1992} is given in Appendix~\ref{apx:invchisqr}.
We define the normalized false alarm threshold $z\ua$ to be the difference between $s\ua$ and the mean of a central $\chisqr$ distribution with $N\us \nu$ degrees of freedom, in units of the distribution's standard deviation:
\begin{equation}
\label{eq:za-threshold}
z\ua = \frac{ s\ua - N\us \nu }{ \sqrt{2 N\us \nu} } \,.
\end{equation}
In the limit of large $N\us$, $z\ua \approx \sqrt2 \erfc^{-1} (2 p\ua/N\ut)$ and is independent of $N\us$ (see Appendix~\ref{apx:invchisqr}).
Note that this limiting value for $z\ua$ is identical to the expression obtained if the central $\chisqr$ distribution is approximated by a normal distribution, as is commonly done (e.g.\ in \cite{Krishnan.etal.2004}).
Figure~\ref{fig:za_thresh} plots $z\ua$ as a function of $p\ua/N\ut$ and $N\us$, and illustrates its convergence to the normal distribution approximation for large $N\us$.

\subsection{False dismissal probability}\label{sec:estim-fdp}

Having determined the threshold $s\ua$ appropriate for a desired false alarm probability $p\ua$, we now attempt to solve Eq.~\eqref{eq:false-dism-prob} for some quantity which usefully characterizes the sensitivity of the search.
The quantity most often used for this purpose has been the dimensionless expression $h_0 \sqrt{T\us / S_h}$, which is sometimes referred to as the \emph{statistical factor} \cite{Watts.etal.2008,Wette.etal.2008}.
The statistical factor quantifies the sensitivity of a search to a population of signals of amplitude $h_0$, relative to the performance of the detector given by its noise power spectral density $S_h$, and the length of (coherently) analyzed data $T\us$.

In this paper, we propose instead to use the root-mean-square SNR, $\sqrt{\avg{\rho^2}}$, to quantify sensitivity.
Here, $\avg{}$ denotes averaging over $\alpha$, $\sin\delta$, $\psi$ [see Eq.~\eqref{eq:sky-avg-antenna-patt}], and $\xi \equiv \cos\iota$, given by $\avg{\rho^2}_{\xi} = (1/2) \int_{-1}^{1} \dd\xi \, \rho^2$.
This choice of averaging implies that the population of gravitational-wave pulsars being searched for are isotropically distributed over the sky, and are isotropically oriented.
Using Eq.~\eqref{eq:signal-to-noise-ratio}, we have
\begin{multline}
\avg{\rho^2} = \frac{ h_0^2 T\us }{ S_h } \big( \avg{a\up^2}_{\xi} \avg{F\up^2}_{\alpha,\sin\delta,\psi,t} \\
+ \avg{a\ux^2}_{\xi} \avg{F\ux^2}_{\alpha,\sin\delta,\psi,t} \big) \,. \!
\end{multline}
The averages of $F\up^2$ and $F\ux^2$ over $\alpha$, $\sin\delta$, and $\psi$ are given by Eq.~\eqref{eq:skydet-avg-antenna-patt} (where we assume $\zeta = \pi/2$); after this averaging $F\up^2$ and $F\ux^2$ are independent of time.
We assume a signal generated by a non-axisymmetrically deformed neutron star, for which (e.g. \cite{Jaranowski.etal.1998,VanDenBroeck.2005}):
\begin{subequations}
\begin{align}
\label{eqs:nonax-signal-amplitudes}
a\up &= \frac{1 + \xi^2}{2} \,,& \avg{a\up^2}_{\xi} &= \frac{7}{15} \,; \\
a\ux &= \xi \,,                & \avg{a\ux^2}_{\xi} &= \frac{1}{3} \,.
\end{align}
\end{subequations}
Finally we have
\begin{equation}
\label{eq:sqrtavgrho-statfac}
\sqrt{\avg{\rho^2}} = \frac{2}{5} h_0 \sqrt{\frac{ T\us }{ S_h }} \,,
\end{equation}
i.e. $\sqrt{\avg{\rho^2}}$ is directly proportional to the statistical factor.
Unlike the statistical factor, however, $\sqrt{\avg{\rho^2}}$ relates directly to a property of the population of signals being searched for (i.e. their mean SNR), and hence is a more directly physical quantity.
It also has a clearer interpretation as a measure of sensitivity: for example, to improve the sensitivity of a search, we must make the search able to detect signals with weaker SNR (at the same false alarm and dismissal probabilities), and hence we must lower the mean SNR, i.e. $\sqrt{\avg{\rho^2}}$, of the population of signals which the search can detect.
It is convenient to write $\rho$ in terms of $\sqrt{\avg{\rho^2}}$ and a factor $R$, defined such that
$\rho = \sqrt{\avg{\rho^2}} R$, which implies $\avg{R^2} = 1$.

\begin{figure}
\includegraphics[width=\linewidth]{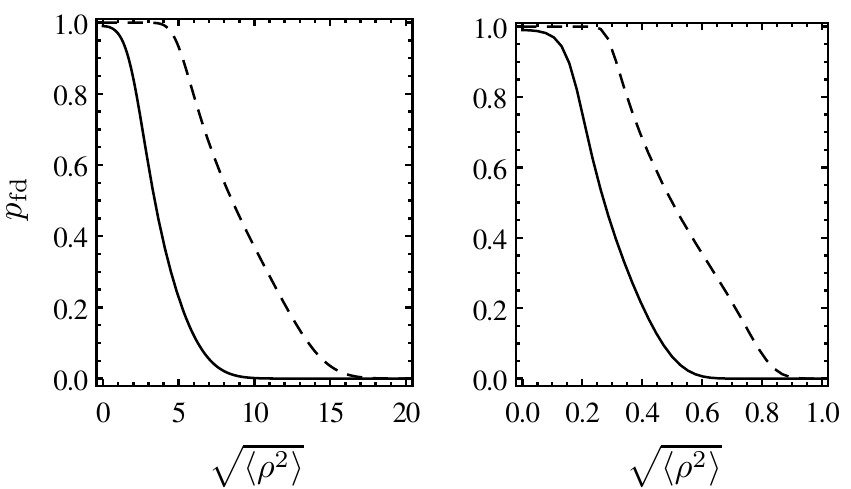}
\caption{\label{fig:false_dism_dist}
False dismissal probability $p\ud$ as a function of $\sqrt{\avg{\rho^2}}$, given by Eq.~\eqref{eq:fdp}, with $\nu = 4$, and for $p\ua/N\ut = 0.01$ (solid) and $10^{-12}$ (dashed), and $N\us = 1$ (left) and $10^4$ (right).
}
\end{figure}

Assuming that a signal with parameters $\vec p$ is present, the probability that the detection statistic $s$ (in the neighborhood of the signal, as discussed in Section~\ref{sec:gwave-sens}) falls below $s\ua$ is $p\ud(\vec p) = p \big( s \le s\ua \big| s \sim \chisqr[N\us \nu, N\us \avg{\rho^2} R^2] \big)$.
Since $h_0$, $S_h$, and $T\us$ are taken to be constants, $\avg{\rho^2}$ is also a constant.
The only quantity which depends on the signals parameters is therefore $R$, which is a function of $\vec p = (\alpha, \delta, \psi, \xi)$.
By Eq.~\eqref{eq:false-dism-prob}, the overall false dismissal probability $p\ud$ is
\begin{multline}
\label{eq:fdp}
p\ud = \avg[\Big]{ p \big( s \le s\ua \big| s \sim \chisqr[N\us \nu, \\
N\us \avg{\rho^2} R^2] \big) }_{\alpha,\sin\delta,\psi,\xi} \,.
\end{multline}
Equation~\eqref{eq:fdp} is plotted in Fig.~\ref{fig:false_dism_dist} for different choices of $p\ua/N\ut$ and $N\us$.

To proceed, we must now solve Eq.~\eqref{eq:fdp} for $\avg{\rho^2}$.
Unfortunately, an analytic solution is difficult to obtain, due to the complicated dependence of $R$ on the signal parameters.
Two approaches to solving Eq.~\eqref{eq:fdp} analytically are presented in Sections~\ref{sec:estim-csnr} and~\ref{sec:estim-iso}.

\subsection{Sensitivity to constant-SNR signal populations}\label{sec:estim-csnr}

To solve Eq.~\eqref{eq:fdp}, it is common to assume that every signal, in the population of signals being searched for, has the \emph{same} SNR, which we denote by $\rhob$.
An alternative interpretation of this approximation is that the population of signals can be replaced by a \emph{single} signal, whose SNR is $\rhob = \sqrt{\avg{\rho^2}}$.
While this assumption does not accurately model a physically reasonable population of gravitational-wave pulsar signals (see Section~\ref{sec:estim-iso}), it does allow Eq.~\eqref{eq:fdp} to be readily solved for $\rhob$: because the signals all have the same SNR, no averaging over signal parameters is required, and $\avg{\rho^2} R^2$ is simply replaced by $\rhob^2$.
The accuracy of this approximation is examined in Section~\ref{sec:estim-iso}.

Another common simplification is to approximate the non-central $\chisqr$ distribution by a normal distribution $\normal(\mu,\sigma)$ with the same mean and standard deviation, which are $\mu = N\us(\nu + \rhob^2)$ and $\sigma = \sqrt{ 2 N\us ( \nu + 2\rhob^2 ) }$ respectively.
This approximation introduces an error in $\rhob$ of $\lesssim 2.5\%$ (at $p\ua/N\ut = 0.01$, $N\us = 1$) which decreases with decreasing $p\ua/N\ut$ and increasing $N\us$.
With this approximation, Eq.~\eqref{eq:fdp} reduces to
\begin{align}
p\ud &\approx p \big( s \le s\ua \big| s \sim \normal[ \mu, \sigma ] \big) \\
\label{eq:fdp-rhob-1}
&= \frac{1}{2} \erfc \left( \frac{ N\us \rhob^2 - z\ua \sqrt{ 2 N\us \nu } }{ 2 \sqrt{ N\us ( \nu + 2\rhob^2 ) } } \right) \,,
\end{align}
where we have substituted $s\ua$ with the normalized threshold $z\ua$ defined by Eq.~\eqref{eq:za-threshold}, and $\erfc$ is the complementary error function.
The solution to this equation is
\begin{equation}
\rhob^2 = \sqrt{ \frac{2\nu}{N\us} } z\ua + \frac{ 2 q^2 }{ N\us }
\left[ 1 + ( 1 + \mathcal{Q} )^{\frac{1}{2}} \right] \,,
\end{equation}
where $\mathcal{Q} = ( N\us\nu + z\ua \sqrt{ 8 N\us \nu } )/( 2 q^2 )$, $q = \sqrt2 \erfc^{-1} 2p\ud$, and $\erfc^{-1}$ is the inverse complementary error function.
For $p\ua/N\ut \le 0.01$, $p\ud \ge 0.05$, $N\us \ge 1$, and $\nu \ge 2$, $\mathcal{Q} \gtrsim 5$, and it is reasonable to approximate $\sqrt{1 + \mathcal{Q}}$ with $\sqrt{\mathcal{Q}}$.
This permits further simplification to
\begin{equation}
\label{eq:rhobar}
\rhob = \left[ \frac{2\nu}{N\us} \right]^{\frac{1}{4}} \left[
z\ua \!+\! q \Bigg( 1 \!+\! \frac{ z\ua \sqrt8 }{ \sqrt{N\us\nu} } \Bigg)^{\frac{1}{2}} \!+\! \frac{ q^2 \sqrt2 }{ \sqrt{N\us\nu} }
\right]^{\frac{1}{2}} .
\end{equation}

\begin{figure}
\includegraphics[width=\linewidth]{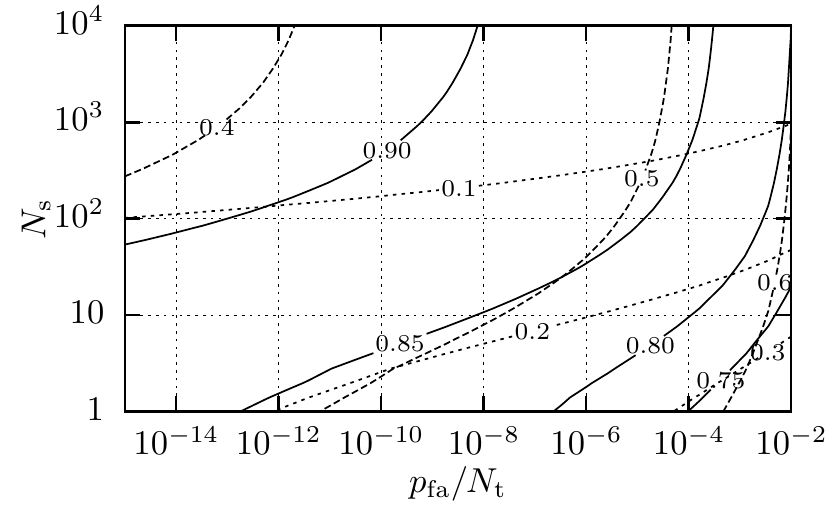}
\caption{\label{fig:rhob_terms_ratios}
Contours of contributions of the terms $\mathcal{T}$ (see the text) to the second factor of Eq.~\eqref{eq:rhobar}: $\sqrt{\mathcal{T}_1 / \sum_i\mathcal{T}_i}$ (solid contours), $\sqrt{\mathcal{T}_2 / \sum_i\mathcal{T}_i}$ (dashed contours), and $\sqrt{\mathcal{T}_3 / \sum_i\mathcal{T}_i}$ (dotted contours).
}
\end{figure}

The first factor of $\rhob$ scales with $N\us^{-1/4}$, which is a well-known property of hierarchical searches \cite{Krishnan.etal.2004,Mendell.Landry.2005}.
The second factor contains, inside the square root, a constant term, $\mathcal{T}_1 = z\ua$, a term which scales approximately with $N\us^{-1/4}$, $\mathcal{T}_2 = q ( 1 + z\ua \sqrt{ 8 / N\us\nu } )^{1/2}$, and a term which scales with $N\us^{-1/2}$, $\mathcal{T}_3 = q^2 \sqrt{ 2 / N\us\nu }$.
These additional terms appear because we do not employ the weak-signal approximation of \cite{Krishnan.etal.2004}, which Taylor-expands $p\ud$ [Eq.~\eqref{eq:fdp-rhob-1}] to first order in $\rhob$.
This approximation is not valid here because we are interested in small false dismissal probabilities, which implies that a large fraction of signals will be strong, i.e. enough to cross the detection threshold.
Each of the terms $\mathcal{T}$ contribute to the value of $\rhob$ over the ranges $p\ua/N\ut \le 0.01$ and $N\us \ge 1$, as shown in Fig.~\ref{fig:rhob_terms_ratios}; over the plotted ranges of $p\ua/N\ut$ and $N\us$, the contributions of each term to the sum $\sqrt{\sum_i\mathcal{T}_i}$ is 70--90\% for $\sqrt{\mathcal{T}_1}$, 40--60\% for $\sqrt{\mathcal{T}_2}$, and $\le 40\%$ for $\sqrt{\mathcal{T}_3}$.
For small $p\ua/N\ut$ and large $N\us$, $\sqrt{\mathcal{T}_1}$ dominates the sum, and $\rhob$ begins to scale purely with $N\us^{-1/4}$.
A general power-law scaling of $\rhob$ with $N\us$ is utilised in \cite{Prix.Shaltev.2011}.

\subsection{Sensitivity to isotropically-distributed signals}\label{sec:estim-iso}

This section presents a new, more accurate sensitivity estimator than that presented in Section~\ref{sec:estim-csnr}.
The important difference is that we will no longer assume that all signals being searched for have the same SNR, and instead perform the correct averaging of $p\ud(\vec p)$ as given in Eq.~\eqref{eq:fdp}.
This implies that we are searching for a population of signals isotropically distributed in sky position $(\alpha,\delta)$ and orientation $(\psi,\xi)$ parameters.
Unlike the assumption of constant SNR, this is a more physically reasonable assumption.
The observed distribution of nearby (millisecond) radio pulsars is roughly isotropic in the sky \cite{Lorimer.2009}, and the angular momentum of neutron stars (which determines $\psi$ and $\xi$) is not expected to have a preferred direction.

\begin{figure}
\includegraphics[width=\linewidth]{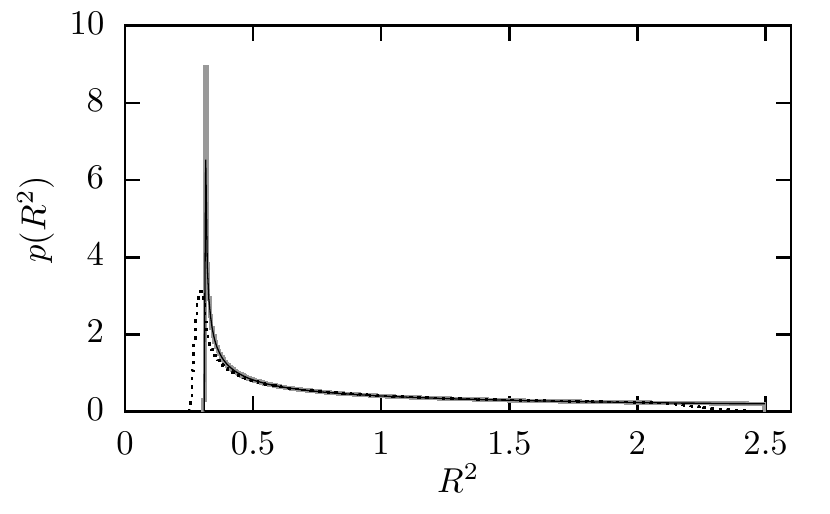}
\caption{\label{fig:Rsqr_histograms}
The distribution of $R^2\uiso$, calculated analytically using Eq.~\eqref{eq:Rsqr_dist} (thin black line), and numerically (thick gray line).
The black dashed line plots the distribution of $R^2\uisomism = R^2\uiso ( 1 - \mu )$, where the mismatch $\mu$ is drawn from a distribution appropriate for a lattice-based template bank (see Section~\ref{sec:assum-mism}).
}
\end{figure}

We also assume that we are searching data from a network comprising a large number of gravitational-wave detectors, evenly distributed over the Earth, such the network is isotropically sensitive to gravitational wave arriving from all directions.
(The validity of this assumption is investigated in Section~\ref{sec:assum-isodet}.)
We can therefore approximate $R^2$ by [see Eqs.~\ref{eq:sky-avg-antenna-patt} and~\ref{eq:det-avg-antenna-patt}]
\begin{equation}
\label{eq:Rsqr-xi}
\begin{split}
R^2\uiso(\xi) &= \avg{R^2}_{\Phi\us,\sin\lambda,\gamma} = \avg{R^2}_{\alpha,\sin\delta,\psi} \\
&= \frac{5}{16} ( \xi^4 + 6\xi^2 + 1 ) \,.
\end{split}
\end{equation}
Substituting into Eq.~\eqref{eq:fdp}, we see that we now need only to average $p\ud(\vec p)$ over $\xi$.
Figure~\ref{fig:Rsqr_histograms} plots the probability distribution of $R^2\uiso$ assuming a uniform distribution in $\xi$, given analytically by
\begin{equation}
\label{eq:Rsqr_dist}
p(R^2\uiso)^{-2} = \frac{1}{10} ( 50 + 20 R^2\uiso )^{\frac{3}{2}} - \frac{3}{4} ( 50 + 20 R^2\uiso ) \,,
\end{equation}
where $R^2\uiso$ ranges from $5/16$ ($\xi = 0$) to $5/2$ ($|\xi| = 1$).
Note that the most probable values of $R^2\uiso$ are those from linearly-polarized signals ($\xi = 0$), with a rapid fall-off towards circularly-polarized ($|\xi| = 1$) signals.
It is clear from this plot that the assumption that all signals have the same SNR (i.e.\ that $R^2\uiso = 1$), as assumed in Section~\ref{sec:estim-csnr}, is not a reasonable one.

We next approximate the non-central $\chisqr$ distribution by a normal distribution $\normal$, as per Section~\ref{sec:estim-csnr}.
Here, the approximation introduces a smaller error of $\lesssim 0.5\%$ (at $p\ua/N\ut = 0.01$, $N\us = 1$) which also decreases with decreasing $p\ua/N\ut$ and increasing $N\us$.
Equation~\eqref{eq:fdp} now reads
\begin{equation}
\label{eq:fdp-rhoh-1}
p\ud = \avg[\Big]{p\ud(\xi)}_{\xi} \,,
\end{equation}
where
\begin{equation}
\label{eq:fdp-rhoh-psd}
p\ud(\xi) = \frac{1}{2} \erfc \left( 
\frac{ N\us (\nu + \rhoh^2 R^2\uiso) - s\ua }{ 2 \sqrt{ N\us ( \nu + 2\rhoh^2 R^2\uiso ) } }
 \right) \,,
\end{equation}
and we define $\rhoh \equiv \sqrt{\avg{\rho^2}}$.

\begin{figure}
\includegraphics[width=\linewidth]{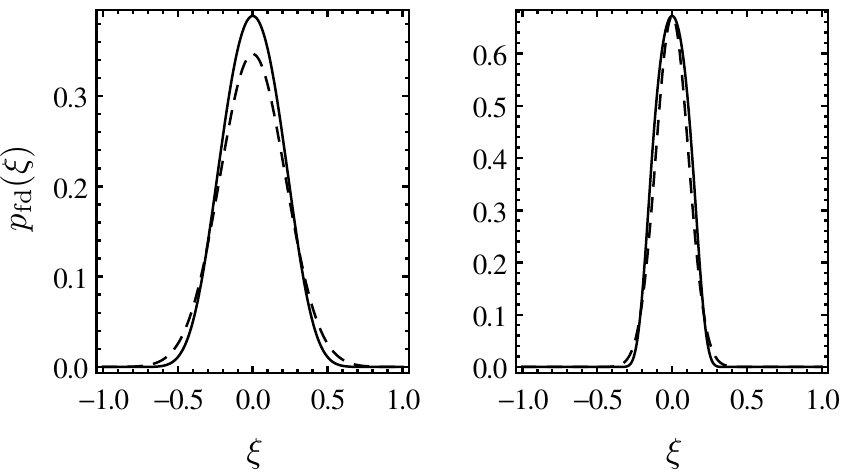}
\caption{\label{fig:false_dism_xi_gauss}
False dismissal probability $p\ud(\xi)$ of a single signal observed in an isotropically-sensitive detector network, as a function of the cosine of the signal's inclination angle, $\xi = \cos\iota$.
The exact result of Eq.~\eqref{eq:fdp-rhoh-psd} (solid line) is compared against the Gaussian function approximation of Eq.~\eqref{eq:fdp-rhoh-gauss} (dashed line), for $p\ud = 1\%$, $\nu = 4$, and: $p\ua / N\ut = 0.01$, $N\us$ = 1 (left), and $p\ua / N\ut = 10^{-12}$, $N\us = 10^4$ (right).
}
\end{figure}

We find that $p\ud(\xi)$ is well-approximated by a Gaussian function:
\begin{equation}
\label{eq:fdp-rhoh-gauss}
p\ud(\xi) \approx A e^{-B \xi^2} \,,
\end{equation}
for suitable choices of $A$ and $B$.
This is illustrated in Fig.~\ref{fig:false_dism_xi_gauss}, where we plot Eq.~\eqref{eq:fdp-rhoh-psd} against Eq.~\eqref{eq:fdp-rhoh-gauss} for two different choices of $p\ua / N\ut$ and $N\us$.
We note that it is important only that the integral of Eq.~\eqref{eq:fdp-rhoh-gauss} accurately approximate that of Eq.~\eqref{eq:fdp-rhoh-psd}; thus, the slight underestimation of $p\ud(\xi)$ by Eq.~\eqref{eq:fdp-rhoh-gauss} at $\xi \approx 0$ is partly offset by its overestimation at $|\xi| \gtrsim 0.5$.
We note that $p\ud(\xi) \approx 0$ at $|\xi| = 1$; numerical investigations confirm that this property holds true for $p\ua/N\ut \lesssim 0.01$ and $N\us \gtrsim 1$.
We can therefore replace the integral of Eq.~\eqref{eq:fdp-rhoh-gauss} over $-1 \le \xi \le 1$ with one over $-\infty \le \xi \le \infty$, since the integral over $|\xi| > 1$ contributes little to the value of $p\ud$.
The integral of Eq.~\eqref{eq:fdp-rhoh-gauss} then simplifies to
\begin{equation}
\label{eq:fdp-rhoh-2}
p\ud \approx \frac{1}{2} \int_{-\infty}^{\infty} \dd\xi \, A e^{-B \xi^2}
 = \frac{ A }{ 2 } \sqrt{ \frac{ \pi }{ B } } \,.
\end{equation}
We choose $A$ by setting $\xi = 0$ in Eq.~\eqref{eq:fdp-rhoh-gauss}, and obtain
\begin{equation}
\label{eq:fdp-rhoh-A}
A = \frac{1}{2} \erfc \left( \frac{ N\us \rhoh^2 R_0^2 - z\ua \sqrt{ 2 N\us \nu } }{ 2 \sqrt{ N\us ( \nu + 2\rhoh^2 R_0^2 ) } } \right) \,,
\end{equation}
where $R^2_0 = R^2\uiso(\xi=0)$.
Next, we choose $\xi_1 > 0$ to be the value of $\xi$ such that $p\ud(\xi=\xi_1)$ [Eq.~\eqref{eq:fdp-rhoh-psd}] equals the target false dismissal probability $p\ud$.
By equating Eqs.~\eqref{eq:fdp-rhob-1} and~\eqref{eq:fdp-rhoh-psd}, we deduce that $\xi_1$ satisfies
$\rhoh^2 R^2\uiso(\xi=\xi_1) = \rhob^2$, and is given by
\begin{equation}
\label{eq:fdp-rhoh-xi1}
\xi_1 = \left[ 2 \sqrt{ 2 + \frac{4}{5} \left(\frac{ \rhob }{ \rhoh }\right)^{2} } - 3 \right]^{\frac{1}{2}} \,.
\end{equation}
Substituting $\xi_1$ into Eq.~\eqref{eq:fdp-rhoh-gauss} gives the solution for
\begin{equation}
\label{eq:fdp-rhoh-B}
B = \frac{1}{ \xi_1^2 } \ln \left( \frac{ A }{ p\ud } \right) \,.
\end{equation}

Combining Eqs.~\eqref{eq:fdp-rhoh-2},~\eqref{eq:fdp-rhoh-A}, and~\eqref{eq:fdp-rhoh-B}, and taking logarithms, we have
\begin{equation}
\label{eq:fdp-rhoh-3}
\ln \left( \frac{ 2 p\ud }{ \xi_1 \sqrt\pi } \right) = \ln A - \frac{1}{2} \ln \left[ \ln \left( \frac{ A }{ p\ud } \right) \right] \,.
\end{equation}
The terms on the right-hand side of Eq.~\eqref{eq:fdp-rhoh-3} may be Taylor-expanded to second order in $x = \erfc^{-1} 2A$:
\begin{gather}
\label{eq:fdp-rhoh-TElnA}
\ln A = -\ln2 - \frac{2}{\sqrt\pi} x - \frac{2}{\pi} x^2 + \mathcal{O}(x^3) \,, \\
\label{eq:fdp-rhoh-TElnlnA}
\begin{split}
\frac{1}{2} \ln \left[ \ln \left( \frac{ A }{ p\ud } \right) \right] &=
\ln \sqrt{ |\ln2p\ud| } + \frac{1}{ \sqrt\pi \ln2p\ud } x \\ & - 
\frac{ 1 - \ln2p\ud }{ \pi (\ln2p\ud)^2 } x^2 + \mathcal{O}(x^3) \,,
\end{split}
\end{gather}
where the second expansion is valid only for $p\ud < 1/2$.
For $p\ua/N\ut \le 0.01$, $p\ud \ge 0.05$, $N\us \ge 1$, and $\nu \ge 2$, these expansions introduce errors of $\lesssim 1\%$ and $\lesssim 16\%$ respectively.
Substituting Eqs.~\eqref{eq:fdp-rhoh-TElnA} and~\eqref{eq:fdp-rhoh-TElnlnA} into Eq.~\eqref{eq:fdp-rhoh-3}, and using Eq.~\eqref{eq:fdp-rhoh-A}, we solve for
\begin{equation}
\label{eq:fdp-rhoh-4}
\begin{split}
x &= \frac{ N\us \rhoh^2 R_0^2 - z\ua \sqrt{ 2 N\us \nu } }{ 2 \sqrt{ N\us ( \nu + 2\rhoh^2 R_0^2 ) } } \\
&= \frac{\sqrt\pi}{2} \Gamma^{-1} \left( \sqrt{ 1 - 2 (1 - \Delta) \ln (2p\ud\Xi) } - 1 \right) \,,
\end{split}
\end{equation}
where
\begin{subequations}
\label{eqs:rhohat-GDX}
\begin{align}
\label{eq:rhohat-Gamma}
\Gamma &= 1 - \frac{1}{ \ln2p\ud } + \frac{2}{ 1 + 2\ln2p\ud } \,, \\
\label{eq:rhohat-Delta}
\Delta &= \frac{1}{ 1 + 2\ln2p\ud } + \frac{2}{ ( 1 + 2\ln2p\ud )^2 } \,, \\
\label{eq:rhohat-Xi}
\Xi   &= \frac{2}{ \xi_1 } \sqrt{ \frac{ |\ln2p\ud| }{\pi} } \,.
\end{align}
\end{subequations}
Now, suppose that $p\ud\Upr = (1/2) \erfc y$, and Taylor-expand $\ln p\ud\Upr$ to second order in $y$, as in Eq.~\eqref{eq:fdp-rhoh-TElnA}; solving for $y$ gives an expression (valid only for $p\ud\Upr < \sqrt{e} / 2$) for
\begin{equation}
\label{eq:fdp-rhoh-5}
y = \erfc^{-1} 2p\ud\Upr = \frac{\sqrt\pi}{2} \left( \sqrt{ 1 - 2\ln2p\ud\Upr } - 1 \right) \,.
\end{equation}
Note the similarity of the right-hand sides of Eqs.~\eqref{eq:fdp-rhoh-4} and~\eqref{eq:fdp-rhoh-5}.
We now define the \emph{effective} false dismissal probability $p\ud\Upr$, normalized false alarm threshold $z\ua\Upr$, and number of segments $N\us\Upr$:
\begin{align}
\label{eqs:rhohat-eff}
p\ud\Upr &= p\ud \, \frac{ \Xi }{ ( 2 p\ud \Xi )^{\Delta} } \,, &
z\ua\Upr &= z\ua \, \Gamma \,, &
N\us\Upr &= N\us \, \Gamma^2 \,.
\end{align}
Using these quantities, we equate Eqs.~\eqref{eq:fdp-rhoh-4} and~\eqref{eq:fdp-rhoh-5}, obtaining
\begin{equation}
\label{eq:fdp-rhoh-6}
\frac{ N\us\Upr \rhoh^2 R_0^2 - z\ua\Upr \sqrt{ 2 N\us\Upr \nu } }{ 2 \sqrt{ N\us\Upr ( \nu + 2\rhoh^2 R_0^2 ) } } = \erfc^{-1} 2p\ud\Upr \,.
\end{equation}
Finally, by noting that Eq.~\eqref{eq:fdp-rhoh-6} is similar in form to Eq.~\eqref{eq:fdp-rhob-1}, it follows that its solution for $\rhoh$ is given by analogy to Eq.~\eqref{eq:rhobar}:
\begin{multline}
\label{eq:rhohat}
\rhoh = \frac{1}{R_0} \left[ \frac{2\nu}{N\us\Upr} \right]^{\frac{1}{4}} \left[
z\ua\Upr \!+\! q\Upr \Bigg( 1 \!+\! \frac{ z\ua\Upr \sqrt8 }{ \sqrt{N\us\Upr\nu} } \Bigg)^{\frac{1}{2}} \!+\! \frac{ {q\Upr}^2 \sqrt2 }{ \sqrt{N\us\Upr\nu} }
\right]^{\frac{1}{2}} .
\end{multline}
where $q\Upr = \sqrt2 \erfc^{-1} 2p\ud\Upr$.

Evaluation of Eq.~\eqref{eq:rhohat} is complicated by the fact that its right-hand side is itself a function of $\rhoh$, through $p\ud\Upr$ [Eq.~\eqref{eqs:rhohat-eff}], $\Xi$ [Eq.~\eqref{eq:rhohat-Xi}], and $\xi_1$ [Eq.~\eqref{eq:fdp-rhoh-xi1}].
Nevertheless, Eq.~\eqref{eq:rhohat} may be iteratively solved for $\rhoh$ using the following scheme.
First, a reasonable initial guess, $\rhoh_{0}$ is chosen: an appropriate choice is $\rhoh_{0} \approx 1.4 \rhob$ (see Fig.~\ref{fig:rhob_rhohA_ratio}).
Next, $\rhoh_{0}$ is used to compute an updated value, $\rhoh_{1}$, by substituting into Eq.~\eqref{eq:rhohat}: $\rhoh_{1} = \rhoh (\rhoh_{0})$.
Thereafter, new values of $\rhoh$ are obtained using the mean of the previous two values, i.e. the $n$th value of $\rhoh$ is
\begin{equation}
\label{eq:rhohat-iter}
\rhoh_{n} = \rhoh \left( \frac{ \rhoh_{n-1} + \rhoh_{n-2} }{ 2 } \right) \,.
\end{equation}
Using this scheme, the sequence of values $\{\rhoh_{n}\}$ reliably converges to an accurate value of $\rhoh$: typically, 20--80 iterations are required to achieve sufficient accuracy.
The use of the mean of the previous two $\rhoh$ in Eq.~\eqref{eq:rhohat-iter} suppresses divergent oscillations in the sequence $\{\rhoh_{n}\}$.

\begin{figure*}
\includegraphics[width=\linewidth]{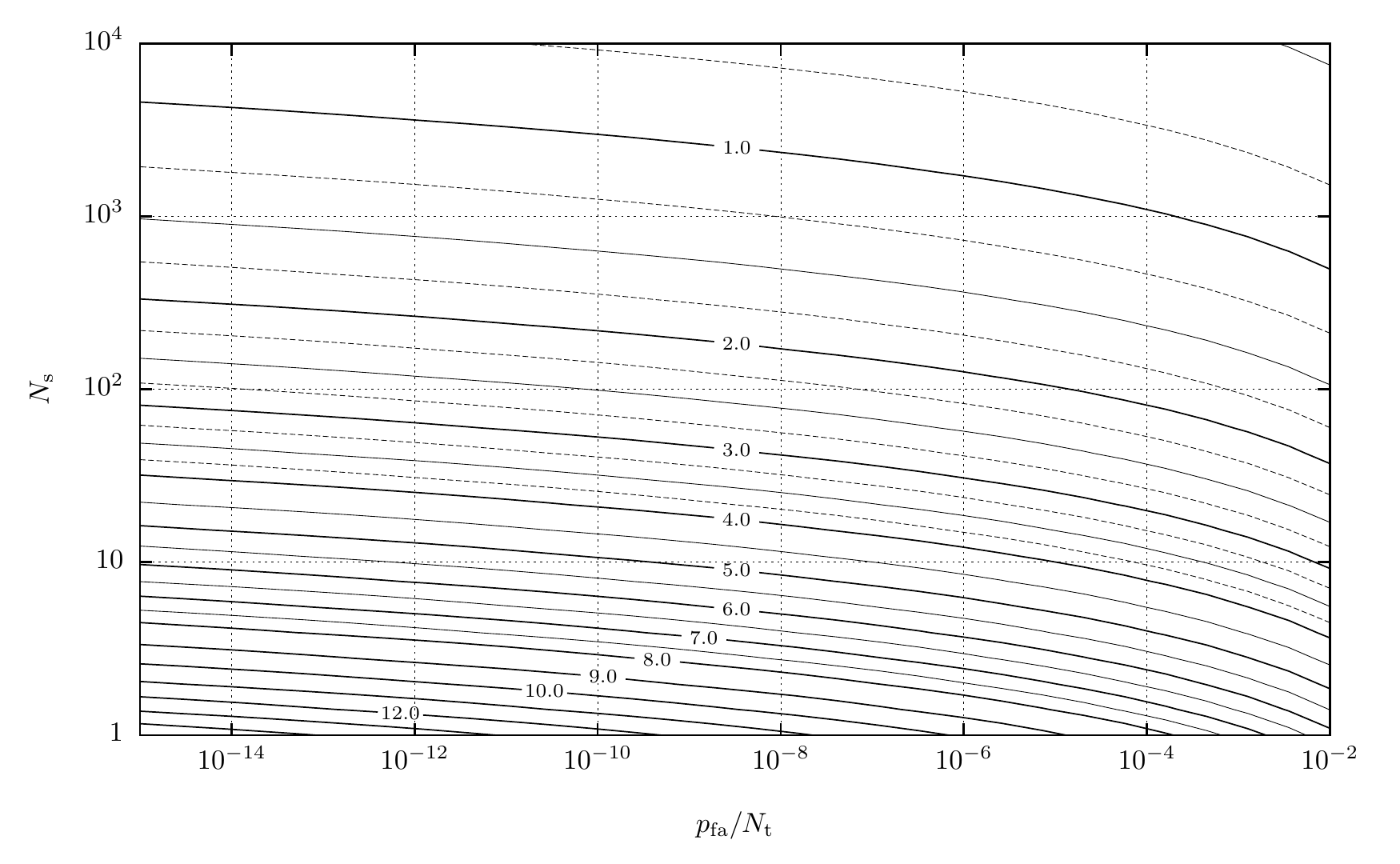}
\caption{\label{fig:rhohA}
Contours of $\rhoh$ as a function of $p\ua/N\ut$ and $N\us$, with $p\ud = 0.1$ and $\nu = 4$.
Thick solid contours are in units of 1.0, thin solid contours are in units of 0.5, and thin dashed contours are in units of 0.25.
}
\end{figure*}

Figure~\ref{fig:rhohA} plots $\rhoh$  as a function of $p\ua/N\ut$ and $N\us$, with $p\ud = 0.1$, for a detection statistic with $\nu=4$ degrees of freedom (e.g. the $\F$-statistic).
For a single-template search at 1\% false alarm and 10\% false dismissal probabilities, we see that $p\ua = 0.01$, $N\ut = N\us = 1$, and $\rhoh = 6.3$, which should be interpreted as the \emph{average} sensitivity of a collection of single-template searches which cover the parameter space uniformly in sky position $(\alpha,\delta)$ and orientation $(\psi,\xi)$.
This value of $\rhoh$ is equivalent to a statistical factor of 15.7.
This differs from the often-quoted (e.g.\ \cite{Abbott.etal.2004c}) statistical factor for a single-template search of 11.4, because that statistical factor is calculated assuming a signal with an \emph{average} SNR, i.e.\ it is calculated from $\rhob$ instead of $\rhoh$.
For the search for Cassiopeia~A presented in \cite{Abadie.etal.2010a}, $p\ua = 0.01$, $p\ud = 0.05$, $N\ut \sim 1.8 \times 10^{10}$~\cite{Wette.2009}, $N\us = 1$, and $\rhoh \approx 14.5$, which implies a statistical factor of $\sim 36$ consistent with that quoted in~\cite{Wette.etal.2008,Wette.2009}.

\begin{figure}
\includegraphics[width=\linewidth]{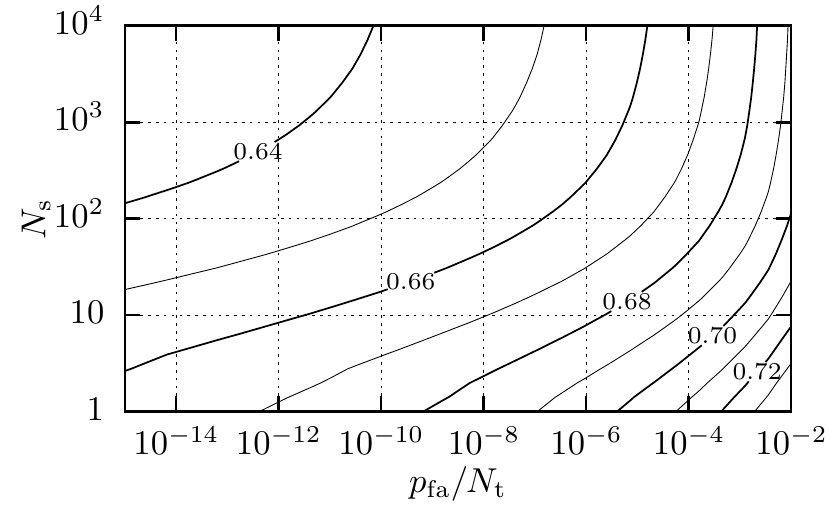}
\caption{\label{fig:rhob_rhohA_ratio}
Contours of the ratio of $\rhob/\rhoh$ as a function of $p\ua/N\ut$ and $N\us$, with $p\ud = 0.1$ and $\nu = 4$.
Unlabeled thin contours are in units of 0.01.
}
\end{figure}

Figure~\ref{fig:rhob_rhohA_ratio} plots the ratio of $\rhob$ to $\rhoh$, as a function of $p\ua/N\ut$ and $N\us$.
Relative to $\rhoh$, we see that $\rhob$ underestimates the mean SNR detectable by a search by $\sim 29 \pm 5$\%, i.e. it overestimates the search's sensitivity by the same amount.
For comparison, the typically amplitude calibration error of the LIGO detectors is $\sim 10$\% \cite{Abadie.etal.2010b}, and the ratio of the best upper limits on gravitational waves from Cassiopeia~A to the indirect limits from energy conservation is $\sim 60$\% \cite{Abadie.etal.2010a}.
Thus, an error of $\sim 30$\% in estimating a search's sensitivity is a significant discrepancy.
Note, however, that the change in $\rhob/\rhoh$ is small over the ranges of $p\ua/N\ut$ and $N\us$ plotted in Fig.~\ref{fig:rhob_rhohA_ratio}.
We conclude from this that, while $\rhob$ does not predict the correct sensitivity, it does capture the correct scaling of sensitivity with the false alarm probability, template count, and number of segments.
This conclusion also follows from the similarity in form between Eqs.~\eqref{eq:rhobar} and~\eqref{eq:rhohat}.

\begin{figure}
\includegraphics[width=\linewidth]{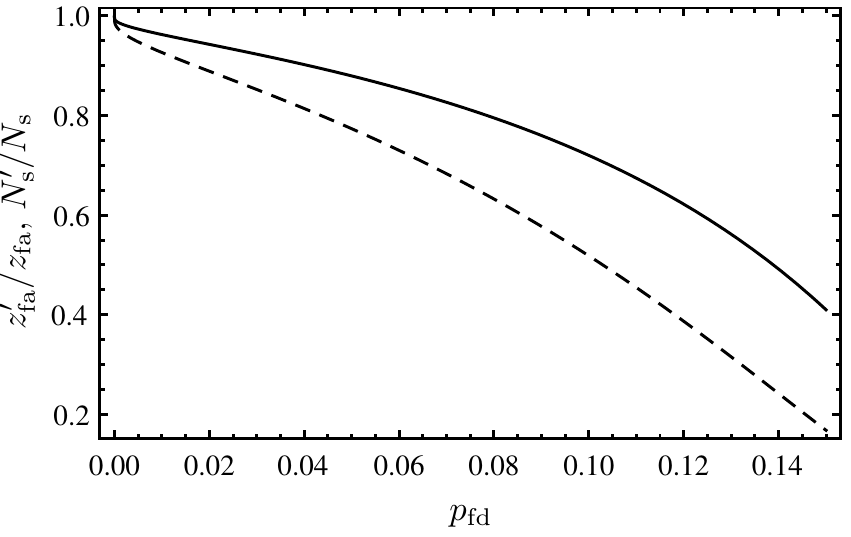}
\caption{\label{fig:effective_za_Ns}
Ratios of the effective false alarm threshold $z\ua\Upr/z\ua$ (solid) and number of segments $N\us\Upr/N\us$ (dashed) as functions of $p\ud$.
}
\end{figure}

\begin{figure}
\includegraphics[width=\linewidth]{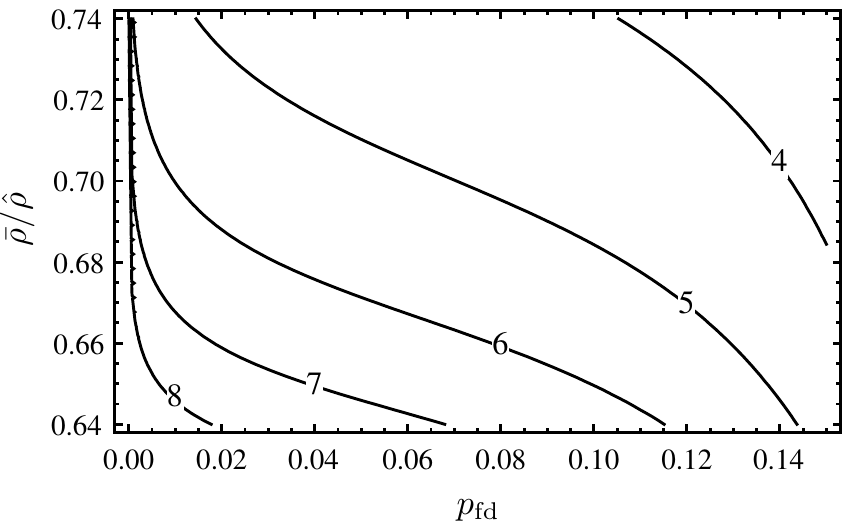}
\caption{\label{fig:effective_pd}
Contours of the effective false dismissal probability $p\ud\Upr / p\ud$ as a function of $p\ud$ and $\rhob / \rhoh$.
}
\end{figure}

To illustrate the relationship between $\rhoh$ and $\rhob$, Fig.~\ref{fig:effective_za_Ns} plots $z\ua\Upr/z\ua$ and $N\us\Upr/N\us$ as functions of $p\ud$, and Fig.~\ref{fig:effective_pd} plots $p\ud\Upr / p\ud$ as a function of $p\ud$ and $\rhob / \rhoh$.
We see that both $z\ua\Upr/z\ua$ and $N\us\Upr/N\us$ are less than unity for $p\ud > 0$, while $p\ud\Upr$ is generally greater than $p\ud$ by a factor of $\sim 4$--8.
Note too that the right-hand side of Eq.~\eqref{eq:rhohat} is divided by $R_0 \approx 0.56$ [Eq.~\eqref{eq:Rsqr-xi}].
This implies that $\rhoh$ is dominated by contributions from linearly-polarised signals, as seen in the distribution of $R^2\uiso$ plotted in Fig.~\ref{fig:Rsqr_histograms}.
Therefore, we may think of $\rhoh$ as estimating the sensitivity of a search to a population of constant-SNR signals (as for $\rhob$), but where the signals are linearly polarised (hence the division by $R_0$), and where the search is performed with reduced false alarm threshold $z\ua\Upr$, a reduced number of segments $N\us\Upr$, and a greatly increased false dismissal probability $p\ud\Upr$.

We note that Eqs.~\eqref{eq:fdp-rhoh-TElnlnA} and~\eqref{eq:fdp-rhoh-5} impose restrictions on permissible values of $p\ud < 0.5$ and $p\ud\Upr < \sqrt{e}/2 \approx 0.82$ respectively.
The latter restriction and Fig.~\ref{fig:effective_pd} implies that, depending on the value of $\rhob/\rhoh$, $p\ud$ is further restricted to be less than $0.82/4$--$0.82/8 \approx 0.1$--$0.2$.
In general, therefore, the use of Eq.~\eqref{eq:rhohat} is restricted to values of $p\ud \lesssim 10$--20\%.
In practice this is not an onerous restriction, as we are generally only interested in small false dismissal probabilities of either 10\% or 5\%.
Gravitational-wave pulsar searches of LIGO and Virgo data (see Section~\ref{sec:intro} for references) have set upper limits with corresponding upper limit confidences of 90\% and 95\% respectively.

\section{Accuracy of analytic sensitivity estimator}\label{sec:accur}

In this section, we validate the accuracy of the analytic sensitivity estimator $\rhoh$, derived in Section~\ref{sec:estim-iso}, against the sensitivity calculated using two methods: numerically solving Eq.~\eqref{eq:fdp}, in Section~\ref{sec:accur-num}; and performing software injections, in Section~\ref{sec:accur-inj}.
The validation is performed using a detection statistic with $\nu = 4$, and for a target false dismissal probability of $p\ud = 0.1$.

\subsection{Numerical solution of false dismissal equation}\label{sec:accur-num}

\begin{figure}
\includegraphics[width=\linewidth]{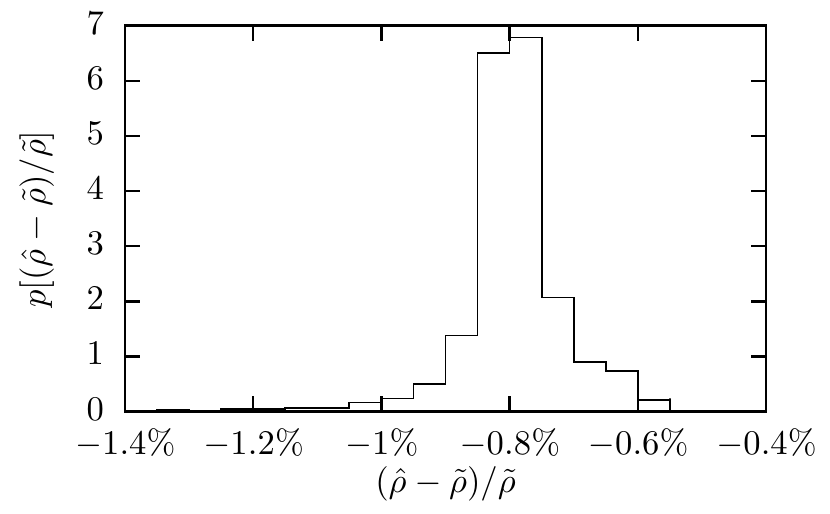}
\caption{\label{fig:rhohA_rhoh_error}
Histogram of the relative difference between $\rhoh$, given by Eq.~\eqref{eq:rhohat}, and the numerical solution to Eq.~\eqref{eq:fdp}, $\rhot$,
over the ranges $10^{-15} \le p\ua/N\ut \le 10^{-2}$ and $1 \le N\us \le 10^{4}$.
}
\end{figure}

We first compare the sensitivity predicted by $\rhoh$ against the sensitivity $\sqrt{\avg{\rho^2}}$ calculated by solving Eq.~\eqref{eq:fdp} numerically.
We denote by $\rhot$ the value of $\sqrt{\avg{\rho^2}}$ which solves Eq.~\eqref{eq:fdp} for a target false dismissal probability $p\ud$.
We denote by $p\ud(\rhot\utrial)$ the result of computing Eq.~\eqref{eq:fdp} for a given trial value of $\rhot$, denoted $\rhot\utrial$.
For $\rhot\utrial = 0$, $p\ud(\rhot\utrial) > p\ud$, otherwise Eq.~\eqref{eq:fdp} has no solution for the chosen $p\ud$.
We start by determining a $\rhot\umax$ such that $p\ud(\rhot\umax) < p\ud$, thus bracketing $\rhot$ to between 0 and $\rhot\umax$.
We then use a simple bifurcation search to converge to $\rhot$, which terminates when the relative error between $p\ud(\rhot)$ and the target $p\ud$ is less than $10^{-3}$.
Equation~\eqref{eq:fdp} is solved for a grid of logarithmically-spaced values of $p\ua/N\ut$ and $N\us$: 30 values in the range $10^{-15} \le p\ua/N\ut \le 10^{-2}$, and 28 values in the range $1 \le N\us \le 10^{4}$.

Figure~\ref{fig:rhohA_rhoh_error} shows the relative difference between the analytic $\rhoh$, and the numerically calculated $\rhot$.
The maximum relative error between $\rhoh$ and $\rhot$ is $\lesssim 1.4\%$ over the given range of $p\ua/N\ut$ and $N\us$.
Note that $\rhoh$ is consistently smaller than $\rhot$, i.e.\ $\rhoh$ slightly overestimates the sensitivity calculated using Eq.~\eqref{eq:fdp}.

\subsection{Software injections}\label{sec:accur-inj}

We next determine whether the sensitivity predicted by $\rhoh$, and calculated numerically from Eq.~\eqref{eq:fdp} in the previous section, correctly predict the performance of a real gravitational-wave pulsar search pipeline.
To do so, we perform software injection studies similar (with some simplications) to the Monte Carlo simulations used to set upper limits for gravitational-wave searches of LIGO and Virgo data (see Section~\ref{sec:intro} for references).
We use software from the \textsf{LALSuite}\footnote{%
Available from \url{https://www.lsc-group.phys.uwm.edu/daswg/projects/lalsuite.html}.} repository.

First, the \textsf{Makefakedata\_v4} program is used to generate gravitational wave strain data, of timespan $N\us T\us$ (see Table~\ref{tab:injections}), containing Gaussian noise (with a power spectral density of $S_h = 1$), and a simulated gravitational-wave pulsar signal, as it would be observed in the LIGO Livingston detector.
The strain amplitude $h_0$ is given by [c.f.\ Eq.~\eqref{eq:sqrtavgrho-statfac}]
\begin{equation}
\label{eq:injection-h0}
h_0 = \frac{5}{2} \sqrt{\frac{ S_h }{ T\us }} \rhoh
 = \frac{5}{2} \sqrt{\frac{ S_h }{ T\us }} \rhot \,,
\end{equation}
where $S_h = 1$, $T\us$ is given values from Table~\ref{tab:injections}, and either $\rhoh$ or $\rhot$ are calculated for $p\ud = 0.1$ and values for $p\ua/N\ut$ and $N\us$ given in Table~\ref{tab:injections}.
Other parameters of the simulated signal are chosen uniformly from the following ranges: $-1 \le \xi \le 1$, $0 \le \psi < 2\pi$, $0 \le \phi_0 < 2\pi$, $0 \le \alpha < 2\pi$, $-1 \le \sin\delta \le 1$, $f = 100~\text{Hz}$, and $-10^{-8}~\text{Hz s}^{-1} \le \dot{f} \le 0$ (with higher-order spindowns set to zero).

Next, the \textsf{ComputeFStatistic\_v2} program, an implementation of the $\F$-statistic, is used to perform a single-template search of the data generated by \textsf{Makefakedata\_v4}, at precisely the sky position $(\alpha,\delta)$ and frequency evolution $(f,\dot{f})$ of the simulated signal.
(We consider the loss in sensitivity due to mismatch between the search template and signal in Section~\ref{sec:assum-mism}.)
A single value of the $\F$-statistic is returned by \textsf{ComputeFStatistic\_v2} and stored.
For $N\us > 1$, since \textsf{ComputeFStatistic\_v2} cannot combine searches of multiple data segments, we instead run \textsf{ComputeFStatistic\_v2} $N\us$ times on successive segments, of timespans $T\us$, and add together the $N\us$ returned $\F$-statistic values.
Thus, the (summed) $\F$-statistic values will follow $\chisqr$ distributions with $N\us\nu = 4N\us$ degrees of freedom.

Finally, we repeat the injection procedure 5000 times.
We record the number of (summed) $\F$-statistic values which are below the false alarm threshold $s\ua$, as calculated in Section~\ref{sec:estim-fap}.
This fraction equals the false dismissal probability as determined by the software injections, which we denote $p\udinj$.
If the sensitivity of \textsf{ComputeFStatistic\_v2} is accurately estimated by $\rhoh$ and/or $\rhot$, $p\udinj$ should be close to the target false dismissal probability of $p\ud = 0.1$.
Different choices of $p\ua/N\ut$ are used to set different thresholds $s\ua$, and thus simulate the sensitivity of a wide-parameter-space search over $N\ut$ templates.

Ideally, we would then refine the injected value of $h_0$, increasing it if $p\udinj > p\ud$, decreasing it if $p\udinj < p\ud$, and then repeat the entire injection procedure, until $p\udinj = p\ud$.
The relative error between the value of $h_0$ calculated using Eq.~\eqref{eq:injection-h0}, and the value of $h_0$ arrived at by repeating the injection procedure, would then be equal to the relative error in the estimation of the sensitivity of \textsf{ComputeFStatistic\_v2}.
Since the injection procedure is time-consuming and computationally intensive, however, we instead re-compute $\rhoh$ or $\rhot$, as appropriate, using the false dismissal probability $p\udinj$ determined by the injections.
We denote by $\Delta \rhoh = |\rhoh(p\udinj) - \rhoh(p\ud)|/\rhoh(p\ud)$ the relative error between the value of $\rhoh$ calculated using $p\udinj$, and the value of $\rhoh$ calculate using $p\ud$; similarly for $\Delta \rhot$.
These quantities serve as a reasonable estimate of the error in the sensitivities estimated by $\rhoh$ and $\rhot$ respectively.

\begin{table}
\caption{\label{tab:injections}
Validation of the analytic sensitivity estimator $\rhoh$, and the numerically-computed sensitivity $\rhot$, using software injections.
The injections are performed for three values of $p\ua/N\ut$ (listed in row~1), five combination of $N\us$ and $T\us$ (listed in columns~1 and~2), and once each using either $\rhoh$ (top panel) or $\rhot$ (bottom panel).
Values of $\Delta \rhoh$ (or $\Delta \rhot$) and $p\udinj$ are given, for each of the three values of $p\ua/N\ut$, in columns~3--4, 5--6, and~7--8 respectively.
}
\begin{tabular}{r r r@{.}l r@{.}l r@{.}l r@{.}l r@{.}l r@{.}l}
$N\us$ & $T\us$ &
\multicolumn{4}{c}{$p\ua/N\ut = 10^{-2}$} &
\multicolumn{4}{c}{$p\ua/N\ut = 10^{-6}$} &
\multicolumn{4}{c}{$p\ua/N\ut = 10^{-10}$} \\
\hline
\multicolumn{14}{c}{Analytic $\rhoh$} \\
& &
\multicolumn{2}{c}{$\Delta \rhoh$} & \multicolumn{2}{c}{$p\udinj$} &
\multicolumn{2}{c}{$\Delta \rhoh$} & \multicolumn{2}{c}{$p\udinj$} &
\multicolumn{2}{c}{$\Delta \rhoh$} & \multicolumn{2}{c}{$p\udinj$} \\
\hline
1 & 1 & $-5$&$\%$ & $13$&$2\%$ & $-2$&$5\%$ & $12$&$4\%$ & $-5$&$\%$ & $14$&$9\%$ \\
1 & 5 & $-2$&$4\%$ & $11$&$8\%$ & $-2$&$8\%$ & $12$&$6\%$ & $-2$&$9\%$ & $13$&$\%$ \\
1 & 10 & $-4$&$9\%$ & $13$&$2\%$ & $-4$&$7\%$ & $14$&$\%$ & $-2$&$5\%$ & $12$&$7\%$ \\
10 & 1 & $-1$&$3\%$ & $11$&$1\%$ & $-1$&$7\%$ & $11$&$9\%$ & $-1$&$7\%$ & $12$&$2\%$ \\
25 & 1 & $-0$&$16\%$ & $10$&$5\%$ & $-0$&$93\%$ & $11$&$4\%$ & $-1$&$5\%$ & $12$&$2\%$ \\
\hline
\multicolumn{14}{c}{Numerical $\rhot$} \\
& &
\multicolumn{2}{c}{$\Delta \rhot$} & \multicolumn{2}{c}{$p\udinj$} &
\multicolumn{2}{c}{$\Delta \rhot$} & \multicolumn{2}{c}{$p\udinj$} &
\multicolumn{2}{c}{$\Delta \rhot$} & \multicolumn{2}{c}{$p\udinj$} \\
\hline
1 & 1 & $-3$&$\%$ & $11$&$5\%$ & $-4$&$1\%$ & $12$&$9\%$ & $-3$&$8\%$ & $13$&$1\%$ \\
1 & 5 & $-4$&$7\%$ & $12$&$4\%$ & $-3$&$5\%$ & $12$&$4\%$ & $-3$&$4\%$ & $12$&$8\%$ \\
1 & 10 & $-2$&$9\%$ & $11$&$4\%$ & $-3$&$\%$ & $12$&$\%$ & $-2$&$6\%$ & $12$&$1\%$ \\
10 & 1 & $-0$&$2\%$ & $10$&$1\%$ & $-1$&$6\%$ & $11$&$2\%$ & $-1$&$8\%$ & $11$&$5\%$ \\
25 & 1 & $0$&$36\%$ & $9$&$8\%$ & $-1$&$2\%$ & $10$&$9\%$ & $-0$&$99\%$ & $10$&$9\%$ \\
\hline
\end{tabular}
\end{table}

The results of the software injections are shown in Table~\ref{tab:injections}.
Both $\rhoh$ and $\rhot$ underestimate, by $\lesssim 5\%$ and $\lesssim 4.7\%$ respectively, the $h_0$ required to achieve a false dismissal probability of 10\%; except in one instance (last row of column~4), $p\udinj > 10\%$.
Nevertheless, these errors are still within the $\sim 10$\% typical calibration error of gravitational wave detectors, e.g.\ LIGO \cite{Abadie.etal.2010b}, and hence can be considered small.
The difference between the predicted and actual sensitivity of \textsf{ComputeFStatistic\_v2} is likely because the $\F$-statistic values returned by \textsf{ComputeFStatistic\_v2} do not strictly follow a $\chisqr$ distribution; small errors, on the order of a few percent, are introduced due to implementation details of the code \cite{Prix.2010,Prix.2011}; see Section~\ref{sec:assum-statindp} and Figs.~\ref{fig:stat_indp_allsky} and~\ref{fig:stat_indp_spdown}.

\section{Validity of assumptions}\label{sec:assum}

In the previous two sections, we have derived an analytic estimator of the sensitivity of wide-parameter-space gravitational-wave pulsar searches (Section~\ref{sec:estim}), and validated its accuracy (Section~\ref{sec:accur}).
In doing so, we made certain assumptions: that the values of the detection statistic returned by the search are statistically independent, that the network of gravitational wave detectors being searched is sensitive to gravitational waves from all sky locations, and that any signal present in the data is perfectly matched by at least one of the searched templates.
In this section, we investigate to what extent these assumptions are valid for real gravitational-wave pulsar search pipelines.

\subsection{Statistical independence of templates}\label{sec:assum-statindp}

It was assumed, in deriving an expression for the false alarm probability in Section~\ref{sec:estim-fap}, that the $N\ut$ values of the detection statistic $s$ returned by the search are statistically independent.
This is not necessarily the case in practice.
Gravitational-wave pulsar searches typically use template banks with small, e.g. 20\% mismatches; templates nearby in parameter space will therefore have similarly-shaped waveforms.
Matched filtering of nearby templates will therefore combine the same data with nearly-identical waveforms to produce corresponding values of the detection statistic.
There is the potential, therefore, for values of the detection statistic computed from nearby templates to be correlated with each other.

To assess to what extent this effect is important, we perform four fully-coherent searches, using \textsf{ComputeFStatistic\_v2} (see Section~\ref{sec:accur-inj}) of 3~days of computer-generated Gaussian noise.
The search parameter space is sky and frequency, with higher spindowns set to zero.
Template banks were generated using the \textsf{gridType=2} option, which places templates over the sky using an adaptive mesh.
The first three searches are performed using template banks with the following mismatches: an unrealistically-small mismatch of 1\%, a realistic mismatch of 20\%, and an unrealistically-high mismatch of 500\%.
The fourth search repeats the first search using the 1\% mismatch template bank, with the following modification to the \textsf{ComputeFStatistic\_v2} program: before each value of the $\F$-statistic is computed, the input Gaussian noise data is regenerated, so that each $\F$-statistic is computed from independent instances of Gaussian noise.
The frequency bands of the searches are chosen such that the searches return $4.2 \times 10^{6}$ values of the $\F$-statistic.

We partition the returned $\F$-statistic values into 2100 blocks of $N = 2000$ values, contiguous in frequency, and select the maximum value of the $\F$-statistic in each block.
If all the values of the $\F$-statistic are mutually independent, the distribution of the 2100 $\F$-statistic maxima is expected to be \cite{Wette.2009}
\begin{multline}
\label{eq:max-twoF-dist}
p(2\F\umax) = N p\big( 2\F = 2\F\umax \big| 2\F \sim \chisqr[4] \big) \\ \times
\Big[ p\big( 2\F < 2\F\umax \big| 2\F \sim \chisqr[4] \big) \Big]^{N-1} \,.
\end{multline}
If, however, the values of the $\F$-statistic exhibit some mutual correlation, then we expect the distribution of the 2100 $\F$-statistic maxima to be well-modeled by Eq.~\eqref{eq:max-twoF-dist}, but with an \emph{effective} number of statistically independent templates $N \le 2000$.
This procedure was used by the gravitational-wave search for Cassiopeia~A \cite{Abadie.etal.2010a} to estimate the statistical correlation of the template bank.

\begin{figure}
\includegraphics[width=\linewidth]{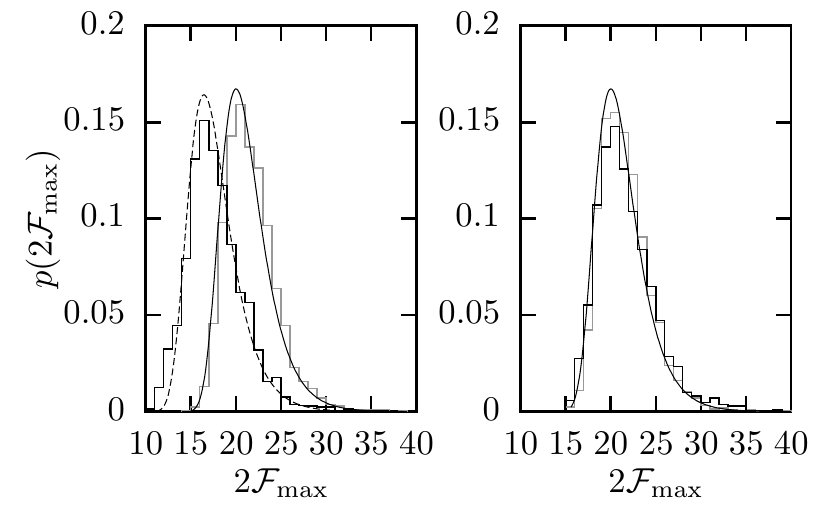}
\caption{\label{fig:stat_indp_allsky}
Histograms of the maximum values of the $\F$-statistic obtained in blocks of $2000$ templates returned by an all-sky--frequency search of 3~days of computer-generated Gaussian noise.
(Left) Histogram of the 1\% mismatch template bank, computed from the same Gaussian noise (black), and from regenerated independent Gaussian noise (gray).
The theoretical distribution is plotted for $N = 2000$ (solid black curve), and $N = 400$ (dashed black curve).
(Right) Histograms of the 20\% mismatch template bank (black), and the 500\% mismatch template bank (gray).
The theoretical distribution is plotted for $N = 2000$ (solid black curve).
}
\end{figure}

Figure~\ref{fig:stat_indp_allsky} plots histograms of the $\F$-statistic maxima obtained from the four searches.
In the left-hand plot of Fig.~\ref{fig:stat_indp_allsky}, we see that the $\F$-statistic values from the 1\% mismatch search (without regenerating the input data) are highly correlated; they are best fitted by Eq.~\eqref{eq:max-twoF-dist} with $N = 400$ effectively statistically-independent templates.
When the input data is regenerated before computing each $\F$-statistic value, the distribution is much closer to the expected distribution with $N = 2000$ statistically independent templates.
This demonstrates that the origin of the correlations is that the same data are being used to compute the $\F$-statistic values of nearby templates.
The small differences between the computed and expected distributions is likely due to the implementation details of \textsf{ComputeFStatistic\_v2} (see Section~\ref{sec:accur-inj}).
In the right-hand plot of Fig.~\ref{fig:stat_indp_allsky}, we see that the distribution of the $\F$-statistic maxima, for both the 20\% and 500\% mismatch template banks, are both close to the expected distribution with $N = 2000$.
This suggests that, while statistical correlation between $\F$-statistic values is noticeable for very closely-spaced templates, it is less significant for realistic template bank mismatches.

\begin{figure}
\includegraphics[width=\linewidth]{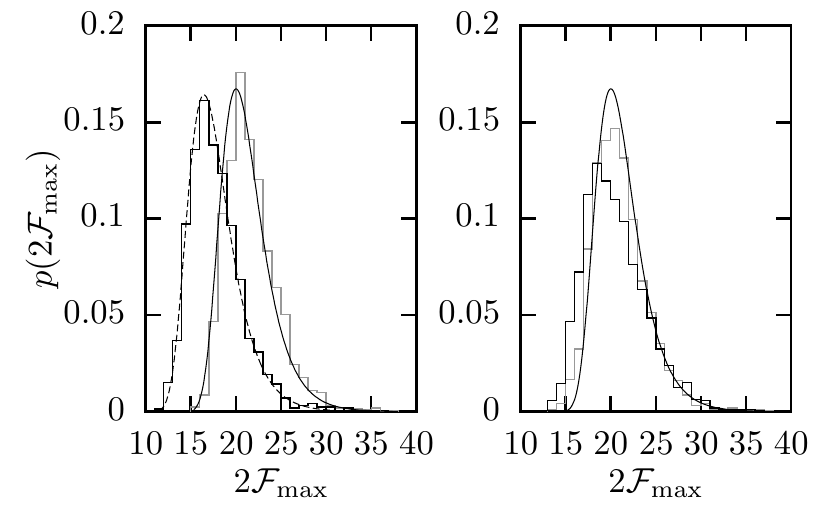}
\caption{\label{fig:stat_indp_spdown}
Histograms of the maximum values of the $\F$-statistic obtained in blocks of $2000$ templates returned by an frequency--spindown search of 7~days of computer-generated Gaussian noise.
Details are the same as Fig.~\ref{fig:stat_indp_allsky}.
}
\end{figure}

We repeat the above four searches, instead using 7~days of data and searching over frequency and spindown.
Template banks were generated using the \textsf{gridType=8} option to \textsf{ComputeFStatistic\_v2}, which places templates over frequency and spindown using a lattice.
The resulting distributions of the $\F$-statistic maxima, plotted in Fig.~\ref{fig:stat_indp_spdown}, are similar to those presented in Fig.~\ref{fig:stat_indp_allsky}, except that the 20\% mismatch distribution is less well-fitted by the $N = 2000$ distribution.
The same template bank was used in the search for Cassiopeia~A \cite{Abadie.etal.2010a}, which found the number of statistically-independent templates to be $\sim 88\%$ of the total number of templates.

While it is difficult to precisely quantify the effect of statistical correlations between templates on search sensitivity, for template banks with realistic mismatches the effect is likely to be small.
As may be deduced from Fig.~\ref{fig:rhohA}, a change in $N\ut$ of e.g.\ 10\% does not significantly alter the predicted $\rhoh$.
Given that, as shown in Table~\ref{tab:injections}, both analytic and numerical sensitivity estimators accurately predict the sensitivity of a real gravitational-wave search pipeline to $\sim 5\%$, it is reasonable to neglect the smaller effect due to statistical correlations between templates.

\subsection{Isotropic sensitivity of detector network}\label{sec:assum-isodet}

We now investigate whether it is reasonable to assume that a real gravitational-wave detector network, e.g. that of the LIGO and Virgo detectors, can be modeled by a network which is isotropically sensitive to gravitational waves arriving from all directions.
While this assumption is reasonable for all-sky searches, since averaging over sky position is equivalent to averaging over detector orientation (see Appendix~\ref{apx:antenna}), it may not be reasonable for searches targeting a single sky position.

In Section~\ref{sec:accur-num}, we numerically solved Eq.~\eqref{eq:fdp} for $\rhot = \sqrt{\avg{\rho^2}}$, assuming an isotropically-sensitive detector network, and averaging over the sky position $(\alpha,\delta)$ and polarization angle $\psi$; we denote the values of $\rhoh$ thus obtained by $\rhot\uiso$.
We now solve Eq.~\eqref{eq:fdp}, using the same algorithm detailed in Section~\ref{sec:accur-num}, for three detector networks: the LIGO Livingston detector, the two LIGO (Livingston and Hanford) detectors, and the three-detector network comprising the LIGO and Virgo detectors; detector locations and orientations are taken from \cite{Schutz.2011}.
We assume that all detectors in the network are equally sensitive, since we are concerned only with the effect of the geographic configuration of the network.
We consider the following five search scenarios:
\begin{enumerate}
\item
\label{item:rho_diff}
A search using a long data segment length of $T\us \gg$~days, covering the entire sky.
For large $T\us$, the signal SNR becomes independent of $\alpha$ (see Appendix~\ref{apx:antenna}).
We compute $\rhot$ averaged over the remaining parameters, $\sin\delta$ and $\psi$.
\item
\label{item:rho_diff_sdelta}
A search using a long data segment length of $T\us \gg$~days, and targeting a source at a known sky position but with an unknown polarization.
We compute $\rhot$ for a grid of 41 linearly-spaced values of $\sin\delta$ in the range $-1 \le \sin\delta \le 1$, and average over $\psi$.
\item
\label{item:rho_diff_sdelta_psi}
A search using a long data segment length of $T\us \gg$~days, and targeting a source with both a known sky position and polarization.
We compute $\rhot$ for the same grid of $\sin\delta$ values as in scenario~\ref{item:rho_diff_sdelta}, and a grid of 10 linearly-spaced values of $\psi$ in the range $-\pi/4 \le \psi < \pi/4$.
\item
\label{item:rho_diff_alpha_sdelta}
A search using a short data segment length of $T\us = 0.5$~days, and targeting a source at a known sky position but with an unknown polarization.
For $T\us$ shorter than a day, the signal SNR is a function of $\alpha$, $\sin\delta$, and $\psi$.
We compute $\rhot$ for a grid of 20 linearly-spaced values of $\alpha$ in the range $0 \le \alpha < 2\pi$, 19 linearly-spaced values of $\sin\delta$ in the range $-1 < \sin\delta < 1$, and average over $\psi$.
\item
\label{item:rho_diff_alpha_sdelta_psi}
A search using a short data segment length of $T\us = 0.5$~days, and targeting a source with both a known sky position and polarization.
We compute $\rhot$ for the same grids of $\alpha$ and $\sin\delta$ values as in scenario~\ref{item:rho_diff_alpha_sdelta}, and the same grid of $\psi$ values as in scenario~\ref{item:rho_diff_sdelta_psi}.
\end{enumerate}
For each of the above search scenarios, we record the minimum, mean, standard deviation, and maximum of the relative error $|\rhot - \rhot\uiso|/\rhot\uiso$ and relative difference $(\rhot - \rhot\uiso)/\rhot\uiso$, over the grids of $\alpha$, $\sin\delta$, and $\psi$ values given above, and over the grids of $p\ua/N\ut$ and $N\us$ values given in Section~\ref{sec:accur-num}.

\begin{table}
\caption{\label{tab:iso_det_net}
Relative errors and differences in the sensitivity estimated assuming an isotropically-sensitive detector network, under five search scenarios (see the text for details).
For each detector network (column 1), the mean and standard deviation of the relative error $|\rhot - \rhot\uiso|/\rhot\uiso$ (column 2), and the minimum and maximum difference $(\rhot - \rhot\uiso)/\rhot\uiso$ (columns 3 and 4) are listed.
The abbreviations \emph{L}, \emph{LH}, and \emph{LHV} indicate the LIGO Livingston detector, the LIGO detector network, and the LIGO--Virgo detector network respectively.
Each block of the table corresponds to a search scenario, and a summary of each scenario is given just above each block.
}
\begin{tabular}{l r@{.}l@{}c@{}r@{.}l r@{.}l r@{.}l}
\hline
Network &
\multicolumn{5}{c}{Error $|\rhot - \rhot\uiso|/\rhot\uiso$} &
\multicolumn{4}{c}{Difference $(\rhot - \rhot\uiso)/\rhot\uiso$} \\
&
\multicolumn{2}{r}{mean} & $\pm$ & \multicolumn{2}{l}{stdv.} &
\multicolumn{2}{c}{minimum} &
\multicolumn{2}{c}{maximum} \\
\hline
\multicolumn{10}{c}{1.~$T\us \rightarrow \infty$; averaged $\sin\delta$, and $\psi$} \\
\hline
L   & $0$&$0058$&$\pm$&$0$&$0037$\% & $-0$&$018$\% & $+0$&$017$\% \\
LH  & $0$&$0062$&$\pm$&$0$&$0042$\% & $-0$&$02$\% & $+0$&$017$\% \\
LHV & $0$&$0063$&$\pm$&$0$&$0044$\% & $-0$&$02$\% & $+0$&$014$\% \\
\hline
\multicolumn{10}{c}{2.~$T\us \rightarrow \infty$; known $\sin\delta$; averaged $\psi$} \\
\hline
L   & $3$&$4$&$\pm$&$1$&$8$\% & $-4$&$5$\% & $+6$&$5$\% \\
LH  & $2$&$6$&$\pm$&$1$&$3$\% & $-4$&$9$\% & $+3$&$8$\% \\
LHV & $3$&$7$&$\pm$&$2$&\% & $-7$&$5$\% & $+5$&$4$\% \\
\hline
\multicolumn{10}{c}{3.~$T\us \rightarrow \infty$; known $\sin\delta$ and $\psi$} \\
\hline
L   & $4$&$3$&$\pm$&$2$&$4$\% & $-11$&\% & $+6$&$5$\% \\
LH  & $2$&$8$&$\pm$&$2$&\% & $-4$&$9$\% & $+8$&$7$\% \\
LHV & $3$&$8$&$\pm$&$2$&$4$\% & $-7$&$5$\% & $+9$&$3$\% \\
\hline
\multicolumn{10}{c}{4.~$T = 0.5$ days; known $\alpha$, $\sin\delta$; averaged $\psi$} \\
\hline
L   & $5$&$1$&$\pm$&$3$&$6$\% & $-6$&$5$\% & $+16$&\% \\
LH  & $4$&$7$&$\pm$&$3$&\% & $-10$&\% & $+10$&\% \\
LHV & $4$&$1$&$\pm$&$2$&$3$\% & $-8$&$8$\% & $+7$&$4$\% \\
\hline
\multicolumn{10}{c}{5.~$T = 0.5$ days; known $\alpha$, $\sin\delta$, and $\psi$} \\
\hline
L   & $6$&$3$&$\pm$&$4$&$7$\% & $-14$&\% & $+20$&\% \\
LH  & $5$&$6$&$\pm$&$3$&$7$\% & $-10$&\% & $+18$&\% \\
LHV & $4$&$4$&$\pm$&$3$&$1$\% & $-9$&$9$\% & $+15$&\% \\
\hline
\end{tabular}
\end{table}

\begin{figure}
\includegraphics[width=\linewidth]{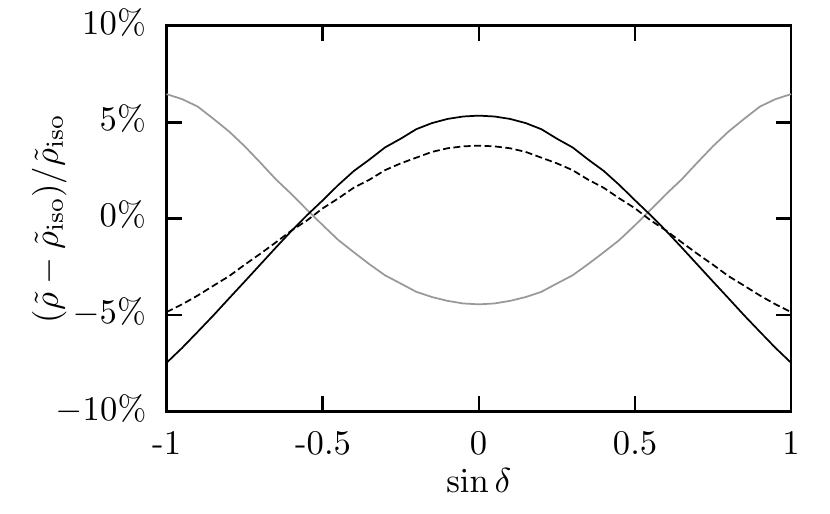}
\caption{\label{fig:rho_diff_mean_sdelta}
Mean relative difference $(\rhot - \rhot\uiso)/\rhot\uiso$ in the sensitivity estimated assuming an isotropically-sensitive detector network, for search scenario~\ref{item:rho_diff_sdelta} (see the text), as a function of $\sin\delta$, for the LIGO Livingston detector (gray), the LIGO detector network (black, dashed), and the LIGO--Virgo detector network (black). 
}
\end{figure}

\begin{figure}
\includegraphics[width=\linewidth]{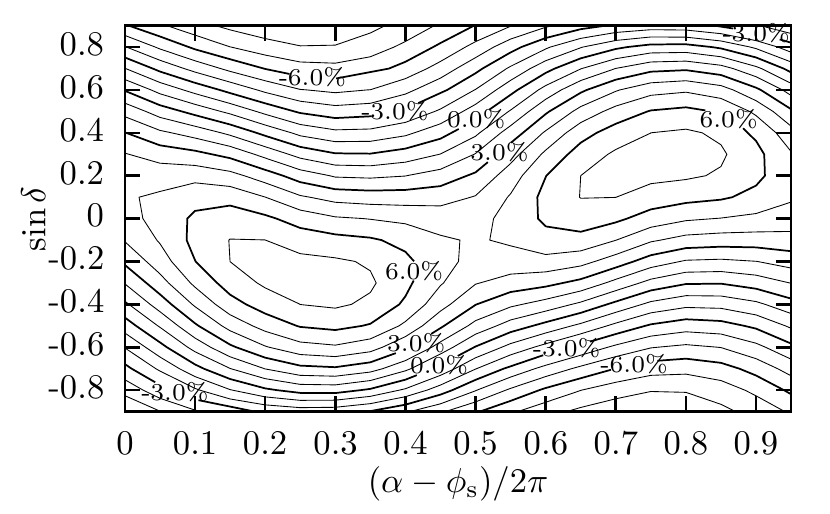}
\caption{\label{fig:rho_diff_mean_alpha_sdelta_LHV}
Mean relative difference $(\rhot - \rhot\uiso)/\rhot\uiso$ in the sensitivity estimated assuming an isotropically-sensitive detector network, for scenario~\ref{item:rho_diff_alpha_sdelta} (see the text), as a function of $\alpha$ and $\sin\delta$, for the LIGO--Virgo detector network.
The sidereal time at Greenwich at the mid-point of the observation time is denoted by $\phi\us$.
}
\end{figure}

Table~\ref{tab:iso_det_net} shows the mean and standard deviation of the relative error, and the minimum and maximum relative difference obtained under the five search scenarios.
For an all-sky search (scenario~\ref{item:rho_diff}), the assumption of an isotropic detector network is an excellent one.
For the remaining search scenarios, the error in assuming an isotropically-sensitive detector network increases as $T\us$ is reduced, and as more parameters are set to fixed values.
While the mean error is limited to $\lesssim 6.3\%$ for all four scenarios, the maximum difference can be up to 20\% for a single detector, although it reduces to 15\% for a three-detector network.
There are only a few (potential) gravitational-wave sources for which the polarization angle may be determined with any accuracy; noted examples are the Crab and Vela pulsars, where X-ray observations of the pulsar wind nebula provide information on the pulsar's orientation \cite{Abbott.etal.2010,Abadie.etal.2011}.
Thus, scenarios~\ref{item:rho_diff_sdelta_psi} and~\ref{item:rho_diff_alpha_sdelta_psi} are less likely to arise in practice (but see Section~\ref{sec:discussion} for a discussion of these scenarios in relation to the PowerFlux upper limit procedure).
We conclude that, while the error in assuming an isotropically-sensitive detector network may be acceptable in many cases, it can be significant for particular choices of fixed search parameters.

Figure~\ref{fig:rho_diff_mean_sdelta} plots the mean relative difference $(\rhot - \rhot\uiso)/\rhot\uiso$ for scenario~\ref{item:rho_diff_sdelta}, as a function of $\sin\delta$, for the three detector networks considered.
The difference between $\rhot$ and $\rhot\uiso$ is smallest at $\sin\delta \sim 0.5$--0.55, the approximate latitudes of the three detectors ($\sim 30$--$46^\circ$); a signal originating from these declinations would therefore be located at the approximate maximum sensitivities of the detectors.
For the same reason, a single detector at low latitude (LIGO Livingston) is more sensitive to signals arriving at low declinations, explaining the increased sensitivity relative to that of an isotropic detector network.
The addition of two detectors at higher latitudes (LIGO Hanford and Virgo) shifts the network configuration towards improved sensitivity at higher declinations.

Figure~\ref{fig:rho_diff_mean_alpha_sdelta_LHV} plots the mean relative difference $(\rhot - \rhot\uiso)/\rhot\uiso$ for scenario~\ref{item:rho_diff_alpha_sdelta}, as a function of $\alpha$ and $\sin\delta$, for the LIGO--Virgo network.
Consistent with Fig.~\ref{fig:rho_diff_mean_sdelta}, the network is more sensitive to signals arriving at the poles, and less sensitive to signals arriving at the equator, than an isotropically-sensitive network.
The change in sensitivity as a function of $\alpha$ and $\delta$ is comparable to previous studies of gravitational-wave detector network configuration; see e.g.\ Fig~3 in \cite{Searle.etal.2002}.

\subsection{Template bank mismatch}\label{sec:assum-mism}

\begin{figure}
\includegraphics[width=\linewidth]{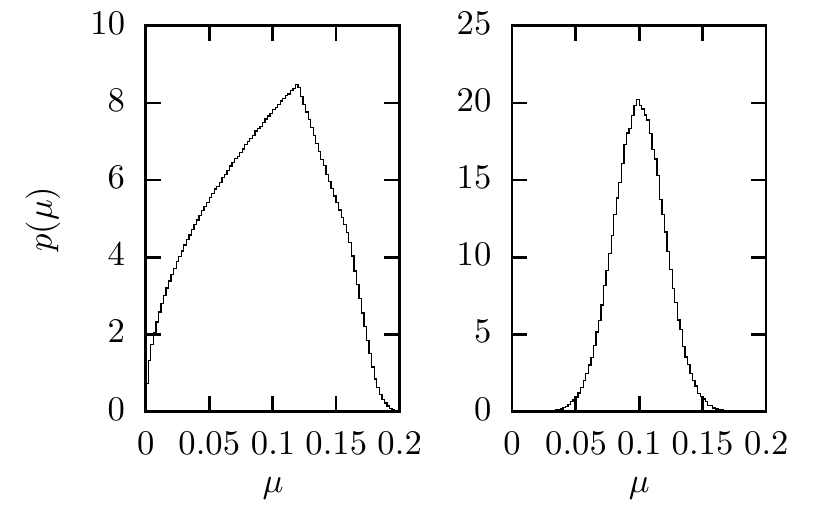}
\caption{\label{fig:mismatch_histograms}
Histograms of the mismatch distribution of a 3-dimensional body-centered cubic lattice template bank (left), and an \emph{ad-hoc} Gaussian mismatch distribution (right).
}
\end{figure}

\begin{figure}
\includegraphics[width=\linewidth]{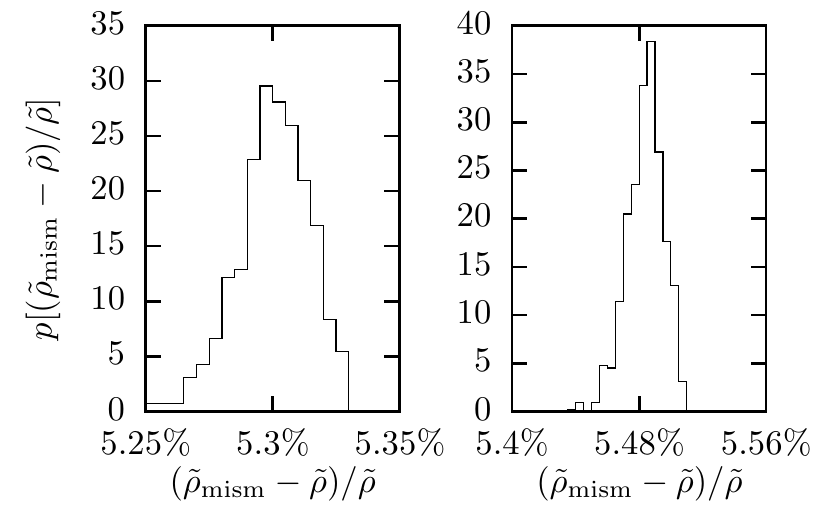}
\caption{\label{fig:rhot_mism_histograms}
Histograms of the relative difference between $\rhot\umism$, calculated using the lattice template bank (left) and Gaussian mismatch distributions (right), and $\rhot$ (zero mismatch) over the ranges $10^{-15} \le p\ua/N\ut \le 10^{-2}$ and $1 \le N\us \le 10^{4}$.
}
\end{figure}

Finally, we consider the reduction in sensitivity due to mismatch between the searched templates and any signal which may be present.
The mismatch $\mu$ is related to the difference between the SNR $\rho^2(\vec p)$ of a signal with parameters $\vec p$, and the SNR $\rho^2(\vec p\Upr)$ recovered by a search template with mismatched parameters $\vec p\Upr$.
It is given by (e.g.\ \cite{Prix.2007})
\begin{equation}
\label{eq:mismatch}
\mu = \frac{ \rho^2(\vec p) - \rho^2(\vec p\Upr) }{ \rho^2(\vec p) } \,.
\end{equation}
Substituting $\sqrt{\avg{\rho^2}} R$ for $\rho(\vec p)$ and $\sqrt{\avg{\rho^2}} R\umism$ for $\rho(\vec p\Upr)$, we find that $R^2\umism = R^2 ( 1 - \mu )$; thus the effect of mismatch is to reduce $R^2$ by a factor $1 - \mu$.
For a large template bank, $\mu$ can be considered as a random variable drawn from a distribution $p(\mu)$ characterizing the geometric arrangement of the templates in parameter space.
Equation~\eqref{eq:fdp} is then modified to additionally average over mismatch:
\begin{multline}
\label{eq:fdp-mism}
p\ud = \avg[\Big]{ p \big( s \le s\ua \big| s \sim \chisqr[N\us \nu, \\
N\us \avg{\rho^2} R^2( 1 - \mu )  ] \big) }_{\alpha,\sin\delta,\psi,\xi,\mu} \,,
\end{multline}
where $\avg{\cdots}_{\mu} = \int_{0}^{\mu\umax} \dd\mu \, p(\mu) \cdots$, and $\mu\umax$ is the maximum allowed mismatch.

We solve Eq.~\eqref{eq:fdp-mism} numerically for $\rhot\umism = \sqrt{\avg{\rho^2}}$, using a modified version of the algorithm described in Section~\ref{sec:accur-num}.
We compute $\rhot\umism$ for two examples of $p(\mu)$: the mismatch distribution for a template bank constructed using a 3-dimensional body-centered cubic lattice with $\mu\umax = 20\%$, from \cite{Wette.2009}, and an \emph{ad hoc} Gaussian mismatch distribution with a mean of 10\% and standard deviation of 2\%, restricted to the range $0 \le \mu \le 20\%$.
Histograms of these distributions are plotted in Fig.~\ref{fig:mismatch_histograms}; mismatch distributions for lattice template banks in other dimensions are plotted in \cite{Messenger.etal.2009}.
The lattice template bank mismatch distribution has a mean mismatch of 10\% and standard deviation of 4\%.

We find that taking mismatch into account increases $\rhot\umism$, relative to the equivalent (zero-mismatch) $\rhot$ computed in Section~\ref{sec:accur-num}, by on average $5.3 \pm 0.10$\% for the lattice template bank mismatch distribution, and $5.5 \pm 0.10$\% for the Gaussian distribution.
This effect is of the same magnitude as the error in the analytic/numerical estimators compared to software injections (see Table~\ref{tab:injections}).
We note that, by substituting $R_0 \sqrt{ 1 - \avg{\mu} }$ (where $\avg{\mu}$ is the mean of the mismatch distribution) for $R_0$ in Eq.~\eqref{eq:rhohat}, we can predict the observed relative increase in $\rhot\umism$ reasonably accurately; for $\avg{\mu} = 10$\%, $\rhoh$ is increased by 5.4\%.
This suggests that it is reasonable to model the effect of template bank mismatch as a uniform reduction in recovered SNR (by $\sqrt{ 1 - \avg{\mu} }$), and that the mean of the mismatch distribution is sufficient to quantify this reduction; this approach is also taken in \cite{Prix.Shaltev.2011}.
Figure~\ref{fig:rhot_mism_histograms} plots histograms of the relative difference between $\rhot\umism$ and $\rhot$; the histograms are narrow, with no long tails, confirming that the most typical reduction in SNR is close to the average reduction.

Figure~\ref{fig:Rsqr_histograms} plots the distribution of $R^2\uisomism = R^2\uiso ( 1 - \mu )$, where $\mu$ are drawn from the lattice template bank mismatch distribution.
Relative to the zero-mismatch distribution, $R^2\uisomism$ is reduced for circularly-polarized signals (at high $R^2\uisomism$), and exhibits a broader peak at linearly-polarized signals (at low $R^2\uisomism$).

\section{Discussion}\label{sec:discussion}

In this paper, we develop a new analytic method of estimating the sensitivity of wide-parameter space searches for gravitational-wave pulsars.
The new method avoids the unphysical assumption of constant-SNR signals, used by previous analytic sensitivity estimation methods, and hence can accurately (to $\lesssim 5\%$) reproduce the sensitivity estimated using Monte Carlo simulations, but without their computational cost.
Additional effects, not included in the analytic model but which may be important for real gravitational-wave pulsar searches, are investigated: the statistical correlation between values of the detection statistic due to closely-spaced template banks; the deviation of the directional sensitivity of real gravitational wave detector networks from an ideal, isotropically-sensitive network; and the loss in SNR due to mismatch between a signal and the searched templates.
While the first of these effects can be neglected, the remaining two may need to be accounted for where accurate sensitivity estimates are required.

It is important to note that an accurate prediction of the \emph{absolute} sensitivity of a search may not always be necessary.
For studies of the optimal configuration of a hierarchical search (e.g.\ \cite{Cutler.etal.2005,Prix.Shaltev.2011}), the \emph{relative} sensitivity of different search schemes is usually more important.
For these studies, use of the constant-SNR sensitivity estimator $\rhob$ may be acceptable; Fig.~\ref{fig:rhob_rhohA_ratio} shows that $\rhob$ reproduces to $\lesssim 5\%$ the correct scaling (i.e.\ that of $\rhoh$) of sensitivity with respect to $p\ua/N\ut$ and $N\us$.
On the other hand, given the similarity between the expressions for $\rhob$ and $\rhoh$ [compare Eqs.~\eqref{eq:rhobar} and~\eqref{eq:rhohat}], these studies may also be able to utilise the more accurate sensitivity scaling of $\rhoh$ with little modification.

This work has restricted its attention to detection statistics which follow $\chisqr$ distributions, which does not cover all gravitational-wave search methods.
In particular, the Hough semi-coherent method~\cite{Krishnan.etal.2004} computes a statistic, the number count, which is the number of coherently-analyzed segments where the coherent detection statistic (which may either be power or the $\F$-statistic) exceeds a set threshold.
It follows that the number count follows a binomial distribution.
Since, however, the Hough method is generally used to analyze a large number of data segments (i.e. $N\us$ is large), the binomial distribution can generally be approximated by a normal distribution (see \cite{Krishnan.etal.2004}), an approximation also used in Section~\ref{sec:estim}.
Thus, the analytic sensitivity estimation method developed here may be applicable to the Hough method with minor modifications.

Searches performed using the PowerFlux method \cite{Abbott.etal.2008,Abbott.etal.2009d,Abadie.etal.2011d} employ a slightly different formulation of the frequentist procedure described in Section~\ref{sec:gwave-sens}.
(To distinguish these two frequentist procedures, in the following discussion we refer to the upper limits produced by the frequentist procedure described in Section~\ref{sec:gwave-sens} as \emph{population-averaged} upper limits.)
An upper limit on $h_0$ is set, using the Feldman-Cousins method of confidence interval construction \cite{Feldman.Cousins.1998}, for every searched parameter (typically sky position, frequency, and first-order spindown), and assuming either linearly-polarized signals ($\xi = 0$, with a range of $\psi$) or a circularly-polarized signal ($|\xi| = 1$).
For each polarization case, the maximum value of $h_0$ over the searched parameters is chosen.
Two upper limits are then quoted: a best-case upper limit, which assumes circular polarization; and a worst-case\footnote{%
Actually, PowerFlux worst-case upper limits, as quoted in e.g.\ \cite{Abadie.etal.2011d}, also account for the worst-case mismatch; in this discussion however we assume that no mismatch is present.} upper limit, where $h_0$ is further maximized over the linear polarizations.

The PowerFlux upper limit procedure differs from the population-averaged upper limit procedure in two respects.
(In the following comparison, we assume that the data being searched is relatively free of non-Gaussian instrumental noise.)
First, while the population-average procedure first selects the maximum value of the detection statistic $s$ over the parameter space, and then computes an upper limit on $h_0$ using $s$, the PowerFlux procedure first computes an upper limit on $h_0$ for every value of $s$, assuming a fixed polarisation, and then selects the maximum upper limit.
By computing upper limits assuming a fixed sky position $(\alpha,\delta)$, corresponding to each value of $s$, and a fixed $\psi$, PowerFlux is essentially performing search scenario~\ref{item:rho_diff_alpha_sdelta_psi} from Table~\ref{tab:iso_det_net} (although with different search parameters, e.g.\ $T\us = 30$~minutes, $p\ud = 0.05$). 
From the maximum relative difference in sensitivity quoted in Table~\ref{tab:iso_det_net} for this scenario, we estimate that the effect of maximising over $h_0$ instead of over $s$ raises PowerFlux upper limits by $\sim 10$--20\% relative to population-averaged upper limits.

Second, the population-average procedure produces a single upper limit averaged over all parameters; the PowerFlux procedure instead quotes upper limits for best-case (circular) and worst-case (linear) polarizations.
By setting upper limits at fixed polarizations (i.e. fixed values of $\xi$), the PowerFlux procedure is setting upper limits on a population of signals with a fixed value of $R^2$, i.e. a fixed SNR.
We can therefore predict the upper limits on $h_0$ set by the PowerFlux procedure by replacing $\avg{\rho^2} R^2$ with $\rho_{\mathrm{P.F.}}^2 R^2\uiso(\xi)$ in Eq.~\eqref{eq:fdp}; it follows that $\rho_{\mathrm{P.F.}}$ is given by $\rhob / R\uiso(\xi)$, where $\rhob$ is given by Eq.~\eqref{eq:rhobar}.
The best- and worst-case sensitivities are then given by
\begin{align}
\rho_{\mathrm{P.F.~best}} &= \frac{ \rhob }{ R\uiso(|\xi|=1) } = \sqrt{\frac{2}{5}} \, \rhob \\
\rho_{\mathrm{P.F.~worst}} &= \frac{ \rhob }{ R\uiso(\xi=0) } = \frac{4}{\sqrt{5}}  \, \rhob
\end{align}
Ratios of Power-Flux best- and worst-case sensitivities to the sensitivity predicted by $\rhoh$ range over $0.39 \le \rho_{\mathrm{P.F.~best}}/\rhoh \le 0.46$, and $1.1 \le \rho_{\mathrm{P.F.~worst}}/\rhoh \le 1.3$, for the range of $\rhob/\rhoh$ plotted in Fig.~\ref{fig:rhob_rhohA_ratio}.
An injection study by Dergachev~\cite{PF_UL_conversion} found ratios consistent with these ranges: $\rho_{\mathrm{P.F.~best}}/\rhoh \sim 0.43$, and $\rho_{\mathrm{P.F.~worst}}/\rhoh \sim 1.2$.
Further investigation into the differences between the PowerFlux and population-averaged upper limits methods would facilitate direct comparison of the upper limits produced by different gravitational wave searches, e.g. between \cite{Abbott.etal.2009d} and \cite{Abbott.etal.2009b}.

Recent searches for gravitational waves from known pulsars \cite{Abbott.etal.2010,Abadie.etal.2011} use Bayesian inference to set upper limits.
The frequentist method described in this paper constructs confidence intervals which are derived from the probability of the \emph{data}, given a particular hypothesis e.g. that a signal is or is not present in the data.
In contrast, Bayesian inference directly calculates the probability of the \emph{hypothesis}, given the particular data that were observed.
Despite their very different interpretations, a study in \cite{Rover.etal.2011} found that the frequentist and Bayesian methods produce similar upper limits, in the limit of large signal amplitudes.
Further research is needed to understand fully the relationship between these two methods.

\begin{acknowledgments}
I thank Bruce Allen, Joseph Betzwieser, Vladimir Dergachev, Evan Goetz, David Keitel, Ben Owen, Chris Messenger, Reinhard Prix, Letizia Sammut, and members of the LIGO Scientific Collaboration's Continuous Wave Working Group for many informative discussions, and for helpful comments on the manuscript.
Numerical simulations were performed on the ATLAS computer cluster of the Max-Planck-Institut f\"ur Gravitationsphysik.
This paper has document numbers LIGO-P1100151-v4 and AEI-2011-104.
\end{acknowledgments}

\appendix

\section{Antenna-pattern functions and their averages}\label{apx:antenna}

We present expressions for the antenna-pattern functions $F\up(t)$ and $F\ux(t)$ of an interferometric detector, and the averages of $F\up(t)^2$ and $F\ux(t)^2$ over various parameters; see also the references in Section~\ref{sec:gwave-signal}.

The antenna-pattern functions can be written in terms of a time-dependent detector response matrix $\mat R(t)$ and constant polarization matrices $\mat H\up$ and $\mat H\ux$:
\begin{align}
\label{eqs:antenna-patt-def}
F\up(t) &= \trace \big( \mat R(t)^{T} \mat H\up \big) \,, &
F\ux(t) = \trace \big( \mat R(t)^{T} \mat H\ux \big) \,,
\end{align}
where $\trace$ is the matrix trace and $^{T}$ denotes transposition.
We implicitly assume a coordinate system at rest with respect to the gravitational radiation, e.g.\ the Solar System barycenter.

Let $\uvec n_1(t)$ and $\uvec n_2(t)$ be unit vectors along the interferometer's arms, such that the cross product $\uvec n_1(t) \times \uvec n_2(t)$ points toward zenith.
The directions along which the interferometer is insensitive to any gravitational radiation are given by the unit vectors $\pm\uvec a(t)$ and $\pm\uvec b(t)$, where
\begin{align}
\uvec a(t) &= \frac{ \uvec n_1(t) - \uvec n_2(t) }{ 2\sin(\zeta/2) } \,, &
\uvec b(t) &= \frac{ \uvec n_1(t) + \uvec n_2(t) }{ 2\cos(\zeta/2) } \,,
\end{align}
and $\zeta$ is the angle between the interferometer arms.
Note that $\uvec a(t)$ and $\uvec b(t)$ are orthogonal, i.e.\ $\uvec a(t) \cdot \uvec b(t) = 0$.
Assuming that the gravitational wavelength is much larger than the interferometer arm-length, as is the case for ground-based detectors,
the response matrix $\mat R$ is given in terms of these vectors by (e.g.\ \cite{Finn.2009})
\begin{equation}
\label{eq:response-mat}
\mat R(t) = \frac{\sin\zeta}{2} \big[ \uvec a(t) \otimes \uvec b(t) + \uvec b(t) \otimes \uvec a(t) \big] \,,
\end{equation}
where $\otimes$ is the vector outer product.
The vectors can be expressed in terms of time-independent components:
\begin{equation}
\uvec a(t) = \uvec a_1 \cos \Omega\us (t - t_0) + \uvec a_2 \sin \Omega\us (t - t_0) + \uvec a_3 \,,
\end{equation}
where the Earth rotates about the unit vector $\uvec\Omega\us$ in a right-handed sense with angular frequency $\Omega\us$, and $\uvec a_1 = \uvec a_2 \times \uvec \Omega\us$, $\uvec a_2 = \uvec \Omega\us \times \uvec a_0$, $\uvec a_3 = ( \uvec \Omega\us \cdot \uvec a_0 ) \uvec \Omega\us$, and $\uvec a_0 = \uvec a(t = t_0)$; similarly for $\uvec b(t)$.

The polarization matrices $\mat H\up$ and $\mat H\ux$ may also be written in terms of certain vectors \cite{Prix.Whelan.2007,Prix.2010}.
The directions along which a gravitational wave creates no space-time perturbation are given by the mutually orthogonal unit vectors $\pm\uvec x\up$ and $\pm\uvec y\up$, for a purely plus-polarized wave, and $\pm\uvec x\ux$ and $\pm\uvec y\ux$, for a purely cross-polarized wave.
The two sets of vectors are related by
\begin{align}
\uvec x\up \sqrt2 &= \uvec x\ux - \uvec y\ux \,, &
\uvec y\up \sqrt2 &= \uvec x\ux + \uvec y\ux \,.
\end{align}
The cross-polarisation vectors are given by
\begin{align}
\label{eq:x-y-cross}
\uvec x\ux &= \uvec \xi  \cos\psi + \uvec \eta \sin\psi \,, &
\uvec y\ux &= \uvec \eta \cos\psi - \uvec \xi  \sin\psi \,,
\end{align}
where $\uvec \xi = (\uvec n \times \uvec \Omega\us) / \|\uvec n \times \uvec \Omega\us\|$, $\uvec \eta = \uvec \xi \times \uvec n$, and $-\uvec n = \uvec x\up \times \uvec y\up = \uvec x\ux \times \uvec y\ux$ is the direction of propagation of the wave.
The polarization matrices are then given by
\begin{equation}
\label{eq:polar-mat}
\mat H = \uvec x \otimes \uvec y + \uvec y \otimes \uvec x \,,
\end{equation}
where we henceforth take $F(t)$, $\mat H$, $\uvec x$, $\uvec y$ to mean either $F\up(t)$, $\mat H\up$, $\uvec x\up$, $\uvec y\up$ or $F\ux(t)$, $\mat H\ux$, $\uvec x\ux$, $\uvec y\ux$, as appropriate.

Combining Eqs.~\eqref{eqs:antenna-patt-def},~\eqref{eq:response-mat}, and~\eqref{eq:polar-mat}, we find that the antenna-pattern functions $F$ can be written as:
\begin{equation}
\label{eq:antenna-patt}
\frac{ F(t) }{ \sin\zeta } = \big[\uvec a(t) \cdot \uvec x\big]\big[\uvec b(t) \cdot \uvec y\big] + \big[\uvec a(t) \cdot \uvec y\big]\big[\uvec b(t) \cdot \uvec x\big] \,.
\end{equation}
Note that when $\uvec a(t)$,~$\uvec b(t)$ and $\uvec x$,~$\uvec y$ are parallel to each other (e.g. when $\uvec a(t)$ is parallel to $\uvec x$, implying $\uvec b(t)$ is parallel to $\uvec y$), $F(t)$ achieves its maximal values of $\pm \sin\zeta$.

The averages of $F(t)^2$ over isotropic source sky position $(\alpha,\delta)$ and orientation $\psi$, and over isotropic detector location $(\Phi\us,\lambda)$ and orientation $\gamma$, are given by:
\begin{align}
\label{eq:sky-avg-antenna-patt}
\avg{F^2}_{\alpha,\sin\delta,\psi} &=
\int_{0}^{2\pi} \!\!\! \dd\alpha
\int_{-1}^{1} \!\!\! \dd(\sin\delta)
\int_{0}^{2\pi} \!\!\! \dd\psi \,
\frac{F^2}{8\pi^2} \,, \\
\label{eq:det-avg-antenna-patt}
\avg{F^2}_{\Phi\us,\sin\lambda,\gamma} &=
\int_{0}^{2\pi} \!\!\! \dd\Phi\us
\int_{-1}^{1} \!\!\! \dd(\sin\lambda)
\int_{0}^{2\pi} \!\!\! \dd\gamma \,
\frac{F^2}{8\pi^2} \,,
\end{align}
where $\Phi\us = \Omega\us (t - t_0)$ is the local sidereal time at the detector.
Note that these two equations can be transformed into each other if one makes the exchanges $\alpha \leftrightarrow \Phi\us$, $\delta \leftrightarrow \lambda$, and $\psi \leftrightarrow \gamma$.
Note too that, since $\avg{F^2}_{\alpha,\sin\delta,\psi}$ averages over all possible orientations of $\uvec x$ and $\uvec y$, it must be invariant to the orientation of $\uvec a(t)$ and $\uvec b(t)$; for the same reason, $\avg{F^2}_{\Phi\us,\sin\lambda,\gamma}$ must be invariant to the orientation of $\uvec x$ and $\uvec y$.
Finally, note that $F(t)$ is invariant if one exchanges $\uvec a(t)$ and $\uvec b(t)$ for $\uvec x$ and $\uvec y$, and vice versa.
We conclude that the averages $\avg{F^2}_{\alpha,\sin\delta,\psi}$ and $\avg{F^2}_{\Phi\us,\sin\lambda,\gamma}$ must be equal.
To calculate e.g.\ $\avg{F^2}_{\alpha,\sin\delta,\psi}$, we choose $\uvec a(t) = (1,0,0)$, $\uvec b(t) = (0,1,0)$ for convenience, and
substituting $\uvec n = (\cos\alpha\cos\delta, \sin\alpha\cos\delta, \sin\delta)$ into Eqs.~\eqref{eq:x-y-cross} obtain
\begin{equation}
\label{eq:skydet-avg-antenna-patt}
\avg{F^2}_{\alpha,\sin\delta,\psi} = \avg{F^2}_{\Phi\us,\sin\lambda,\gamma} = \frac{1}{5} \sin^2\zeta \,.
\end{equation}

To compute the average of $F(t)^2$ over time, $\avg{F^2}_{t}$, we write $\uvec a(t) = \sum_{i=1}^3 c_i(t) \uvec a_i$, where $c_1(t) = \cos \Omega\us (t - t_0)$, $c_2(t) = \sin \Omega\us (t - t_0)$, $c_3(t) = 1$, and $t_0$ is chosen to be the mid-point of the observation time, at which $\uvec a_0$ is calculated; similarly for $\uvec b(t)$.
We can now write
\begin{equation}
\label{eq:time-avg-antenna-patt-int}
\avg{F^2}_{t} = \sin^2\zeta \sum_{i,j=1}^3 \frac{1}{T} \int_{-T/2}^{T/2} \dd t \, \big[ c_i(t) c_j(t) B_{ij} \big]^2 \,,
\end{equation}
where we define
\begin{equation}
B_{ij} = \big(\uvec a_i \cdot \uvec x\big)\big(\uvec b_j \cdot \uvec y\big) + \big(\uvec a_i \cdot \uvec y\big)\big(\uvec b_j \cdot \uvec x\big) \,.
\end{equation}
To write down the result of the integration in Eq.~\eqref{eq:time-avg-antenna-patt-int}, we first define
\begin{subequations}
\label{eqs:JKS-expressions}
\begin{align}
J_{\pm i} &= B_{i\Upr i\Upr} \pm B_{i\Uprpr i\Uprpr} \,, \\
K_{\pm i} &= B_{i\Upr i\Uprpr} \pm B_{i\Uprpr i\Upr} \,, \\
S_{\pm i} &= \sqrt{ B_{i\Upr i\Uprpr} B_{i\Uprpr i\Upr} \pm B_{i\Upr i\Upr} B_{i\Uprpr i\Uprpr} } \,,
\end{align}
\end{subequations}
where $i = 1,2,3$, and $\prime$ denotes a permutation operation: $1 = 3\Upr = 2\Uprpr$, $2 = 1\Upr = 3\Uprpr$, and $3 = 2\Upr = 1\Uprpr$.
We also note the following identities:
\begin{align}
\label{eqs:JF-identities}
J_{-1} &= -B_{11} \,, & J_{-2} &= B_{22} \,, & J_{+3} = B_{33} \,.
\end{align}
Employing Eqs.~\eqref{eqs:JKS-expressions} and~\eqref{eqs:JF-identities}, we find that
\begin{equation}
\label{eq:time-avg-antenna-patt}
\avg{F^2}_{t} = \sin^2\zeta \sum_{n=0}^4 C_n \sinc \frac{n \Omega\us T}{2} \,,
\end{equation}
where
\begin{subequations}
\label{eqs:avg-antenna-patt-C}
\begin{align}
C_{0} &= \frac{ 19 J_{+3}^2 + K_{-3}^2 }{8} + \frac{ K_{+1}^2 + K_{+2}^2 + S_{-3}^2 }{2} \,, \\
C_{1} &= \left( \frac{5}{2} J_{+3} - J_{-1} \right) K_{+2} + \frac{1}{2} K_{+1} K_{+3} \,, \\
C_{2} &= \frac{3}{2} J_{-3} J_{+3} + \frac{1}{2} \big(K_{+2}^2 - K_{+1}^2 \big) \,, \\
C_{3} &= \frac{1}{2} \big( J_{-3} K_{+2} - K_{+1} K_{+3} \big) \,, \\
C_{4} &= \frac{1}{8} \big( J_{-3}^2 - K_{+3}^2 \big) \,.
\end{align}
\end{subequations}

For $T \gtrapprox 1$~sidereal day, $\avg{F^2}_{t} \approx C_0$, which may also be written as
\begin{multline}
\label{eq:time-avg-antenna-patt-C0}
C_0 = 9 a\ull^2 b\ull^2 x\ull^2 y\ull^2 + 8 a\upp^2 b\upp^2 x\upp^2 y\upp^2 \\
+ 2 ( a\ull^2 b\upp^2 + a\upp^2 b\ull^2 - a\ull^2 b\ull^2 ) \\ \times
( x\ull^2 y\upp^2 + x\upp^2 y\ull^2 - x\ull^2 y\ull^2 ) \,,
\end{multline}
where
\begin{align}
a\ull &= \vec \Omega\us \cdot \uvec a_0 \,, &
2a\upp^2 &= 1 - a\ull^2 \,, \\
x\ull &= \vec \Omega\us \cdot \uvec x \,, &
2x\upp^2 &= 1 - x\ull^2 \,,
\end{align}
and similarly for $b\ull$, $b\upp$, $y\ull$ and $y\upp$.
Note that $a\ull$, $b\ull$, $x\ull$, and $y\ull$ (and consequentially $a\upp$, $b\upp$, $x\upp$, and $y\upp$) are independent of the Earth's angular displacement.
We deduce that $C_0$ cannot depend on parameters defined relative to the Earth's angular displacement, which are the detector's local sidereal time $\Phi\us$, and the source's right ascension $\alpha$.

\section{Limited inverse of the central $\chisqr$ distribution}\label{apx:invchisqr}

The right-hand side of Eq.~\eqref{eq:fap} is equal to the normalized upper incomplete gamma function:
\begin{equation}
p \big( s > s\ua \big| s \sim \chisqr[N\us \nu, 0] \big) = \frac{ \Gamma \big( N\us \nu/2, s\ua/2 \big) }{ \Gamma \big( N\us \nu/2 \big) } \,.
\end{equation}
We use a limiting form of the asymptotic inverse of the incomplete gamma function given in \cite{Temme.1992}, which is sufficiently accurate for small values of $p\ua/N\ut$.
It gives the following expression for $s\ua$:
\begin{equation}
\label{eq:false-alarm-thresh}
s\ua = N\us \nu \lambda( \eta ) \,,
\end{equation}
where
\begin{align}
\eta &= \eta_0 + \frac{2}{ N\us \nu \eta_0 } \ln\left( \frac{ \eta_0 }{ \lambda(\eta_0) - 1 } \right) \,, \\
\eta_0 &= \frac{2}{ \sqrt{ N\us \nu } } \erfc^{-1} ( 2 p\ua/N\ut ) \,, \\
\lambda(x) &= -W_{-1} \left[ -\exp \left( -1 - \frac{x^2}{2} \right) \right] \,,~ x \ge 0 \,,
\end{align}
and $W_{-1}$ is the $-1$ branch of the Lambert $W$-function (see e.g\ \cite{Corless.etal.1996}).
For the purposes of this work, $\lambda(x)$ is well approximated by the following two functions [see \cite{DLMF}, Eqs.~(4.13.6) and~(4.13.11)]:
\begin{align}
\lambda_1(x) &= 1 + x + \frac{x^2}{3} + \frac{x^3}{36} - \frac{x^4}{270}
\intertext{for $x \lesssim 2\sqrt\pi$, and}
\lambda_2(x) &= y + \big( 1 + y^{-1} + y^{-2} \big) \ln y \,,~ y = 1 + \frac{x^2}{2} \,,
\end{align}
as $x \rightarrow \infty$.
A sufficiently accurate piecewise expression for $\lambda(x)$ is
\begin{equation}
\begin{cases}
\lambda_1(x) \,, & x < 2 \,, \\
g(x) \lambda_1(x) + [1 - g(x)] \lambda_2(x) \,, & 2 \le x \le 4 \,,  \\
\lambda_2(x) \,, & x > 4 \,, \\
\end{cases}
\end{equation}
where $g(x) = [1 - \tanh 5(x - 3)]/2$.
Equation~\eqref{eq:false-alarm-thresh} is accurate to $\lesssim 0.3\%$ for values of $p\ua/N\ut \le 0.01$.

As $N\us \rightarrow \infty$, $\eta_0 \rightarrow 0$ and $\lambda(\eta_0) \rightarrow \lambda_1(\eta_0) \approx 1 + x$.
With this approximation, $\eta \approx \eta_0$, $s\ua \approx N\us\nu ( 1 + \eta_0 )$, and the normalized false alarm threshold [Eq.~\eqref{eq:za-threshold}] is $z\ua \approx \sqrt2 \erfc^{-1} (2 p\ua/N\ut)$.
Thus, in the limit of large $N\us$, $z\ua$ is independent of $N\us$.

\bibliography{sensitivity_paper}

\begin{thebibliography}{10}%
\makeatletter
\providecommand \@ifxundefined [1]{%
 \ifx #1\undefined \expandafter \@firstoftwo
 \else \expandafter \@secondoftwo
\fi
}%
\providecommand \@ifnum [1]{%
 \ifnum #1\expandafter \@firstoftwo
 \else \expandafter \@secondoftwo
\fi
}%
\providecommand \enquote [1]{``#1''}%
\providecommand \bibnamefont  [1]{#1}%
\providecommand \bibfnamefont [1]{#1}%
\providecommand \citenamefont [1]{#1}%
\providecommand\href[0]{\@sanitize\@href}%
\providecommand\@href[1]{\endgroup\@@startlink{#1}\endgroup\@@href}%
\providecommand\@@href[1]{#1\@@endlink}%
\providecommand \@sanitize [0]{\begingroup\catcode`\&12\catcode`\#12\relax}%
\@ifxundefined \pdfoutput {\@firstoftwo}{%
 \@ifnum{\z@=\pdfoutput}{\@firstoftwo}{\@secondoftwo}%
}{%
 \providecommand\@@startlink[1]{\leavevmode\special{html:<a href="#1">}}%
 \providecommand\@@endlink[0]{\special{html:</a>}}%
}{%
 \providecommand\@@startlink[1]{%
  \leavevmode
  \pdfstartlink
   attr{/Border[0 0 1 ]/H/I/C[0 1 1]}%
   user{/Subtype/Link/A<</Type/Action/S/URI/URI(#1)>>}%
  \relax
 }%
 \providecommand\@@endlink[0]{\pdfendlink}%
}%
\providecommand \url  [0]{\begingroup\@sanitize \@url }%
\providecommand \@url [1]{\endgroup\@href {#1}{\urlprefix}}%
\providecommand \urlprefix [0]{URL }%
\providecommand \Eprint[0]{\href }%
\@ifxundefined \urlstyle {%
  \providecommand \doi [1]{doi:\discretionary{}{}{}#1}%
}{%
  \providecommand \doi [0]{doi:\discretionary{}{}{}\begingroup
  \urlstyle{rm}\Url }%
}%
\providecommand \doibase [0]{http://dx.doi.org/}%
\providecommand \Doi[1]{\href{\doibase#1}}%
\providecommand \bibAnnote [3]{%
  \BibitemShut{#1}%
  \begin{quotation}\noindent
    \textsc{Key:}\ #2\\\textsc{Annotation:}\ #3%
  \end{quotation}%
}%
\providecommand \bibAnnoteFile [2]{%
  \IfFileExists{#2}{\bibAnnote {#1} {#2} {\input{#2}}}{}%
}%
\providecommand \typeout [0]{\immediate \write \m@ne }%
\providecommand \selectlanguage [0]{\@gobble}%
\providecommand \bibinfo [0]{\@secondoftwo}%
\providecommand \bibfield [0]{\@secondoftwo}%
\providecommand \translation [1]{[#1]}%
\providecommand \BibitemOpen[0]{}%
\providecommand \bibitemStop [0]{}%
\providecommand \bibitemNoStop [0]{.\EOS\space}%
\providecommand \EOS [0]{\spacefactor3000\relax}%
\providecommand \BibitemShut [1]{\csname bibitem#1\endcsname}%
\bibitem{Abbott.etal.2009e}%
  \BibitemOpen
  \bibfield{author}{%
  \bibinfo {author} {\bibfnamefont{B.~P.}\ \bibnamefont{Abbott}} \emph{et~al.}
  (\bibinfo {collaboration} {LIGO Scientific Collaboration}),\ }%
  \bibfield{journal}{%
  \Doi{10.1088/0034-4885/72/7/076901}{\bibinfo {journal} {Rep. Prog. Phys.}}\
  }%
  \textbf{\bibinfo {volume} {72}},\ \bibinfo {pages} {076901} (\bibinfo {year}
  {2009}),\ \Eprint{http://arxiv.org/abs/0711.3041}{arXiv:0711.3041 [gr-qc]}%
  \bibAnnoteFile{NoStop}{Abbott.etal.2009e}%
\bibitem{Accadia.etal.2011}%
  \BibitemOpen
  \bibfield{author}{%
  \bibinfo {author} {\bibfnamefont{T.}~\bibnamefont{Accadia}} \emph{et~al.},\
  }%
  \bibfield{journal}{%
  \Doi{10.1088/0264-9381/28/11/114002}{\bibinfo {journal} {Class. Quant.
  Grav.}}\ }%
  \textbf{\bibinfo {volume} {28}},\ \bibinfo {pages} {114002} (\bibinfo {year}
  {2011})%
  \bibAnnoteFile{NoStop}{Accadia.etal.2011}%
\bibitem{Grote.etal.2010}%
  \BibitemOpen
  \bibfield{author}{%
  \bibinfo {author} {\bibfnamefont{H.}~\bibnamefont{Grote}} \emph{et~al.}
  (\bibinfo {collaboration} {LIGO Scientific Collaboration}),\ }%
  \bibfield{journal}{%
  \Doi{10.1088/0264-9381/27/8/084003}{\bibinfo {journal} {Class. Quant.
  Grav.}}\ }%
  \textbf{\bibinfo {volume} {27}},\ \bibinfo {pages} {084003} (\bibinfo {year}
  {2010})%
  \bibAnnoteFile{NoStop}{Grote.etal.2010}%
\bibitem{Harry.etal.2010}%
  \BibitemOpen
  \bibfield{author}{%
  \bibinfo {author} {\bibfnamefont{G.~M.}\ \bibnamefont{Harry}} \emph{et~al.}
  (\bibinfo {collaboration} {LIGO Scientific Collaboration}),\ }%
  \bibfield{journal}{%
  \Doi{10.1088/0264-9381/27/8/084006}{\bibinfo {journal} {Class. Quant.
  Grav.}}\ }%
  \textbf{\bibinfo {volume} {27}},\ \bibinfo {pages} {084006} (\bibinfo {year}
  {2010})%
  \bibAnnoteFile{NoStop}{Harry.etal.2010}%
\bibitem{Kuroda.etal.2010}%
  \BibitemOpen
  \bibfield{author}{%
  \bibinfo {author} {\bibfnamefont{K.}~\bibnamefont{Kuroda}} \emph{et~al.}
  (\bibinfo {collaboration} {LCGT Collaboration}),\ }%
  \bibfield{journal}{%
  \Doi{10.1088/0264-9381/27/8/084004}{\bibinfo {journal} {Class. Quant.
  Grav.}}\ }%
  \textbf{\bibinfo {volume} {27}},\ \bibinfo {pages} {084004} (\bibinfo {year}
  {2010})%
  \bibAnnoteFile{NoStop}{Kuroda.etal.2010}%
\bibitem{Prix.2009}%
  \BibitemOpen
  \bibfield{author}{%
  \bibinfo {author} {\bibfnamefont{R.}~\bibnamefont{Prix}},\ }%
  in\ \Doi{10.1007/978-3-540-76965-1_24}{\emph{\bibinfo {booktitle} {{Neutron
  Stars and Pulsars}}}},\ \bibinfo {series} {Astrophysics and Space Science
  Library}, Vol.\ \bibinfo {volume} {357},\ \bibinfo {editor} {edited by\
  \bibinfo {editor} {\bibfnamefont{W.}~\bibnamefont{Becker}}}\ (\bibinfo
  {publisher} {Springer},\ \bibinfo {address} {Berlin/Heidelberg},\ \bibinfo
  {year} {2009})\ p.\ \bibinfo {pages} {651},\ \bibinfo {note} {preprint
  available from
  \url{https://dcc.ligo.org/cgi-bin/DocDB/ShowDocument?docid=P060039}}%
  \bibAnnoteFile{NoStop}{Prix.2009}%
\bibitem{Bildsten.1998}%
  \BibitemOpen
  \bibfield{author}{%
  \bibinfo {author} {\bibfnamefont{L.}~\bibnamefont{Bildsten}},\ }%
  \bibfield{journal}{%
  \Doi{10.1086/311440}{\bibinfo {journal} {Astrophys.~J}}\ }%
  \textbf{\bibinfo {volume} {501}},\ \bibinfo {pages} {L89} (\bibinfo {year}
  {1998}),\
  \Eprint{http://arxiv.org/abs/astro-ph/9804325}{arXiv:astro-ph/9804325}%
  \bibAnnoteFile{NoStop}{Bildsten.1998}%
\bibitem{Melatos.Payne.2005}%
  \BibitemOpen
  \bibfield{author}{%
  \bibinfo {author} {\bibfnamefont{A.}~\bibnamefont{Melatos}}\ and\ \bibinfo
  {author} {\bibfnamefont{D.~J.~B.}\ \bibnamefont{Payne}},\ }%
  \bibfield{journal}{%
  \Doi{10.1086/428600}{\bibinfo {journal} {Astrophys.~J}}\ }%
  \textbf{\bibinfo {volume} {623}},\ \bibinfo {pages} {1044} (\bibinfo {year}
  {2005}),\
  \Eprint{http://arxiv.org/abs/astro-ph/0503287}{arXiv:astro-ph/0503287}%
  \bibAnnoteFile{NoStop}{Melatos.Payne.2005}%
\bibitem{Vigelius.Melatos.2009}%
  \BibitemOpen
  \bibfield{author}{%
  \bibinfo {author} {\bibfnamefont{M.}~\bibnamefont{Vigelius}}\ and\ \bibinfo
  {author} {\bibfnamefont{A.}~\bibnamefont{Melatos}},\ }%
  \bibfield{journal}{%
  \Doi{10.1111/j.1365-2966.2009.14690.x}{\bibinfo {journal} {M.~Not. R.~Astron.
  Soc}}\ }%
  \textbf{\bibinfo {volume} {395}},\ \bibinfo {pages} {1972} (\bibinfo {year}
  {2009}),\ \Eprint{http://arxiv.org/abs/0902.4264}{arXiv:0902.4264
  [astro-ph.HE]}%
  \bibAnnoteFile{NoStop}{Vigelius.Melatos.2009}%
\bibitem{Abbott.etal.2010}%
  \BibitemOpen
  \bibfield{author}{%
  \bibinfo {author} {\bibfnamefont{B.~P.}\ \bibnamefont{Abbott}} \emph{et~al.}
  (\bibinfo {collaboration} {LIGO Scientific Collaboration et al.}),\ }%
  \bibfield{journal}{%
  \Doi{10.1088/0004-637X/713/1/671}{\bibinfo {journal} {Astrophys.~J}}\ }%
  \textbf{\bibinfo {volume} {713}},\ \bibinfo {pages} {671} (\bibinfo {year}
  {2010}),\ \Eprint{http://arxiv.org/abs/0909.3583}{arXiv:0909.3583
  [astro-ph.HE]}%
  \bibAnnoteFile{NoStop}{Abbott.etal.2010}%
\bibitem{Abadie.etal.2011}%
  \BibitemOpen
  \bibfield{author}{%
  \bibinfo {author} {\bibfnamefont{J.}~\bibnamefont{Abadie}} \emph{et~al.}
  (\bibinfo {collaboration} {LIGO Scientific Collaboration, Virgo Collaboration
  et al.}),\ }%
  \bibfield{journal}{%
  \Doi{10.1088/0004-637X/737/2/93}{\bibinfo {journal} {Astrophys.~J}}\ }%
  \textbf{\bibinfo {volume} {737}},\ \bibinfo {pages} {93} (\bibinfo {year}
  {2011}),\ \Eprint{http://arxiv.org/abs/1104.2712}{arXiv:1104.2712
  [astro-ph.HE]}%
  \bibAnnoteFile{NoStop}{Abadie.etal.2011}%
\bibitem{Abadie.etal.2010a}%
  \BibitemOpen
  \bibfield{author}{%
  \bibinfo {author} {\bibfnamefont{J.}~\bibnamefont{Abadie}} \emph{et~al.}
  (\bibinfo {collaboration} {LIGO Scientific Collaboration}),\ }%
  \bibfield{journal}{%
  \Doi{10.1088/0004-637X/722/2/1504}{\bibinfo {journal} {Astrophys.~J}}\ }%
  \textbf{\bibinfo {volume} {722}},\ \bibinfo {pages} {1504} (\bibinfo {year}
  {2010}),\ \Eprint{http://arxiv.org/abs/1006.2535}{arXiv:1006.2535 [gr-qc]}%
  \bibAnnoteFile{NoStop}{Abadie.etal.2010a}%
\bibitem{Abbott.etal.2009b}%
  \BibitemOpen
  \bibfield{author}{%
  \bibinfo {author} {\bibfnamefont{B.~P.}\ \bibnamefont{Abbott}} \emph{et~al.}
  (\bibinfo {collaboration} {LIGO Scientific Collaboration et al.}),\ }%
  \bibfield{journal}{%
  \Doi{10.1103/PhysRevD.80.042003}{\bibinfo {journal} {Phys. Rev.~D}}\ }%
  \textbf{\bibinfo {volume} {80}},\ \bibinfo {pages} {042003} (\bibinfo {year}
  {2009}),\ \Eprint{http://arxiv.org/abs/0905.1705}{arXiv:0905.1705 [gr-qc]}%
  \bibAnnoteFile{NoStop}{Abbott.etal.2009b}%
\bibitem{Abadie.etal.2011d}%
  \BibitemOpen
  \bibfield{author}{%
  \bibinfo {author} {\bibfnamefont{J.}~\bibnamefont{Abadie}} \emph{et~al.}
  (\bibinfo {collaboration} {LIGO Scientific Collaboration and Virgo
  Collaboration}),\ }%
  \enquote{\bibinfo {title} {{All-sky Search for Periodic Gravitational Waves
  in the Full S5 LIGO Data}},}\  (\bibinfo {year} {2011}),\
  \Eprint{http://arxiv.org/abs/1110.0208}{arXiv:1110.0208 [gr-qc]}%
  \bibAnnoteFile{NoStop}{Abadie.etal.2011d}%
\bibitem{Knispel.Allen.2008}%
  \BibitemOpen
  \bibfield{author}{%
  \bibinfo {author} {\bibfnamefont{B.}~\bibnamefont{Knispel}}\ and\ \bibinfo
  {author} {\bibfnamefont{B.}~\bibnamefont{Allen}},\ }%
  \bibfield{journal}{%
  \Doi{10.1103/PhysRevD.78.044031}{\bibinfo {journal} {Phys. Rev.~D}}\ }%
  \textbf{\bibinfo {volume} {78}},\ \bibinfo {pages} {044031} (\bibinfo {year}
  {2008}),\ \Eprint{http://arxiv.org/abs/0804.3075}{arXiv:0804.3075 [gr-qc]}%
  \bibAnnoteFile{NoStop}{Knispel.Allen.2008}%
\bibitem{Abbott.etal.2007c}%
  \BibitemOpen
  \bibfield{author}{%
  \bibinfo {author} {\bibfnamefont{B.}~\bibnamefont{Abbott}} \emph{et~al.}
  (\bibinfo {collaboration} {LIGO Scientific Collaboration}),\ }%
  \bibfield{journal}{%
  \Doi{10.1103/PhysRevD.76.082001}{\bibinfo {journal} {Phys. Rev.~D}}\ }%
  \textbf{\bibinfo {volume} {76}},\ \bibinfo {pages} {082001} (\bibinfo {year}
  {2007}),\ \Eprint{http://arxiv.org/abs/gr-qc/0605028}{arXiv:gr-qc/0605028}%
  \bibAnnoteFile{NoStop}{Abbott.etal.2007c}%
\bibitem{Abbott.etal.2007d}%
  \BibitemOpen
  \bibfield{author}{%
  \bibinfo {author} {\bibfnamefont{B.}~\bibnamefont{Abbott}} \emph{et~al.}
  (\bibinfo {collaboration} {LIGO Scientific Collaboration}),\ }%
  \bibfield{journal}{%
  \Doi{10.1103/PhysRevD.76.082003}{\bibinfo {journal} {Phys. Rev.~D}}\ }%
  \textbf{\bibinfo {volume} {76}},\ \bibinfo {pages} {082003} (\bibinfo {year}
  {2007}),\
  \Eprint{http://arxiv.org/abs/astro-ph/0703234}{arXiv:astro-ph/0703234}%
  \bibAnnoteFile{NoStop}{Abbott.etal.2007d}%
\bibitem{Abbott.etal.2004c}%
  \BibitemOpen
  \bibfield{author}{%
  \bibinfo {author} {\bibfnamefont{B.}~\bibnamefont{Abbott}} \emph{et~al.}
  (\bibinfo {collaboration} {LIGO Scientific Collaboration}),\ }%
  \bibfield{journal}{%
  \Doi{10.1103/PhysRevD.69.082004}{\bibinfo {journal} {Phys. Rev.~D}}\ }%
  \textbf{\bibinfo {volume} {69}},\ \bibinfo {pages} {082004} (\bibinfo {year}
  {2004}),\ \Eprint{http://arxiv.org/abs/gr-qc/0308050}{arXiv:gr-qc/0308050}%
  \bibAnnoteFile{NoStop}{Abbott.etal.2004c}%
\bibitem{Abbott.etal.2005a}%
  \BibitemOpen
  \bibfield{author}{%
  \bibinfo {author} {\bibfnamefont{B.}~\bibnamefont{Abbott}} \emph{et~al.}
  (\bibinfo {collaboration} {LIGO Scientific Collaboration}),\ }%
  \bibfield{journal}{%
  \Doi{10.1103/PhysRevD.72.102004}{\bibinfo {journal} {Phys. Rev.~D}}\ }%
  \textbf{\bibinfo {volume} {72}},\ \bibinfo {pages} {102004} (\bibinfo {year}
  {2005}),\ \Eprint{http://arxiv.org/abs/gr-qc/0508065}{arXiv:gr-qc/0508065}%
  \bibAnnoteFile{NoStop}{Abbott.etal.2005a}%
\bibitem{Abbott.etal.2008}%
  \BibitemOpen
  \bibfield{author}{%
  \bibinfo {author} {\bibfnamefont{B.}~\bibnamefont{Abbott}} \emph{et~al.}
  (\bibinfo {collaboration} {LIGO Scientific Collaboration}),\ }%
  \bibfield{journal}{%
  \Doi{10.1103/PhysRevD.77.022001}{\bibinfo {journal} {Phys. Rev.~D}}\ }%
  \textbf{\bibinfo {volume} {77}},\ \bibinfo {pages} {022001} (\bibinfo {year}
  {2008}),\ \Eprint{http://arxiv.org/abs/0708.3818}{arXiv:0708.3818 [gr-qc]}%
  \bibAnnoteFile{NoStop}{Abbott.etal.2008}%
\bibitem{Abbott.etal.2009}%
  \BibitemOpen
  \bibfield{author}{%
  \bibinfo {author} {\bibfnamefont{B.}~\bibnamefont{Abbott}} \emph{et~al.}
  (\bibinfo {collaboration} {LIGO Scientific Collaboration}),\ }%
  \bibfield{journal}{%
  \Doi{10.1103/PhysRevD.79.022001}{\bibinfo {journal} {Phys. Rev.~D}}\ }%
  \textbf{\bibinfo {volume} {79}},\ \bibinfo {pages} {022001} (\bibinfo {year}
  {2009}),\ \Eprint{http://arxiv.org/abs/0804.1747}{arXiv:0804.1747 [gr-qc]}%
  \bibAnnoteFile{NoStop}{Abbott.etal.2009}%
\bibitem{Abbott.etal.2009d}%
  \BibitemOpen
  \bibfield{author}{%
  \bibinfo {author} {\bibfnamefont{B.~P.}\ \bibnamefont{Abbott}} \emph{et~al.}
  (\bibinfo {collaboration} {LIGO Scientific Collaboration}),\ }%
  \bibfield{journal}{%
  \Doi{10.1103/PhysRevLett.102.111102}{\bibinfo {journal} {Phys. Rev. Lett}}\
  }%
  \textbf{\bibinfo {volume} {102}},\ \bibinfo {pages} {111102} (\bibinfo {year}
  {2009}),\ \Eprint{http://arxiv.org/abs/0810.0283}{arXiv:0810.0283 [gr-qc]}%
  \bibAnnoteFile{NoStop}{Abbott.etal.2009d}%
\bibitem{Jaranowski.etal.1998}%
  \BibitemOpen
  \bibfield{author}{%
  \bibinfo {author} {\bibfnamefont{P.}~\bibnamefont{Jaranowski}}, \bibinfo
  {author} {\bibfnamefont{A.}~\bibnamefont{Kr\'olak}},\ and\ \bibinfo {author}
  {\bibfnamefont{B.~F.}\ \bibnamefont{Schutz}},\ }%
  \bibfield{journal}{%
  \Doi{10.1103/PhysRevD.58.063001}{\bibinfo {journal} {Phys. Rev.~D}}\ }%
  \textbf{\bibinfo {volume} {58}},\ \bibinfo {pages} {063001} (\bibinfo {year}
  {1998}),\ \Eprint{http://arxiv.org/abs/gr-qc/9804014}{arXiv:gr-qc/9804014}%
  \bibAnnoteFile{NoStop}{Jaranowski.etal.1998}%
\bibitem{Brady.Creighton.2000}%
  \BibitemOpen
  \bibfield{author}{%
  \bibinfo {author} {\bibfnamefont{P.~R.}\ \bibnamefont{Brady}}\ and\ \bibinfo
  {author} {\bibfnamefont{T.}~\bibnamefont{Creighton}},\ }%
  \bibfield{journal}{%
  \Doi{10.1103/PhysRevD.61.082001}{\bibinfo {journal} {Phys. Rev.~D}}\ }%
  \textbf{\bibinfo {volume} {61}},\ \bibinfo {pages} {082001} (\bibinfo {year}
  {2000}),\ \Eprint{http://arxiv.org/abs/gr-qc/9812014}{arXiv:gr-qc/9812014}%
  \bibAnnoteFile{NoStop}{Brady.Creighton.2000}%
\bibitem{Jaranowski.Krolak.2000}%
  \BibitemOpen
  \bibfield{author}{%
  \bibinfo {author} {\bibfnamefont{P.}~\bibnamefont{Jaranowski}}\ and\ \bibinfo
  {author} {\bibfnamefont{A.}~\bibnamefont{Kr\'olak}},\ }%
  \bibfield{journal}{%
  \Doi{10.1103/PhysRevD.61.062001}{\bibinfo {journal} {Phys. Rev.~D}}\ }%
  \textbf{\bibinfo {volume} {61}},\ \bibinfo {pages} {062001} (\bibinfo {year}
  {2000}),\ \Eprint{http://arxiv.org/abs/gr-qc/9901013}{arXiv:gr-qc/9901013}%
  \bibAnnoteFile{NoStop}{Jaranowski.Krolak.2000}%
\bibitem{Krishnan.etal.2004}%
  \BibitemOpen
  \bibfield{author}{%
  \bibinfo {author} {\bibfnamefont{B.}~\bibnamefont{Krishnan}}, \bibinfo
  {author} {\bibfnamefont{A.~M.}\ \bibnamefont{Sintes}}, \bibinfo {author}
  {\bibfnamefont{M.~A.}\ \bibnamefont{Papa}}, \bibinfo {author}
  {\bibfnamefont{B.~F.}\ \bibnamefont{Schutz}}, \bibinfo {author}
  {\bibfnamefont{S.}~\bibnamefont{Frasca}},\ and\ \bibinfo {author}
  {\bibfnamefont{C.}~\bibnamefont{Palomba}},\ }%
  \bibfield{journal}{%
  \Doi{10.1103/PhysRevD.70.082001}{\bibinfo {journal} {Phys. Rev.~D}}\ }%
  \textbf{\bibinfo {volume} {70}},\ \bibinfo {pages} {082001} (\bibinfo {year}
  {2004}),\ \Eprint{http://arxiv.org/abs/gr-qc/0407001}{arXiv:gr-qc/0407001}%
  \bibAnnoteFile{NoStop}{Krishnan.etal.2004}%
\bibitem{Cutler.etal.2005}%
  \BibitemOpen
  \bibfield{author}{%
  \bibinfo {author} {\bibfnamefont{C.}~\bibnamefont{Cutler}}, \bibinfo {author}
  {\bibfnamefont{I.}~\bibnamefont{Gholami}},\ and\ \bibinfo {author}
  {\bibfnamefont{B.}~\bibnamefont{Krishnan}},\ }%
  \bibfield{journal}{%
  \Doi{10.1103/PhysRevD.72.042004}{\bibinfo {journal} {Phys. Rev.~D}}\ }%
  \textbf{\bibinfo {volume} {72}},\ \bibinfo {pages} {042004} (\bibinfo {year}
  {2005}),\ \Eprint{http://arxiv.org/abs/gr-qc/0505082}{arXiv:gr-qc/0505082}%
  \bibAnnoteFile{NoStop}{Cutler.etal.2005}%
\bibitem{Mendell.Landry.2005}%
  \BibitemOpen
  \bibfield{author}{%
  \bibinfo {author} {\bibfnamefont{G.}~\bibnamefont{Mendell}}\ and\ \bibinfo
  {author} {\bibfnamefont{M.}~\bibnamefont{Landry}},\ }%
  \emph{\bibinfo {title} {{StackSlide and Hough Search SNR and Statistics}}},\
  \bibinfo {type} {Tech. Rep.}\ \bibinfo {number} {T050003-x0}\ (\bibinfo
  {institution} {LIGO},\ \bibinfo {year} {2005})\
  \url{https://dcc.ligo.org/cgi-bin/DocDB/ShowDocument?docid=T050003}%
  \bibAnnoteFile{NoStop}{Mendell.Landry.2005}%
\bibitem{Krishnan.Sintes.2007}%
  \BibitemOpen
  \bibfield{author}{%
  \bibinfo {author} {\bibfnamefont{B.}~\bibnamefont{Krishnan}}\ and\ \bibinfo
  {author} {\bibfnamefont{A.~M.}\ \bibnamefont{Sintes}},\ }%
  \emph{\bibinfo {title} {{Hough search with improved sensitivity}}},\ \bibinfo
  {type} {Tech. Rep.}\ \bibinfo {number} {T070124-x0}\ (\bibinfo {institution}
  {LIGO},\ \bibinfo {year} {2007})\
  \url{https://dcc.ligo.org/cgi-bin/DocDB/ShowDocument?docid=T070124}%
  \bibAnnoteFile{NoStop}{Krishnan.Sintes.2007}%
\bibitem{Watts.etal.2008}%
  \BibitemOpen
  \bibfield{author}{%
  \bibinfo {author} {\bibfnamefont{A.~L.}\ \bibnamefont{Watts}}, \bibinfo
  {author} {\bibfnamefont{B.}~\bibnamefont{Krishnan}}, \bibinfo {author}
  {\bibfnamefont{L.}~\bibnamefont{Bildsten}},\ and\ \bibinfo {author}
  {\bibfnamefont{B.~F.}\ \bibnamefont{Schutz}},\ }%
  \bibfield{journal}{%
  \Doi{10.1111/j.1365-2966.2008.13594.x}{\bibinfo {journal} {M.~Not. R.~Astron.
  Soc}}\ }%
  \textbf{\bibinfo {volume} {389}},\ \bibinfo {pages} {839} (\bibinfo {year}
  {2008}),\ \Eprint{http://arxiv.org/abs/0803.4097}{arXiv:0803.4097
  [astro-ph]}%
  \bibAnnoteFile{NoStop}{Watts.etal.2008}%
\bibitem{Prix.Shaltev.2011}%
  \BibitemOpen
  \bibfield{author}{%
  \bibinfo {author} {\bibfnamefont{R.}~\bibnamefont{Prix}}\ and\ \bibinfo
  {author} {\bibfnamefont{M.}~\bibnamefont{Shaltev}},\ }%
  \enquote{\bibinfo {title} {{Continuous gravitational waves searches at fixed
  computing cost: The single stage of hierarchical search}},}\  (\bibinfo
  {year} {2011}),\ \Eprint{http://arxiv.org/abs/1201.4321}{arXiv:1201.4321}%
  \bibAnnoteFile{NoStop}{Prix.Shaltev.2011}%
\bibitem{Betzwieser.2007}%
  \BibitemOpen
  \bibfield{author}{%
  \bibinfo {author} {\bibfnamefont{J.}~\bibnamefont{Betzwieser}},\ }%
  \emph{\bibinfo {title} {{Analysis of spatial mode sensitivity of a
  gravitational wave interferometer and a targeted search for gravitational
  radiation from the Crab pulsar}}},\ Ph.D. thesis,\ \bibinfo {school}
  {Massachusetts Institute of Technology} (\bibinfo {year} {2007}),\
  \url{http://hdl.handle.net/1721.1/45422}%
  \bibAnnoteFile{NoStop}{Betzwieser.2007}%
\bibitem{Wette.2009}%
  \BibitemOpen
  \bibfield{author}{%
  \bibinfo {author} {\bibfnamefont{K.~W.}\ \bibnamefont{Wette}},\ }%
  \emph{\bibinfo {title} {{Gravitational waves from accreting neutron stars and
  Cassiopeia A}}},\ Ph.D. thesis,\ \bibinfo {school} {The Australian National
  University} (\bibinfo {year} {2009}),\ \url{http://hdl.handle.net/1885/7354}%
  \bibAnnoteFile{NoStop}{Wette.2009}%
\bibitem{LSC.VC.2011}%
  \BibitemOpen
  \bibfield{author}{%
  \bibinfo {author} {\bibnamefont{LSC}}\ and\ \bibinfo {author}
  {\bibnamefont{VC}} (\bibinfo {collaboration} {LIGO Scientific Collaboration
  and Virgo Collaboration}),\ }%
  \emph{\bibinfo {title} {{The 2011-2012 data analysis, software and computing,
  detector characterization white paper}}},\ \bibinfo {type} {Tech. Rep.}\
  \bibinfo {number} {T1100322-v3}\ (\bibinfo {institution} {LSC-VC},\ \bibinfo
  {year} {2011})\
  \url{https://dcc.ligo.org/cgi-bin/DocDB/ShowDocument?docid=T1100322}%
  \bibAnnoteFile{NoStop}{LSC.VC.2011}%
\bibitem{VanDenBroeck.2005}%
  \BibitemOpen
  \bibfield{author}{%
  \bibinfo {author} {\bibfnamefont{C.}~\bibnamefont{Van Den~Broeck}},\ }%
  \bibfield{journal}{%
  \Doi{10.1088/0264-9381/22/9/022}{\bibinfo {journal} {Class. Quant. Grav.}}\
  }%
  \textbf{\bibinfo {volume} {22}},\ \bibinfo {pages} {1825} (\bibinfo {year}
  {2005}),\ \Eprint{http://arxiv.org/abs/gr-qc/0411030}{gr-qc/0411030}%
  \bibAnnoteFile{NoStop}{VanDenBroeck.2005}%
\bibitem{Brady.etal.1998}%
  \BibitemOpen
  \bibfield{author}{%
  \bibinfo {author} {\bibfnamefont{P.~R.}\ \bibnamefont{Brady}}, \bibinfo
  {author} {\bibfnamefont{T.}~\bibnamefont{Creighton}}, \bibinfo {author}
  {\bibfnamefont{C.}~\bibnamefont{Cutler}},\ and\ \bibinfo {author}
  {\bibfnamefont{B.~F.}\ \bibnamefont{Schutz}},\ }%
  \bibfield{journal}{%
  \Doi{10.1103/PhysRevD.57.2101}{\bibinfo {journal} {Phys. Rev.~D}}\ }%
  \textbf{\bibinfo {volume} {57}},\ \bibinfo {pages} {2101} (\bibinfo {year}
  {1998}),\ \Eprint{http://arxiv.org/abs/gr-qc/9702050}{arXiv:gr-qc/9702050}%
  \bibAnnoteFile{NoStop}{Brady.etal.1998}%
\bibitem{Jaranowski.Krolak.1999}%
  \BibitemOpen
  \bibfield{author}{%
  \bibinfo {author} {\bibfnamefont{P.}~\bibnamefont{Jaranowski}}\ and\ \bibinfo
  {author} {\bibfnamefont{A.}~\bibnamefont{Kr\'olak}},\ }%
  \bibfield{journal}{%
  \Doi{10.1103/PhysRevD.59.063003}{\bibinfo {journal} {Phys. Rev.~D}}\ }%
  \textbf{\bibinfo {volume} {59}},\ \bibinfo {pages} {063003} (\bibinfo {year}
  {1999}),\ \Eprint{http://arxiv.org/abs/gr-qc/9809046}{arXiv:gr-qc/9809046}%
  \bibAnnoteFile{NoStop}{Jaranowski.Krolak.1999}%
\bibitem{Wette.etal.2008}%
  \BibitemOpen
  \bibfield{author}{%
  \bibinfo {author} {\bibfnamefont{K.}~\bibnamefont{Wette}} \emph{et~al.},\ }%
  \bibfield{journal}{%
  \Doi{10.1088/0264-9381/25/23/235011}{\bibinfo {journal} {Class. Quant.
  Grav.}}\ }%
  \textbf{\bibinfo {volume} {25}},\ \bibinfo {pages} {235011} (\bibinfo {year}
  {2008}),\ \Eprint{http://arxiv.org/abs/0802.3332}{arXiv:0802.3332 [gr-qc]}%
  \bibAnnoteFile{NoStop}{Wette.etal.2008}%
\bibitem{Schutz.Tinto.1987}%
  \BibitemOpen
  \bibfield{author}{%
  \bibinfo {author} {\bibfnamefont{B.~F.}\ \bibnamefont{Schutz}}\ and\ \bibinfo
  {author} {\bibfnamefont{M.}~\bibnamefont{Tinto}},\ }%
  \bibfield{journal}{%
  \bibinfo {journal} {M.~Not. R.~Astron. Soc}\ }%
  \textbf{\bibinfo {volume} {224}},\ \bibinfo {pages} {131} (\bibinfo {year}
  {1987}),\ \url{http://adsabs.harvard.edu/abs/1987MNRAS.224..131S}%
  \bibAnnoteFile{NoStop}{Schutz.Tinto.1987}%
\bibitem{Jaranowski.Krolak.1994}%
  \BibitemOpen
  \bibfield{author}{%
  \bibinfo {author} {\bibfnamefont{P.}~\bibnamefont{Jaranowski}}\ and\ \bibinfo
  {author} {\bibfnamefont{A.}~\bibnamefont{Krolak}},\ }%
  \bibfield{journal}{%
  \Doi{10.1103/PhysRevD.49.1723}{\bibinfo {journal} {Phys. Rev.~D}}\ }%
  \textbf{\bibinfo {volume} {49}},\ \bibinfo {pages} {1723} (\bibinfo {year}
  {1994})%
  \bibAnnoteFile{NoStop}{Jaranowski.Krolak.1994}%
\bibitem{Bonazzola.Gourgoulhon.1996}%
  \BibitemOpen
  \bibfield{author}{%
  \bibinfo {author} {\bibfnamefont{S.}~\bibnamefont{Bonazzola}}\ and\ \bibinfo
  {author} {\bibfnamefont{E.}~\bibnamefont{Gourgoulhon}},\ }%
  \bibfield{journal}{%
  \bibinfo {journal} {Astron. Astrophys.}\ }%
  \textbf{\bibinfo {volume} {312}},\ \bibinfo {pages} {675} (\bibinfo {year}
  {1996}),\
  \Eprint{http://arxiv.org/abs/astro-ph/9602107}{arXiv:astro-ph/9602107}%
  \bibAnnoteFile{NoStop}{Bonazzola.Gourgoulhon.1996}%
\bibitem{Srivastava.Sahay.2002}%
  \BibitemOpen
  \bibfield{author}{%
  \bibinfo {author} {\bibfnamefont{D.~C.}\ \bibnamefont{Srivastava}}\ and\
  \bibinfo {author} {\bibfnamefont{S.~K.}\ \bibnamefont{Sahay}},\ }%
  \bibfield{journal}{%
  \Doi{10.1046/j.1365-8711.2002.06032.x}{\bibinfo {journal} {M.~Not. R.~Astron.
  Soc}}\ }%
  \textbf{\bibinfo {volume} {337}},\ \bibinfo {pages} {305} (\bibinfo {year}
  {2002}),\ \Eprint{http://arxiv.org/abs/gr-qc/0111050}{arXiv:gr-qc/0111050}%
  \bibAnnoteFile{NoStop}{Srivastava.Sahay.2002}%
\bibitem{Cutler.Schutz.2005}%
  \BibitemOpen
  \bibfield{author}{%
  \bibinfo {author} {\bibfnamefont{C.}~\bibnamefont{Cutler}}\ and\ \bibinfo
  {author} {\bibfnamefont{B.~F.}\ \bibnamefont{Schutz}},\ }%
  \bibfield{journal}{%
  \Doi{10.1103/PhysRevD.72.063006}{\bibinfo {journal} {Phys. Rev.~D}}\ }%
  \textbf{\bibinfo {volume} {72}},\ \bibinfo {pages} {063006} (\bibinfo {year}
  {2005}),\ \Eprint{http://arxiv.org/abs/gr-qc/0504011}{arXiv:gr-qc/0504011}%
  \bibAnnoteFile{NoStop}{Cutler.Schutz.2005}%
\bibitem{Prix.2007}%
  \BibitemOpen
  \bibfield{author}{%
  \bibinfo {author} {\bibfnamefont{R.}~\bibnamefont{Prix}},\ }%
  \bibfield{journal}{%
  \Doi{10.1103/PhysRevD.75.023004}{\bibinfo {journal} {Phys. Rev.~D}}\ }%
  \textbf{\bibinfo {volume} {75}},\ \bibinfo {pages} {023004} (\bibinfo {year}
  {2007}),\ \Eprint{http://arxiv.org/abs/gr-qc/0606088}{arXiv:gr-qc/0606088}%
  \bibAnnoteFile{NoStop}{Prix.2007}%
\bibitem{Prix.2007a}%
  \BibitemOpen
  \bibfield{author}{%
  \bibinfo {author} {\bibfnamefont{R.}~\bibnamefont{Prix}},\ }%
  \bibfield{journal}{%
  \Doi{10.1088/0264-9381/24/19/S11}{\bibinfo {journal} {Class. Quant. Grav.}}\
  }%
  \textbf{\bibinfo {volume} {24}},\ \bibinfo {pages} {S481} (\bibinfo {year}
  {2007}),\ \Eprint{http://arxiv.org/abs/0707.0428}{arXiv:0707.0428 [gr-qc]}%
  \bibAnnoteFile{NoStop}{Prix.2007a}%
\bibitem{Dergachev.2010a}%
  \BibitemOpen
  \bibfield{author}{%
  \bibinfo {author} {\bibfnamefont{V.}~\bibnamefont{Dergachev}},\ }%
  \emph{\bibinfo {title} {{Description of PowerFlux 2 algorithms and
  implementation}}},\ \bibinfo {type} {Tech. Rep.}\ \bibinfo {number}
  {T1000272-v5}\ (\bibinfo {institution} {LIGO},\ \bibinfo {year} {2010})\
  \url{https://dcc.ligo.org/cgi-bin/DocDB/ShowDocument?docid=T1000272}%
  \bibAnnoteFile{NoStop}{Dergachev.2010a}%
\bibitem{Dhurandhar.etal.2008}%
  \BibitemOpen
  \bibfield{author}{%
  \bibinfo {author} {\bibfnamefont{S.}~\bibnamefont{Dhurandhar}}, \bibinfo
  {author} {\bibfnamefont{B.}~\bibnamefont{Krishnan}}, \bibinfo {author}
  {\bibfnamefont{H.}~\bibnamefont{Mukhopadhyay}},\ and\ \bibinfo {author}
  {\bibfnamefont{J.~T.}\ \bibnamefont{Whelan}},\ }%
  \bibfield{journal}{%
  \Doi{10.1103/PhysRevD.77.082001}{\bibinfo {journal} {Phys. Rev.~D}}\ }%
  \textbf{\bibinfo {volume} {77}},\ \bibinfo {pages} {082001} (\bibinfo {year}
  {2008}),\ \Eprint{http://arxiv.org/abs/0712.1578}{arXiv:0712.1578 [gr-qc]}%
  \bibAnnoteFile{NoStop}{Dhurandhar.etal.2008}%
\bibitem{Pletsch.Allen.2009}%
  \BibitemOpen
  \bibfield{author}{%
  \bibinfo {author} {\bibfnamefont{H.~J.}\ \bibnamefont{Pletsch}}\ and\
  \bibinfo {author} {\bibfnamefont{B.}~\bibnamefont{Allen}},\ }%
  \bibfield{journal}{%
  \Doi{10.1103/PhysRevLett.103.181102}{\bibinfo {journal} {Phys. Rev. Lett}}\
  }%
  \textbf{\bibinfo {volume} {103}},\ \bibinfo {pages} {181102} (\bibinfo {year}
  {2009}),\ \Eprint{http://arxiv.org/abs/0906.0023}{arXiv:0906.0023 [gr-qc]}%
  \bibAnnoteFile{NoStop}{Pletsch.Allen.2009}%
\bibitem{Dergachev.2010}%
  \BibitemOpen
  \bibfield{author}{%
  \bibinfo {author} {\bibfnamefont{V.}~\bibnamefont{Dergachev}},\ }%
  \bibfield{journal}{%
  \Doi{10.1088/0264-9381/27/20/205017}{\bibinfo {journal} {Class. Quant.
  Grav.}}\ }%
  \textbf{\bibinfo {volume} {27}},\ \bibinfo {pages} {205017} (\bibinfo {year}
  {2010})%
  \bibAnnoteFile{NoStop}{Dergachev.2010}%
\bibitem{Pletsch.2011}%
  \BibitemOpen
  \bibfield{author}{%
  \bibinfo {author} {\bibfnamefont{H.~J.}\ \bibnamefont{Pletsch}},\ }%
  \bibfield{journal}{%
  \Doi{10.1103/PhysRevD.83.122003}{\bibinfo {journal} {Phys. Rev.~D}}\ }%
  \textbf{\bibinfo {volume} {83}},\ \bibinfo {pages} {122003} (\bibinfo {year}
  {2011}),\ \Eprint{http://arxiv.org/abs/1101.5396}{arXiv:1101.5396 [gr-qc]}%
  \bibAnnoteFile{NoStop}{Pletsch.2011}%
\bibitem{Cutler.2011}%
  \BibitemOpen
  \bibfield{author}{%
  \bibinfo {author} {\bibfnamefont{C.}~\bibnamefont{Cutler}},\ }%
  \enquote{\bibinfo {title} {{An improved, ``phase-relaxed'' F-statistic for
  gravitational-wave data analysis}},}\  (\bibinfo {year} {2011}),\
  \Eprint{http://arxiv.org/abs/1104.2938}{arXiv:1104.2938 [gr-qc]}%
  \bibAnnoteFile{NoStop}{Cutler.2011}%
\bibitem{Neyman.Pearson.1933}%
  \BibitemOpen
  \bibfield{author}{%
  \bibinfo {author} {\bibfnamefont{J.}~\bibnamefont{Neyman}}\ and\ \bibinfo
  {author} {\bibfnamefont{E.~S.}\ \bibnamefont{Pearson}},\ }%
  \bibfield{journal}{%
  \Doi{10.1098/rsta.1933.0009}{\bibinfo {journal} {Phil. Trans. R.~Soc.
  Lond.~A}}\ }%
  \textbf{\bibinfo {volume} {231}},\ \bibinfo {pages} {289} (\bibinfo {year}
  {1933})%
  \bibAnnoteFile{NoStop}{Neyman.Pearson.1933}%
\bibitem{Neyman.1937}%
  \BibitemOpen
  \bibfield{author}{%
  \bibinfo {author} {\bibfnamefont{J.}~\bibnamefont{Neyman}},\ }%
  \bibfield{journal}{%
  \Doi{10.1098/rsta.1937.0005}{\bibinfo {journal} {Phil. Trans. R.~Soc.
  Lond.~A}}\ }%
  \textbf{\bibinfo {volume} {236}},\ \bibinfo {pages} {333} (\bibinfo {year}
  {1937})%
  \bibAnnoteFile{NoStop}{Neyman.1937}%
\bibitem{Johnson.etal.1994}%
  \BibitemOpen
  \bibfield{author}{%
  \bibinfo {author} {\bibfnamefont{N.~L.}\ \bibnamefont{Johnson}}, \bibinfo
  {author} {\bibfnamefont{S.}~\bibnamefont{Kotz}},\ and\ \bibinfo {author}
  {\bibfnamefont{N.}~\bibnamefont{Balakrishnan}},\ }%
  \emph{\bibinfo {title} {{Continuous Univariate Distributions}}},\ \bibinfo
  {edition} {2nd}\ ed.,\ Wiley Series in Probability and Mathematical
  Statistics\ (\bibinfo {publisher} {Wiley},\ \bibinfo {address} {N.Y.},\
  \bibinfo {year} {1994})%
  \bibAnnoteFile{NoStop}{Johnson.etal.1994}%
\bibitem{Temme.1992}%
  \BibitemOpen
  \bibfield{author}{%
  \bibinfo {author} {\bibfnamefont{N.~M.}\ \bibnamefont{Temme}},\ }%
  \bibfield{journal}{%
  \Doi{10.1090/S0025-5718-1992-1122079-8}{\bibinfo {journal} {Math. Comp.}}\ }%
  \textbf{\bibinfo {volume} {58}},\ \bibinfo {pages} {755} (\bibinfo {year}
  {1992})%
  \bibAnnoteFile{NoStop}{Temme.1992}%
\bibitem{Lorimer.2009}%
  \BibitemOpen
  \bibfield{author}{%
  \bibinfo {author} {\bibfnamefont{D.~R.}\ \bibnamefont{Lorimer}},\ }%
  in\ \Doi{10.1007/978-3-540-76965-1_1}{\emph{\bibinfo {booktitle} {{Neutron
  Stars and Pulsars}}}},\ \bibinfo {series} {Astrophysics and Space Science
  Library}, Vol.\ \bibinfo {volume} {357},\ \bibinfo {editor} {edited by\
  \bibinfo {editor} {\bibfnamefont{W.}~\bibnamefont{Becker}}}\ (\bibinfo
  {publisher} {Springer},\ \bibinfo {address} {Berlin/Heidelberg},\ \bibinfo
  {year} {2009})\ p.~\bibinfo {pages} {1}%
  \bibAnnoteFile{NoStop}{Lorimer.2009}%
\bibitem{Abadie.etal.2010b}%
  \BibitemOpen
  \bibfield{author}{%
  \bibinfo {author} {\bibfnamefont{J.}~\bibnamefont{Abadie}} \emph{et~al.}
  (\bibinfo {collaboration} {LIGO Scientific Collaboration}),\ }%
  \bibfield{journal}{%
  \Doi{10.1016/j.nima.2010.07.089}{\bibinfo {journal} {Nucl. Instrum.
  Meth.~A}}\ }%
  \textbf{\bibinfo {volume} {624}},\ \bibinfo {pages} {223} (\bibinfo {year}
  {2010}),\ \Eprint{http://arxiv.org/abs/1007.3973}{arXiv:1007.3973 [gr-qc]}%
  \bibAnnoteFile{NoStop}{Abadie.etal.2010b}%
\bibitem{Prix.2010}%
  \BibitemOpen
  \bibfield{author}{%
  \bibinfo {author} {\bibfnamefont{R.}~\bibnamefont{Prix}},\ }%
  \emph{\bibinfo {title} {{The F-statistic and its implementation in
  ComputeFStatistic\_v2}}},\ \bibinfo {type} {Tech. Rep.}\ \bibinfo {number}
  {T0900149-v3}\ (\bibinfo {institution} {LIGO},\ \bibinfo {year} {2010})\
  \url{https://dcc.ligo.org/cgi-bin/DocDB/ShowDocument?docid=T0900149}%
  \bibAnnoteFile{NoStop}{Prix.2010}%
\bibitem{Prix.2011}%
  \BibitemOpen
  \bibfield{author}{%
  \bibinfo {author} {\bibfnamefont{R.}~\bibnamefont{Prix}},\ }%
  \emph{\bibinfo {title} {{F-statistic bias due to noise-estimator}}},\
  \bibinfo {type} {Tech. Rep.}\ \bibinfo {number} {T1100551-v1}\ (\bibinfo
  {institution} {LIGO},\ \bibinfo {year} {2011})\
  \url{https://dcc.ligo.org/cgi-bin/DocDB/ShowDocument?docid=T1100551}%
  \bibAnnoteFile{NoStop}{Prix.2011}%
\bibitem{Schutz.2011}%
  \BibitemOpen
  \bibfield{author}{%
  \bibinfo {author} {\bibfnamefont{B.~F.}\ \bibnamefont{Schutz}},\ }%
  \bibfield{journal}{%
  \Doi{10.1088/0264-9381/28/12/125023}{\bibinfo {journal} {Class. Quant.
  Grav.}}\ }%
  \textbf{\bibinfo {volume} {28}},\ \bibinfo {pages} {125023} (\bibinfo {year}
  {2011}),\ \Eprint{http://arxiv.org/abs/1102.5421}{arXiv:1102.5421
  [astro-ph.IM]}%
  \bibAnnoteFile{NoStop}{Schutz.2011}%
\bibitem{Searle.etal.2002}%
  \BibitemOpen
  \bibfield{author}{%
  \bibinfo {author} {\bibfnamefont{A.~C.}\ \bibnamefont{Searle}}, \bibinfo
  {author} {\bibfnamefont{S.~M.}\ \bibnamefont{Scott}},\ and\ \bibinfo {author}
  {\bibfnamefont{D.~E.}\ \bibnamefont{McClelland}},\ }%
  \bibfield{journal}{%
  \Doi{10.1088/0264-9381/19/7/331}{\bibinfo {journal} {Class. Quant. Grav.}}\
  }%
  \textbf{\bibinfo {volume} {19}},\ \bibinfo {pages} {1465} (\bibinfo {year}
  {2002}),\ \Eprint{http://arxiv.org/abs/gr-qc/0110053}{arXiv:gr-qc/0110053}%
  \bibAnnoteFile{NoStop}{Searle.etal.2002}%
\bibitem{Messenger.etal.2009}%
  \BibitemOpen
  \bibfield{author}{%
  \bibinfo {author} {\bibfnamefont{C.}~\bibnamefont{Messenger}}, \bibinfo
  {author} {\bibfnamefont{R.}~\bibnamefont{Prix}},\ and\ \bibinfo {author}
  {\bibfnamefont{M.~A.}\ \bibnamefont{Papa}},\ }%
  \bibfield{journal}{%
  \Doi{10.1103/PhysRevD.79.104017}{\bibinfo {journal} {Phys. Rev.~D}}\ }%
  \textbf{\bibinfo {volume} {79}},\ \bibinfo {pages} {104017} (\bibinfo {year}
  {2009}),\ \Eprint{http://arxiv.org/abs/0809.5223}{arXiv:0809.5223 [gr-qc]}%
  \bibAnnoteFile{NoStop}{Messenger.etal.2009}%
\bibitem{Feldman.Cousins.1998}%
  \BibitemOpen
  \bibfield{author}{%
  \bibinfo {author} {\bibfnamefont{G.~J.}\ \bibnamefont{Feldman}}\ and\
  \bibinfo {author} {\bibfnamefont{R.~D.}\ \bibnamefont{Cousins}},\ }%
  \bibfield{journal}{%
  \Doi{10.1103/PhysRevD.57.3873}{\bibinfo {journal} {Phys. Rev.~D}}\ }%
  \textbf{\bibinfo {volume} {57}},\ \bibinfo {pages} {3873} (\bibinfo {year}
  {1998}),\
  \Eprint{http://arxiv.org/abs/physics/9711021}{arXiv:physics/9711021}%
  \bibAnnoteFile{NoStop}{Feldman.Cousins.1998}%
\bibitem{PF_UL_conversion}%
  \BibitemOpen
  \bibfield{author}{%
  \bibinfo {author} {\bibfnamefont{V.}~\bibnamefont{Dergachev}},\ }%
  \enquote{\bibinfo {title} {{Conversion of PowerFlux-style limits to 95\%
  population based limits}},}\  (\bibinfo {year} {2010}),\ \bibinfo {note}
  {unpublished}%
  \bibAnnoteFile{NoStop}{PF_UL_conversion}%
\bibitem{Rover.etal.2011}%
  \BibitemOpen
  \bibfield{author}{%
  \bibinfo {author} {\bibfnamefont{C.}~\bibnamefont{R\"over}}, \bibinfo
  {author} {\bibfnamefont{C.}~\bibnamefont{Messenger}},\ and\ \bibinfo {author}
  {\bibfnamefont{R.}~\bibnamefont{Prix}},\ }%
  \enquote{\bibinfo {title} {{Bayesian versus frequentist upper limits}},}\
  (\bibinfo {year} {2011}),\
  \Eprint{http://arxiv.org/abs/1103.2987}{arXiv:1103.2987 [physics.data-an]}%
  \bibAnnoteFile{NoStop}{Rover.etal.2011}%
\bibitem{Finn.2009}%
  \BibitemOpen
  \bibfield{author}{%
  \bibinfo {author} {\bibfnamefont{L.~S.}\ \bibnamefont{Finn}},\ }%
  \bibfield{journal}{%
  \Doi{10.1103/PhysRevD.79.022002}{\bibinfo {journal} {Phys. Rev.~D}}\ }%
  \textbf{\bibinfo {volume} {79}},\ \bibinfo {pages} {022002} (\bibinfo {year}
  {2009}),\ \Eprint{http://arxiv.org/abs/0810.4529}{arXiv:0810.4529 [gr-qc]}%
  \bibAnnoteFile{NoStop}{Finn.2009}%
\bibitem{Prix.Whelan.2007}%
  \BibitemOpen
  \bibfield{author}{%
  \bibinfo {author} {\bibfnamefont{R.}~\bibnamefont{Prix}}\ and\ \bibinfo
  {author} {\bibfnamefont{J.~T.}\ \bibnamefont{Whelan}},\ }%
  \bibfield{journal}{%
  \Doi{10.1088/0264-9381/24/19/S19}{\bibinfo {journal} {Class. Quant. Grav.}}\
  }%
  \textbf{\bibinfo {volume} {24}},\ \bibinfo {pages} {S565} (\bibinfo {year}
  {2007}),\ \Eprint{http://arxiv.org/abs/0707.0128}{arXiv:0707.0128 [gr-qc]}%
  \bibAnnoteFile{NoStop}{Prix.Whelan.2007}%
\bibitem{Corless.etal.1996}%
  \BibitemOpen
  \bibfield{author}{%
  \bibinfo {author} {\bibfnamefont{R.}~\bibnamefont{Corless}}, \bibinfo
  {author} {\bibfnamefont{G.}~\bibnamefont{Gonnet}}, \bibinfo {author}
  {\bibfnamefont{D.}~\bibnamefont{Hare}}, \bibinfo {author}
  {\bibfnamefont{D.}~\bibnamefont{Jeffrey}},\ and\ \bibinfo {author}
  {\bibfnamefont{D.}~\bibnamefont{Knuth}},\ }%
  \bibfield{journal}{%
  \Doi{10.1007/BF02124750}{\bibinfo {journal} {Adv. Comput. Math.}}\ }%
  \textbf{\bibinfo {volume} {5}},\ \bibinfo {pages} {329} (\bibinfo {year}
  {1996})%
  \bibAnnoteFile{NoStop}{Corless.etal.1996}%
\bibitem{DLMF}%
  \BibitemOpen
  \bibfield{author}{%
  \bibinfo {author} {\bibnamefont{{DLMF}}},\ }%
  \enquote{\bibinfo {title} {{Digital Library of Mathematical Functions}},}\
  \bibinfo {howpublished} {National Institute of Standards and Technology from
  \url{http://dlmf.nist.gov}, released July~1} (\bibinfo {year} {2011})%
  \bibAnnoteFile{NoStop}{DLMF}%
\end{thebibliography}%

\end{document}